\documentclass[10pt]{amsart}

\usepackage{amsmath,amssymb,amsthm}
\usepackage{amsmath,amssymb,amsthm,amscd}
\numberwithin{equation}{section}
\title{The theory of superstring with flux on non-K\"ahler manifolds and the
complex Monge-Amp\`ere equation}

\author{Ji-Xiang Fu}
\author{Shing-Tung Yau}
\thanks{}

\address{School of Mathematical Sciences\\ Fudan University \\ Shanghai 200433\\ China\\ and  Key Laboratory of Mathematics for Nonlinear Sciences (Fudan
University)\\ Ministry of Education
\\ China}
\email{majxfu@fudan.edu.cn}
\address{Department of Mathematics\\ Harvard University\\ Cambridge,
MA 02138\\ USA} \email{yau@math.harvard.edu}
\date{}

\newtheorem{prop}{Proposition}
\newtheorem{theo}[prop]{Theorem}
\newtheorem{lemm}[prop]{Lemma}

\newtheorem{rema}[prop]{Remark}

\def\R{{\mathfrak R}}

\def\and{\quad{\rm and}\quad}

\def\dbar{\bar\partial}

\def\sta{^{\ast}}

\def\lab{\label}

\begin{document}

 \maketitle
 \tableofcontents
\section{Introduction}

The purpose of this paper is twofold. The first one is to solve an
old problem posed by Strominger in constructing smooth models of
superstring theory with flux. These are given by non-K\"ahler
manifolds with torsion. To achieve this, we solve a nonlinear
Monge-Amp\`ere equation which is more complicated than the
equation in the Calabi conjecture. The estimate of the volume form
gives extra complication, for example. The second one is to point
out the connection of the newly constructed geometry based on
Strominger's equations in realizing the proposal of M. Reid
\cite{Reid} on connecting one Calabi-Yau manifold to another one
with different topology. In Reid's proposal, the construction of
Clemens-Friedman (see \cite{Fri}) is needed where a Calabi-Yau
manifold is deformed to complex manifolds diffeomorphic to
connected sums of $S^3\times S^3$. These are non-K\"ahler
manifolds.

There is a rich class of non-K\"ahler complex manifolds for
dimension greater than two. It is therefore important to construct
canonical geometry on such manifolds. Since for non-K\"ahler
geometry, the complex structure is not quite compatible with the
Riemannian metric,  it has been difficult to find a reasonable
class of Hermitian metric that exhibit rich geometry. We believe
that metrics motivated by theoretic physics should have good
properties. This is especially true for those metrics which admit
parallel spinors. The work of Strominger did provide such a
candidate. In this paper, we provide  a smooth solution to the
Strominger system. This is an important open problem in the past
twenty years. Our method is based on a priori estimates which can
be generalized to    elliptic fibration over general Calabi-Yau
manifolds. However, in this paper,  for the sake of importance in
string theory, we shall restrict ourselves to complex
three-dimensional manifolds. The structure of the equations for
higher-dimensional Calabi-Yau manifolds are little bit different.
They are also more relevant to algebraic geometry and hence will
be treated in a later occasion.

The physical context of the solutions is
 discussed in a companion  paper  \cite{BBFTY} written jointly
  with K. Becker, M. Becker and L.-S. Tseng.

{\bf Acknowledgement.} The authors would like to thank K. Becker,
M. Becker
 and L.-S. Tseng for useful discussions. J.-X. Fu would also like to thank  J.
 Li and  X.-P. Zhu for useful  discussions. J.-X. Fu is supported in part by
 NSFC grant 10471026. S.-T. Yau
 is  supported in part by NSF grants DMS-0244462, DMS-0354737
 and DMS-0306600.

\section{Motivation from string theory}
In the original proposal for  compactification of superstring
\cite{CHSW}, Candelas, Horowitz, Strominger and Witten constructed
the metric product of a maximal symmetric four-dimensional
spacetime $M$ with a six-dimensional Calabi-Yau vacuum $X$ as the
ten-dimensional spacetime; they identified the Yang-Mills
connection with the $SU(3)$ connection of the Calabi-Yau metric
and set the dilaton to be a constant.  Adapting the second
author's suggestion of using Uhlenbeck-Yau's theorem on
constructing Hermitian-Yang-Mills connections over stable bundles
\cite{UY}, Witten \cite{W1} and later Horava-Witten \cite{HW}
proposed to use higher rank bundles for strong coupled heterotic
string theory so that the gauge groups can be $SU(4)$ or $SU(5)$.

At around the same time, Strominger \cite{Str} analyzed heterotic
superstring background with spacetime supersymmetry and non-zero
torsion by allowing a scalar ``warp factor'' for the spacetime
metric. He considered a ten-dimensional spacetime that is a warped
product of a maximal symmetric four-dimensional spacetime $M$ and
an internal space $X$; the metric on $M\times X$ takes the form
$$g^0=e^{2D(y)}\left(%
\begin{array}{cc}
g_{\mu\nu}(x)&0\\
0&g_{ij}(y)\\
\end{array}
\right),\qquad x\in M,\quad y\in X;
$$
the connection on an auxiliary bundle is Hermitian-Yang-Mills
connection over $X$:
$$F\wedge\omega^2=0,\quad  F^{2,0}=F^{0,2}=0.
$$
Here $\omega$ is the Hermitian form
$\omega=\frac{\sqrt{-1}}{2}g_{i\bar j}dz^i\wedge d\bar z^j$
defined on the internal space $X$. In this system, the physical
relevant quantities are
$$h=-\sqrt{-1}(\bar\partial-\partial)\omega,
$$
$$\phi=-\frac{1}{2}\log\|\Omega\|+\phi_0,
$$
and
$$g_{ij}^0=e^{2\phi_0}\|\Omega\|^{-1}g_{ij},
$$
for a constant $\phi_0$.

In order for the ansatz to provide a supersymmetric configuration,
one introduces a Majorana-Weyl spinor $\epsilon$ so that
$$
\delta \psi_M=\bigtriangledown_M\epsilon-\frac 1 8
h_{MNP}\gamma^{NP}\epsilon=0,
$$
$$
\delta\lambda = \gamma^M\partial_M\phi\epsilon-\frac 1 {12}
h_{MNP}\gamma^{MNP}\epsilon=0,
$$
$$
\delta \chi = \gamma^{MN}F_{MN}\epsilon=0,
$$
where $\psi_M$ is the gravitino, $\lambda$ is the dilatino, $\chi$
is the gluino, $\phi$ is the dilaton and $h$ is the Kalb-Ramond
field strength obeying
$$dh=\frac{\alpha'}{2}(\text{tr} F\wedge F-\text{tr} R\wedge R).
$$
 Strominger \cite{Str} showed that in order to achieve
spacetime supersymmetry, the internal six manifold $X$ must be a
complex manifold with a non-vanishing holomorphic three-form
$\Omega$; and the anomaly cancellation demands that the Hermitian
form $\omega$ obey\footnote{The curvature $F$ of the vector bundle
$E$ in ref.\cite{Str} is real, i.e., $c_1(E)=\frac{F}{2\pi}$.  But
we are used to taking the curvature $F$ such that
$c_1(E)=\frac{\sqrt{-1}}{2\pi}F$. So this equation  corrects eq.
(2.18) of ref. \cite{Str} by a minus sign.}
$$\sqrt{-1}\partial\bar\partial\omega=\frac{\alpha'}{4}(\text{tr} R\wedge R-\text{tr} F\wedge
F)
$$
and supersymmetry requires\footnote{See eq. (56) of
ref.\cite{Str2}, which corrects eq. (2.30) of ref.\cite{Str} by a
minus sign.}
$$d\sta\omega=\sqrt{-1}(\dbar-\partial)\log\|\Omega\|_\omega.
$$
Accordingly, he proposed  the system
\begin{equation}\label{101}
F_H\wedge\omega^2=0;
\end{equation}
\begin{equation}\label{102}
F^{2,0}_H=F^{0,2}_H=0;
\end{equation}
\begin{equation}\label{103}
\sqrt{-1}\partial\dbar \omega= \frac{\alpha'}{4}(\text{tr} R\wedge
R-\text{tr}F_H\wedge F_H);
\end{equation}
\begin{equation}\label{104}
d\sta \omega=\sqrt{-1}(\dbar-\partial)\ln\|\Omega\|_\omega.
\end{equation}
This system gives a solution of a superstring theory with flux
that allows non-trivial dilaton field and Yang-Mills field. (It
turns out  $D(y)=\phi$ and is the dilaton field.) Here $\omega$ is
the Hermitian form and $R$ is the curvature tensor of the
Hermitian metric $\omega$; $H$ is the Hermitian metric and $F$ is
its curvature of a vector bundle $E$; $\text{tr}$ is the trace of
the endomorphism bundle of either $E$ or $TX$.

In \cite{LY}, Li and Yau observed the following:
\begin{lemm}\lab{lemm 1}
Equation (\ref{104})  is equivalent to
\begin{equation}\label{105}
d(\parallel\Omega\parallel_\omega\omega^2)=0.
\end{equation}
\end{lemm}
In fact, Li and Yau gave  the first irreducible non-singular
solution of the supersymmetric  system of Strominger for $U(4)$
and $U(5)$ principle bundle. They obtained their solutions by
perturbing around the Calabi-Yau vacuum coupled with the sum of
tangent bundle and trivial line bundles. In this paper, we
consider the solution on complex manifolds which do {\sl not}
admit K\"ahler structures. Study of non-K\"{a}hler manifold should
be useful to understand the speculation of M. Reid that all
Calabi-Yau manifolds can be deformed to each other through
conifold transition.

An example of  non-K\"ahler manifolds $X$ is given by
$T^2$-bundles over Calabi-Yau varieties
\cite{BBDG,BBDP,GP,GLM,KST}. Since we demand that the internal six
manifold $X$ is  a complex manifold with a non-vanishing
holomorphic three form $\Omega$,  we  consider the $T^2-$bundle
$(X,\omega,\Omega)$ over a complex surface $(S,\omega_S,\Omega_S)$
with a non-vanishing holomorphic 2-form $\Omega_S$. According to
the classification of complex surfaces by Enriques and Kodaira,
such complex  surfaces must be finite quotients of  K3 surface,
complex torus (K\"ahler) and Kodaira surface (non-K\"ahler). If
$(X,\omega,\Omega)$ satisfies  Strominger's equation (\ref{104}),
Lemma 1 shows that $d(\parallel\Omega\parallel_\omega\omega^2)=0$.
Let $\omega'=\parallel\Omega\parallel_\omega^{\frac{1}{2}}\omega$.
Then $d\omega'^2=0$, i.e., $\omega'$ is a balanced metric
\cite{MI}. The balanced metric  was studied extensively by
Michelsohn. She  proved that the balanced condition  is preserved
under proper holomorphic submersions.  Note that Alessandrini and
Bassanelli \cite{AB} proved that this condition is also preserved
under modifications of complex manifolds. Hence  if a holomorphic
submersion $\pi$ from a balanced manifold $X$ to a complex surface
$S$ is proper, $S$ is also balanced (actually $\pi_\ast\omega'^2$
is the balanced metric on $S$, see proposition 1.9 in \cite{MI}).
When the dimension of complex manifold is two, the conditions of
being balanced and K\"{a}hler coincide. Hence there is no solution
to Strominger's equation (1.4) on $T^2$ bundles over Kodaira
surface and we  consider  $T^2$-bundles over $K3$ surface and
complex torus only.

On the other hand,   duality  from $M$-theory suggests that there
is no  supersymmetric solution when the base manifold is a complex
torus (see \cite{BBFTY}). This class of three manifolds   includes
the Iwasawa manifold. But the  solution to Strominger's system
should exist when the base is $K3$ surface.  In this paper we do
prove the existence of solutions to Strominger's system on such
torus bundles over  $K3$ surfaces.

\section{Statement of main result}
Let $(S,\omega_S,\Omega_S)$ be a K3 surface or a complex
 torus with  a K\"ahler form $\omega_S$ and a non-vanishing holomorphic
(2,0)-form $\Omega_S$. Let $\omega_1$ and $\omega_2$ be
anti-self-dual (1,1)-forms such that $\frac{\omega_1}{2\pi}$ and
$\frac{\omega_2}{2\pi}$ represent integral cohomology classes.
Using these two forms, Goldstein and Prokushkin \cite{GP}
constructed a non-K\"ahler manifold $X$ such that
$\pi:X\rightarrow S$ is a holomorphic $T^2$-fibration over $S$
with a Hermitian form
$\omega_0=\pi^*\omega_S+\frac{\sqrt{-1}}{2}\theta\wedge\bar\theta$
and a holomorphic (3,0)-form $\Omega=\Omega_S\wedge\theta$ (for
the definition of $\theta$, see section 3). Note that
$(\omega_0,\Omega)$ satisfies  equation (\ref{105}).

Let $u$ be any smooth function on $S$ and let
\begin{equation*}
\omega_u=\pi^*(e^u\omega_S)+\frac{\sqrt{-1}}{2}\theta\wedge\bar\theta.
\end{equation*}
Then $(\omega_u,\Omega)$ also satisfies equation (\ref{105}) (see
\cite{GP} or Lemma \ref{lemm 10}), i.e.,  $\omega_u$ is conformal
balanced.  The stability concept can be defined on a vector bundle
over a complex manifold using the  Gauduchon metric \cite{LY1},
and hence for  complex manifolds with balanced metrics. Note that
the stability concept of the vector bundle depends only on the
conformal class of metric.  Let $V\rightarrow X$ be a stable
bundle over $X$ with degree zero with respect to the metric
$\omega_u$. (Such bundles can be obtained by pulling back stable
bundles over a $K3$ surface or a complex torus,  see Lemma 16.)
According to Li-Yau's theorem \cite{LY1}, there is a
Hermitian-Yang-Mills metric $H$ on $V$, which is unique up to
positive constants. The curvature $F_H$ of the Hermitian metric
$H$ satisfies equation (\ref{101}) and (\ref{102}). So $(V,
F_{H},X,\omega_u)$ satisfies  Strominger's equations (\ref{101}),
(\ref{102}) and (\ref{104}). Therefore we   only need  to consider
 equation (\ref{103}). As $\omega_1$ and $\omega_2$ are harmonic,
$\bar\partial\omega_1=\bar\partial\omega_2=0$. According to
$\bar\partial$-Poincar\'e Lemma,  we can write $\omega_1$ and
$\omega_2$ locally as
$$\omega_1=\bar\partial\xi=\bar\partial(\xi_1dz_1+\xi_2dz_2)$$
and
$$\omega_2=\bar\partial\zeta=\bar\partial(\zeta_1dz_1+\zeta_2dz_2),$$
where $(z_1,z_2)$ is a local coordinate on $S$. Let
 $$B=\left(%
\begin{array}{c}
   \xi_1+\sqrt{-1}\zeta_1\\
  \xi_2+\sqrt{-1}\zeta_2  \\
\end{array}%
\right).$$ We can use $B$ to compute $\text{tr}R_0\wedge R_0$ of
the metric $\omega_0$ (see Proposition \ref{prop 6}) and
$\text{tr}R_u\wedge R_u$ of the metric $\omega_u$ (see Lemma
\ref{lemm 12}). Then We  reduce equation (\ref{103}) to
\begin{equation}\lab{107}
\begin{aligned}
&\ \ \sqrt{-1}\partial\bar\partial e^u\wedge
\omega_S-\frac{\alpha'}{2}\partial\dbar(e^{-u}\text{tr}(\dbar
B\wedge
\partial B^*\cdot g^{-1}))-\frac{\alpha'}{2}\partial\dbar u\wedge
\partial\dbar u\\
=&\ \ \frac{\alpha'}{4}\text{tr} R_S\wedge
R_S-\frac{\alpha'}{4}\text{tr}F_{H}\wedge F_{H}-\frac 1 2
(\parallel
\omega_1\parallel_{\omega_S}^2+\parallel\omega_2\parallel_{\omega_S}^2)\frac
{\omega_S^2}{2!},
\end{aligned}
\end{equation}
where $g=(g_{i\bar j})$ is the Ricci-flat metric on $S$ associated
to the K\"{a}hler form $\omega_S$ and $g^{-1}$ is the inverse
matrix of $g$; $R_S$ is the curvature of $g$. Taking wedge product
with $\omega_u$ and  integrating both sides of the  above equation
over $X$, we obtain
\begin{equation}\lab{108}
\alpha'\int_X\{\text{tr} R_S\wedge R_S-\text{tr}F_{H}\wedge
F_{H}\}\wedge \omega_u-2 \int_X(\parallel
\omega_1\parallel_{\omega_S}^2+\parallel\omega_2\parallel_{\omega_S}^2)\frac
{\omega_S^2}{2!}\wedge \omega_u=0.
\end{equation}
When $S=T^4$, $R_S=0$. We obtain immediately
\begin{prop}
There is no solution of Strominger's system on the torus bundle
$X$ over $T^4$ if  the metric has the form $e^u\omega_S+\frac
{\sqrt{-1}}{2}\theta\wedge \bar\theta$.
\end{prop}

This situation is different if the base is a $K3$ surface. If $E$
is a stable bundle over $S$ with degree 0 with respect to the
metric $\omega_S$, then $V=\pi^*E$ is also a stable bundle with
degree 0 over $X$ with respect to the  Hermitian metric
$\omega_u$. In this case, equation (\ref{107}) on $X$ can be
considered as an equation on $S$. Integrating equation (\ref{107})
over $S$,
\begin{equation}\lab{109}
\alpha'\int_S\{\text{tr} R_S\wedge R_S-\text{tr}F_{H}\wedge
F_{H}\}=2\int_S(\parallel
\omega_1\parallel_{\omega_S}^2+\parallel\omega_2\parallel_{\omega_S}^2)\frac
{\omega_S^2}{2!}.
\end{equation}
 As $\int_S\text{tr} R_S\wedge
R_S=8\pi^2c_2(V)=8\pi^2\times 24$, and $\int_S\text{tr} F_H\wedge
F_H=8\pi^2\times (c_2(E)-\frac 1 2 c_1^2(E))\geq 0$,  we can
rewrite  equation (\ref{109}) as
\begin{equation}\lab{110}
\alpha'(24-(c_2(E)-\frac 1 2 c_1^2(E)))=\int_S(\parallel\frac
{\omega_1}{2\pi}\parallel_{\omega_S}^2+\parallel\frac
{\omega_2}{2\pi}\parallel_{\omega_S}^2)\frac {\omega_S^2}{2!}.
\end{equation}
 For a compact, oriented, simply connected four-manifold $S$, the
Poincar\'e duality gives rise to a pairing
\begin{equation*}
Q:H_2(S;\mathbb{Z})\times H_2(S;\mathbb{Z})\rightarrow \mathbb{Z}
\end{equation*}
defined by
\begin{equation*}
Q(\beta,\gamma)=\int_S\beta\wedge\gamma.
\end{equation*}
 We shall  denote $Q(\beta,\beta)$ by $Q(\beta)$. Then  for an integral
anti-self-dual (1,1)-form $\frac{\omega_1}{2\pi}$, the
intersection number $Q(\frac{\omega_1}{2\pi})$ can be expressed as
$-\int_S\parallel\frac
{\omega_1}{2\pi}\parallel^2\frac{\omega_S^2}{2!}$. On the other
hand, the intersection form on $K3$ surface is given by \cite{Do}
\begin{equation*}
3\left(\begin{array}{cc}0& 1\\
1& 0\end{array}\right)\oplus 2(-E_8),
\end{equation*}
where
\begin{equation*}
E_8=\left(\begin{array}{cccccccc}2& 0& -1&  & &&&\\
0&2&0&-1& &&&\\
-1& 0& 2&-1&&&&\\
&-1&-1&2&-1&&&\\
&&&-1&2&-1&&\\
&&&&-1&2&-1&\\
&&&&&-1&2&-1\\
&&&&&&-1&2\end{array}\right).
\end{equation*}
Hence   $Q(\frac{\omega_1}{2\pi})\in \{-2,-4,-6,\cdots\}$.

We shall use the following convention for vector bundles over a
compact oriented four-manifold:
\begin{equation*}
\begin{aligned}
\kappa(E)&=c_2(E)  \ \ \ \ \ \ \ \ \ \ \ \ \ \ \ \ \  \text{for $SU(r)$ bundle $E$},\\
&=c_2(E)-\frac 1 2 c_1^2(E)\ \ \ \ \text{for $U(r)$ bundle $E$},\\
&=-\frac 1 2 p_1(E)\ \ \ \ \ \ \ \ \ \ \ \ \ \text{for $SO(r)$
bundle $E$}.
\end{aligned}
\end{equation*}
Then (\ref{110}) implies
\begin{equation}\lab{1114}
\alpha'(24-\kappa(E))+\left(Q\left(\frac
{\omega_1}{2\pi}\right)+Q\left(\frac{\omega_2}{2\pi}\right)\right)=0,
\end{equation}
which means  that there is a smooth function $\mu$ such that
\begin{equation}\lab{111}
\frac{\alpha'}{4}\text{tr} R_S\wedge
R_S-\frac{\alpha'}{4}\text{tr}F_{H}\wedge F_{H}-\frac 1 2
(\parallel
\omega_1\parallel^2+\parallel\omega_2\parallel^2)\frac{\omega_S^2}{2!}=-\mu\frac
{\omega_S^2}{2!}
\end{equation}
and $\int_S\mu\frac {\omega^2}{2!}=0$. Inserting (\ref{111}) into
(\ref{107}), we obtain the following equation:
\begin{equation}\lab{112}
\sqrt{-1}\partial\bar\partial
e^u\wedge\omega_S-\frac{\alpha'}{2}\partial\dbar(e^{-u}\text{tr}(\dbar
B\wedge
\partial B^*\cdot g^{-1}))-\frac{\alpha'}{2}\partial\dbar u\wedge
\partial\dbar u+\mu\frac {\omega_S^2}{2!}=0,
\end{equation}
where $\text{tr}(\dbar B\wedge \partial B^*\cdot g^{-1})$ is a
smooth well-defined $(1,1)$-form on $S$. In particular, when
$\omega_2=n\omega_1$, $n\in \mathbb{Z}$,
\begin{equation*}
\text{tr}(\bar \partial B\wedge \partial B^*\cdot
g^{-1})=\sqrt{-1}\frac
{1+n^2}{4}\parallel\omega_1\parallel^2_{\omega_S}\omega_S
\end{equation*}
(see Proposition \ref{prop 9}). Hence  if we set $f=\frac
{1+n^2}{4}\parallel\omega_1\parallel_{\omega_S}^2$, we can rewrite
equation (\ref{112}) as the standard complex Monge-Amp\`ere
equation:
\begin{equation}\lab{114}
\Delta(e^{u}-\frac{\alpha'}{2} f e^{-u})+4\alpha' \frac{\det
u_{i\bar j}}{\det g_{i\bar j}}+\mu=0,
\end{equation}
where  $u_{i\bar j}$ denotes $\frac {\partial^2u}{\partial
z_i\partial\bar z_j}$ and $\bigtriangleup=2g^{i\bar
j}\frac{\partial^2}{\partial z_i\partial \bar z_j}$. We shall
solve equation (\ref{112}) by the continuity method \cite{Yau}.
Our main theorem is
\begin{theo}
The  equation (\ref{112}) has a smooth solution $u$ such that
$$\omega'=e^u\omega_S-\frac{\alpha'}{2} \sqrt{-1}e^{-u} \textup{tr}(\bar\partial
B\wedge\partial B^*\cdot
g^{-1})+\alpha'\sqrt{-1}\partial\bar\partial u$$ defines a
Hermitian metric on $S$.
\end{theo}
Our solution $u$ satisfies $\left(\int_S e^{-4u}\right)^{\frac 1
4}=A<<1$. Actually we can prove that $\inf u\geq  -\ln (C_1A)$
(see Proposition \ref{prop 2}) where  $A$ must be {\sl very small}
(see Proposition \ref{prop 3}) and  our solution $u$ must be {\sl
very big}.
\begin{theo}\lab{theo 2}
Let $S$ be a $K3$ surface with a Ricci-flat metric $\omega_S$. Let
$\omega_1$ and $\omega_2$ be anti-self-dual $(1,1)$-forms on $S$
such that $\frac{\omega_1}{2\pi},\frac{\omega_2}{2\pi}\in
H^2(S,\mathbb{Z})$. Let $X$ be
 a $T^2$-bundle over $S$ constructed by $\omega_1$ and $\omega_2$.
 Let $E$ be a stable bundle over $S$ with degree 0.
 Suppose $\omega_1$, $\omega_2$ and $\kappa(E)$ satisfy
 condition (\ref{1114}). Then there exist  a smooth function $u$ on $S$ and a
Hermitian-Yang-Mills metric $H$ on $E$ such that
 $(V=\pi^*E,\pi^*F_H, X, \omega_u)$ is a solution of Strominger's
 system.
 \end{theo}
 Since it is easy to find $(\omega_1,\omega_2, \kappa(E))$ which satisfies
 condition (\ref{1114}),  this theorem  provides first examples of
 solutions to Strominger's
 system on non-K\"{a}hler manifold.

\section{Geometric model} In this section, we take the geometric model of
Goldstein and Prokushkin  for complex non-K\"ahler manifolds with
an $SU(3)$ structure \cite{GP}. We summarize  their   results as
follows:
\begin{theo}\lab{theo 3}\cite{GP}
Let $(S,\omega_S,\Omega_S)$ be a Calabi-Yau\ 2-fold with  a
non-vanishing holomorphic $(2,0)-$form $\Omega_S$. Let $\omega_1$
and $\omega_2$ be anti-self-dual $(1,1)$-forms on $S$ such that
$\frac{\omega_1}{2\pi}\in H^2(S,\mathbb{Z})$ and
$\frac{\omega_2}{2\pi}\in H^2(S,\mathbb{Z})$. Then there is a
Hermitian 3-fold $X$ such that $\pi:X\rightarrow
S$ is a holomorphic $T^2$-fibration over $S$ and the  following holds: \\
1. For any real 1-forms $\alpha_1$ and $\alpha_2$ defined on some
open subset of $S$ that  satisfy $d\alpha_1=\omega_1$ and
$d\alpha_2=\omega_2$, there are local coordinates $x$ and $y$ on
$X$ such that $dx+idy$ is a holomorphic form on $T^2$-fibers and a
metric on $X$ has the following form:
\begin{equation}\lab{202}
g_0=\pi^*g+(dx+\pi^*\alpha_1)^2+(dy+\pi^*\alpha_2)^2,
\end{equation}
where $g$ is a Calabi-Yau metric on $S$ corresponding to the
K\"ahler form
$\omega_S$.\\
2. $X$ admits a nowhere vanishing holomorphic $(3,0)$-form with
unit length:
$$\Omega=((dx+\pi^*\alpha_1)+i(dy+\pi^*\alpha_2))\wedge\pi^*\Omega_S.$$
3. If either $\omega_1$ or $\omega_2$ represents a non-trivial
cohomological class then $X$ admits no K\"ahler metric. \\
4.  $X$  is a balanced manifold. The Hermitian form
\begin{equation}\label{2011}
\omega_0=\pi^*\omega_S+(dx+\pi^*\alpha_1)\wedge(dy+\pi^*\alpha_2)
\end{equation}
corresponding to the metric (\ref{202}) is balanced, i.e., $d\omega^2_0=0$.\\
5. Furthermore,  for any smooth function $u$ on $S$, the Hermitian
metric
\begin{equation*}
\omega_u=\pi^*(e^u\omega_S)+(dx+\pi^*\alpha_1)\wedge(dy+\pi^*\alpha_2)
\end{equation*}
is conformal balanced. Actually  $(\omega_u,\Omega)$ satisfies
equation (\ref{105}).
\end{theo}

Goldstein and Prokushkin also  studied the cohomology of this
non-K\"{a}hler manifold $X$:
$$h^{1,0}(X)=h^{1,0}(S),$$
$$h^{0,1}(X)=h^{0,1}(S)+1;$$
In particular
$$h^{0,1}(X)=h^{1,0}(X)+1.$$
Moreover,
\begin{eqnarray*}
b_1(X)&=&b_1(S)+1,\ \ \text{when}\ \ \omega_2=n\omega_1,\\
b_1(X)&=&b_1(S),\ \ \ \ \ \ \ \text{when}\ \ \omega_2\neq
n\omega_1;\\
 b_2(X)&=&b_2(S)-1,\ \ \text{when}\ \
\omega_2=n\omega_1,\\
b_2(X)&=&b_2(S)-2,\ \  \text{when}\ \ \omega_2\neq n\omega_1
\end{eqnarray*}
and $$ \chi(X)=0.$$

The  above topological results can be explained as follows. Let
$L_1$ be a holomorphic line bundle over $S$ with the first Chern
class $c_1(L_1)=[-\frac{\omega_1}{2\pi}]$. Then we can choose a
Hermitian metric $h_1$ on $L_1$ such that its curvature is
$\sqrt{-1}\omega_1$. Let $S_1=\{v\in L_1\mid h_1(v,v)=1\}$ which
is a  circle bundle over $S$. Locally we write
$\omega_1=d\alpha_{1U}$ for some real 1-form $\alpha_{1U}$ on some
open subset $U$ on $S$. Such $\alpha_{1U}$ define a connection on
$S_1$, i.e., there is a section $\xi_U$ on $S_1$ such that
$$\bigtriangledown\xi_U=\sqrt{-1}\alpha_{1U}\otimes\xi_U.$$
The section $\xi_U$ defines a local coordinate $x_U$ on  fibers of
$S_1\mid_{U}$, i.e., we can describe the circle $S^1$ by
$e^{\sqrt{-1}x_U}\xi_U$. If we write $\omega_1=d\alpha_{1V}$ on
another open set $V$ of $S$, then there is another section $\xi_V$
such that
\begin{equation}\lab{204}
\bigtriangledown\xi_V=\sqrt{-1}\alpha_{1V}\otimes\xi_V
\end{equation}
and this section $\xi_V$ defines another coordinate $x_V$ on fiber
of $S_1\mid_{V}$. On  $U\cap V$,  $d(\alpha_{1U}-\alpha_{1V})=0$
and  there is a function $f_{UV}$ such that
\begin{equation}\lab{20005}
df_{UV}=\alpha_{1U}-\alpha_{1V} .
\end{equation}
On the other hand, on $U\cap V$, there is also a function $g_{UV}$
on $U\cap V$ such that $\xi_V=e^{\sqrt{-1}g_{UV}}\xi_U$. We
compute
\begin{equation*}
\begin{aligned}
\bigtriangledown\xi_V=&\bigtriangledown(e^{\sqrt{-1}g_{UV}}\xi_U)\\
=&(\sqrt{-1}dg_{UV}+\sqrt{-1}\alpha_{1U})\otimes(e^{\sqrt{-1}g_{UV}}\xi_U)\\
=&(\sqrt{-1}dg_{UV}+\sqrt{-1}\alpha_{1U})\otimes\xi_V.
\end{aligned}
\end{equation*}
Comparing the above equality with (\ref{204}), we get
\begin{equation}\lab{207}
-dg_{UV}=\alpha_{1U}-\alpha_{1V}.
\end{equation}
So combining (\ref{20005}). we find
\begin{equation}
g_{UV}=f_{UV}+c_{UV},
\end{equation}
where $c_{UV}$ is some constant on $U\cap V$.
 On $U\cap
V$, from
$$e^{ix_U}\xi_U=e^{ix_V}\xi_V=e^{\sqrt{-1}x_V}e^{\sqrt{-1}g_{UV}}\xi_U,$$
we obtain
\begin{equation}\lab{208}
x_U=x_V+g_{UV}+2k\pi=x_V+f_{UV}+c_{UV}+2k\pi.
\end{equation}
(\ref{20005}) and (\ref{208}) imply
\begin{equation}
dx_U-dx_V=d f_{UV}=-\alpha_{1U}+\alpha_{1V}.
\end{equation}
So $dx_U+\alpha_{1U}$ is a globally defined  1-form on $X$. We
denote it by $dx+\alpha_1$.

We  construct another line bundle $L_2$ with the  first Chern
class $[-\frac{\omega_2}{2\pi}]$. Similarly,
 we write locally $\omega_2=d\alpha_2$, and   define a
coordinate $y$ on fibers such that $dy+\alpha_2$ is a well-defined
1-form  on the circle bundle $S_1$ of $L_2$. On $X$,
$\omega_1=d(dx+\alpha_1)$ and $\omega_2=d(dy+\alpha_2)$, and so
$[\omega_1]=[\omega_2]=0\in H^2(X,\mathbb{R})$. When
$\omega_2=n\omega_1$,  $d(n(dx+\alpha_1)-(dy+\alpha_2))=0$. So
$[n(dx+\alpha_1)-(dy+\alpha_2)]\in H^1(X,\mathbb{R})$. Finally  we
define
\begin{equation*}
\theta=dx+\alpha_1+\sqrt{-1}(dy+\alpha_2).
\end{equation*}
Then $\theta$ is a $(1,0)$-form on $X$ , see \cite{GP} or the next
section. Because $d\bar\theta=\omega_1-\sqrt{-1}\omega_2$ is a
$(1,1)$-form on $X$,  its $(0,2)$-component
$\bar\partial\bar\theta=0$. So $[\bar\theta]\in
H^{0,1}_{\bar\partial}(X)\cong H^1(X,\mathcal{O})$.

\section{The calculation of $\text{tr}R\wedge R$}
In order to calculate the curvature $R$ and $\text{tr}R\wedge R$,
we  express the Hermitian metric (\ref{202}) in terms of a basis
of holomorphic (1,0) vector fields. Hence  we need to  write down
the complex structure on $X$. Let
$\{U,z_j=x_j+\sqrt{-1}y_j,j=1,2\}$ be a local coordinate in $S$.
The horizontal lifts of vector fields $\frac{\partial}{\partial
x_j}$ and $\frac{\partial}{\partial y_j}$, which are in the kernel
of $dx+\pi^*\alpha_1$ and $dy+\pi^*\alpha_2$, are
\begin{equation*}
X_j=\frac{\partial}{\partial
x_j}-\alpha_1\left(\frac{\partial}{\partial
x_j}\right)\frac{\partial}{\partial
x}-\alpha_2\left(\frac{\partial}{\partial
x_j}\right)\frac{\partial}{\partial y}\ \ \ \ \text{for} \ \
j=1,2,
\end{equation*}
\begin{equation*}
Y_j=\frac{\partial}{\partial
y_j}-\alpha_1\left(\frac{\partial}{\partial
y_j}\right)\frac{\partial}{\partial
x}-\alpha_2\left(\frac{\partial}{\partial
y_j}\right)\frac{\partial}{\partial y}\ \ \ \ \text{for} \ \
j=1,2.
\end{equation*}
The complex structure $\tilde{I}$ on $X$ is defined as
\begin{eqnarray*}
\tilde{I}X_j&=&Y_j,\ \ \ \  \tilde{I}Y_j=-X_j, \ \ \ \ \text{for}
\ \ j=1,2,\\
\tilde{I}\frac{\partial}{\partial x}&=&\frac{\partial}{\partial
y}, \ \ \ \ \tilde{I}\frac{\partial}{\partial
y}=-\frac{\partial}{\partial x}.
\end{eqnarray*}
Let
\begin{eqnarray*}
U_j&=&X_j-\sqrt{-1}\tilde{I}X_j=X_j-\sqrt{-1}Y_j,\\
U_0&=&\frac{\partial}{\partial x}
-\sqrt{-1}\tilde{I}\frac{\partial}{\partial x}
=\frac{\partial}{\partial x}-\sqrt{-1}\frac{\partial}{\partial y}.
\end{eqnarray*}
Then $\{U_j,U_0\}$ is the basis of the $(1,0)$ vector fields on
$X$. The metric (\ref{202}) takes the following Hermitian form:
\begin{equation}\label{203}
\left(\begin{array} {cc} (g_{i\bar j}) & 0\\
0 & 1
\end{array}\right)
\end{equation}
as $U_1$ and $U_2$ are in the kernel of $dx+\pi^*\alpha_1$ and
$dy+\pi^*\alpha_2$. Let
\begin{equation}\label{2012}
 \theta=dx+\sqrt{-1}dy+\pi^*(\alpha_1+\sqrt{-1}\alpha_2).
 \end{equation}
It's easy to check that $\{\pi^*d\overline{z}_j,\overline\theta\}$
annihilates the $\{U_j,U_0\}$ and  is the basis of  $(0,1)$-forms
on $X$. So $\{\pi^*dz_j,\theta\}$ are $(1,0)$-forms on $X$.
Certainly $\pi^*dz_j$ are holomorphic $(1,0)$-forms and $\theta$
is not. We need to construct another holomorphic $(1,0)$-form on
$X$. Because $\omega_1$ and $\omega_2$ are harmonic forms on $S$,
$\overline\partial\omega_1=\overline\partial\omega_2=0$. By
$\bar\partial$-Poincar\'e Lemma, locally we can find $(1,0)$-forms
$\xi=\xi_1dz_1+\xi_2dz_2$ and $\zeta=\zeta_1dz_1+\zeta_2dz_2$ on
$S$ , where $\xi_i$ and $\zeta_j$ are smooth complex functions on
some open set of $S$, such that $\omega_1=\overline
\partial\xi$ and $\omega_2=\overline\partial\zeta$. Let
\begin{eqnarray*}
\theta_0&=&\theta-\pi^*(\xi+\sqrt{-1}\zeta)\\
&=&(dx+\sqrt{-1}dy)+\pi^*(\alpha_1+\sqrt{-1}\alpha_2)-\pi^*(\xi+\sqrt{-1}\zeta).
\end{eqnarray*}
We claim that $\theta_0$ is a holomorphic $(1,0)$-form. By our
construction, $\theta_0$ is the $(1,0)$-form.   But
$d\theta=d(dx+\sqrt{-1}dy+\pi^*(\alpha_1+\sqrt{-1}\alpha_2))=\pi^*(\omega_1+\sqrt{-1}\omega_2)$
is a  $(1,1)$-form on $X$. So
\begin{equation}\lab{2004}
\partial\theta=0\ \
\text{and}\ \
\overline\partial\theta=d\theta=\pi^*(\omega_1+i\omega_2).
\end{equation}
Thus
\begin{eqnarray*}
\overline\partial\theta_0&=&\overline\partial\theta-\overline\partial\pi^*(\xi+\sqrt{-1}\zeta)\\
&=&\pi^*(\omega_1+\sqrt{-1}\omega_2)-\pi^*(\omega_1+\sqrt{-1}\omega_2)=0.
\end{eqnarray*}
So $\theta_0$ is a holomorphic $(1,0)$-form and
$\{\pi^*dz_j,\theta_0\}$ forms a basis of holomorphic
$(1,0)$-forms on $X$.   Let
$$\varphi_j=\xi_j+\sqrt{-1}\zeta_j\ \ \ \ \text{for}\ \ \ \ j=1,2$$
and
$$\tilde U_j=U_j+\varphi_jU_0 \ \ \ \ \text{for}\ \ \ \ j=1,2.$$
Then  $\{\tilde U_j,U_0\}$ is dual to $\{\pi^*dz_j,\theta_0\}$
because  $U_j$ is in the kernel of $\theta$. It's the basis of
holomorphic $(1,0)$-vector fields.  The metric $g_0$ then becomes
 the following Hermitian matrix:
\begin{eqnarray}\label{205}
H_X=\left(\begin{array}{ccc} g_{1\bar 1}+\mid \varphi_1\mid^2 &
g_{1\bar2}+\varphi_1\overline \varphi_2& \varphi_1\\
g_{2\bar1}+\varphi_2\overline \varphi_1 & g_{2\bar2}+\mid \varphi_2\mid^2 & \varphi_2\\
\overline \varphi_1& \overline \varphi_2 &
1\end{array}\right)=\left(\begin{array}{cc} g+B\cdot B^* & B\\
B^* & 1 \end{array}\right),
\end{eqnarray}
where $ g$ is the Calabi-Yau metric on $S$ and
$B=(\varphi_1,\varphi_2)^t$.

According to Strominger's explanation in \cite{Str}, when the
manifold is not K\"ahler, we should take the curvature of
Hermitian connection on the holomorphic tangent bundle $T'X$.
Using the metric (\ref{205}), we  compute the curvature to be
$$ R=\overline\partial(\partial H_X\cdot
H_X^{-1})=\left(\begin{array} {cc} R_{1\bar 1}&
R_{1\bar2}\\R_{2\bar1} & R_{2\bar2}
\end{array}\right),$$
where
\begin{eqnarray*}
R_{1\bar1}&=&R_S+\overline\partial B\wedge(\partial B^*\cdot
g^{-1})+B\cdot \overline\partial(\partial B^*\cdot g^{-1}),\\
R_{1\bar2}&=&-R_SB+(\partial g\cdot g^{-1})\wedge\overline
\partial B-\overline\partial B\wedge(\partial B^*\cdot
g^{-1})B,\\
&&-B\overline\partial(\partial B^*\cdot g^{-1})B+B(\partial
B^*\cdot g^{-1})\wedge\overline\partial
B+\overline\partial\partial B,\\
R_{2\bar1}&=& \overline\partial(\partial B^*\cdot g^{-1}),\\
R_{2\bar2}&=& -\overline\partial(\partial B^*\cdot
g^{-1})B+(\partial B^*\cdot g^{-1})\wedge \overline
\partial B,
\end{eqnarray*}
and $R_S$ is the curvature of Calabi-Yau metric $g$ on $S$. It is
easy to check that $\text{tr}(\overline\partial B\wedge(\partial
B^*\cdot g^{-1})+B\cdot \overline\partial(\partial B^*\cdot
g^{-1}))-\overline\partial(\partial B^*\cdot g^{-1})B+(\partial
B^*\cdot g^{-1})\wedge \overline
\partial B=0$. So $\text{tr}R=\pi^*\text{tr}R_S$.
\begin{prop}\lab{prop 4}\cite{GP2}
The Ricci forms of the Hermitian  connections on $X$ and $S$ have
the relation $\textup{tr}R=\pi^*\textup{tr}R_S$.
\end{prop}
\begin{rema}
In the above calculation, we don't use the condition that the
metric $g$ on $S$ is Calabi-Yau. \end{rema}
\begin{prop}\lab{prop 6}
\begin{equation}\label{301}
\textup{tr} R\wedge R=\pi^*(\textup{tr}R_S\wedge
R_S+2\textup{tr}\partial\overline\partial(\overline\partial
B\wedge
\partial B^*\cdot g^{-1})).
\end{equation}
\end{prop}
\begin{proof}
 Fix any point $p\in S$, we  pick $B$ such that $B(p)=0$.
Otherwise,  $B(p)\neq 0$ and  we simply  replace $B$ by $B-B(p)$.
Hence in the calculation of $\text{tr} R\wedge R$ at  $p$, all
terms containing the factor $B$ will vanish. Thus
\begin{eqnarray*}
&&\text{tr} R\wedge R\\
&=&\text{tr} R_S\wedge R_S+2\text{tr}R_S\wedge \overline\partial
B\wedge(\partial B^*\cdot g^{-1})\\
&&+2\text{tr}\partial g\cdot g^{-1}\wedge\overline\partial
B\wedge\dbar(\partial B^*\cdot g^{-1})+2\text{tr}\overline
\partial \partial B\wedge\overline\partial(\partial B^*\cdot g^{-1})\\
&&+\text{tr}\overline\partial B\wedge((\partial B^*\cdot
g^{-1})\wedge\overline\partial B\wedge(\partial B^*\cdot
g^{-1}))\\
&&+((\partial B^*\cdot g^{-1})\wedge\overline \partial
B\wedge(\partial B^*\cdot g^{-1}))\wedge\overline\partial B\\
&=&\text{tr} R_S\wedge R_S+2\text{tr}R_S\wedge
\overline\partial B\wedge(\partial B^*\cdot g^{-1})\\
&&+2\text{tr}\partial g\cdot g^{-1}\wedge\overline\partial
B\wedge\dbar(\partial B^*\cdot g^{-1})+2\text{tr}\overline
\partial\partial B\wedge\overline\partial(\partial B^*\cdot
g^{-1}).
\end{eqnarray*}
 Proposition 8 follows from the next two lemmas.
\end{proof}

\begin{lemm}\lab{lemm 7}
\begin{eqnarray*}
\textup{tr}\partial\overline\partial(\overline\partial B\wedge
\partial B^*\cdot g^{-1})&=&
\textup{tr}R_S\wedge
\overline\partial B\wedge(\partial B^*\cdot g^{-1})\\
&&+\textup{tr}\partial g\cdot g^{-1}\wedge\overline\partial
B\wedge \dbar(\partial
B^*\cdot g^{-1})\\
&&+\textup{tr}\overline
\partial\partial B\wedge\overline\partial(\partial B^*\cdot
g^{-1}).
\end{eqnarray*}
\end{lemm}
\begin{proof}
\begin{eqnarray*}
&&\text{tr}\partial\overline\partial(\overline\partial B\wedge
\partial B^*\cdot g^{-1})\\
&=&-\text{tr}\partial(\overline\partial B\wedge
\overline\partial(\partial
B^*\cdot g^{-1}))\\
&=&\text{tr}\overline\partial\partial B\wedge
\overline\partial(\partial B^*\cdot
g^{-1})+\text{tr}\overline\partial B\wedge
\partial\overline\partial(\partial B^*\cdot g^{-1})\\
&=&\text{tr}\overline\partial\partial B\wedge
\overline\partial(\partial B^*\cdot
g^{-1})+\text{tr}\overline\partial B\wedge
\overline\partial (\partial B^*\wedge\partial g^{-1})\\
&=&\text{tr}\overline\partial\partial B\wedge
\overline\partial(\partial B^*\cdot
g^{-1})-\text{tr}\overline\partial B\wedge
\overline\partial(\partial B^*\cdot g^{-1}\wedge\partial g\cdot g^{-1})\\
&=&\text{tr}\overline\partial\partial B\wedge
\overline\partial(\partial B^*\cdot
g^{-1})-\text{tr}\overline\partial B\wedge
\overline\partial(\partial B^*\cdot g^{-1})\wedge\partial g\cdot
g^{-1}\\
&&+\text{tr}\overline\partial B\wedge
(\partial B^*\cdot g^{-1})\wedge\dbar(\partial g\cdot g^{-1})\\
&=&\text{tr}\overline\partial\partial B\wedge
\overline\partial(\partial B^*\cdot
g^{-1})-\text{tr}\overline\partial B\wedge
\overline\partial(\partial B^*\cdot g^{-1})\wedge\partial g\cdot
g^{-1}\\
&&+\text{tr}\overline\partial B\wedge
(\partial B^*\cdot g^{-1})\wedge R_S\\
&=&\text{tr}(\overline\partial\partial B\wedge
\overline\partial(\partial B^*\cdot
g^{-1}))+\text{tr}(R_S\wedge\overline\partial B\wedge
\partial B^*\cdot
g^{-1}) \\
&&+\text{tr}(\partial g\cdot g^{-1}\wedge\overline\partial B\wedge
\overline\partial(\partial B^*\cdot g^{-1}).
\end{eqnarray*}
\end{proof}

\begin{lemm}\lab{lemm 8}
 $\textup{tr}(\dbar B\wedge \partial B^*\cdot g^{-1})$ is a
well-defined (1,1)-form on $S$.
\end{lemm}
\begin{proof}
We take local coordinates $(U,z_i)$ and $(W,w_j)$ on $S$ such that
$U\cap W\neq\emptyset$. Let $J=(\frac{\partial w_i}{\partial
z_j})$ and
$$(\omega_1+\sqrt{-1}\omega_2)\mid_U=\overline\partial(\varphi_1dz_1+\varphi_2dz_2)=\overline\partial
\varphi_1\wedge dz_1+\dbar \varphi_2\wedge dz_2,$$
$$(\omega_1+\sqrt{-1}\omega_2)\mid_W=\overline\partial(\gamma_1dw_1+\gamma_2dw_2)=\overline\partial
\gamma_1\wedge dw_1+\dbar \gamma_2\wedge dw_2.$$
 Then on $U\cap W$,
\begin{eqnarray*}
\left(\begin{array}{cc}\dbar \gamma_1 &
\dbar \gamma_2\end{array}\right)\wedge\left(\begin{array}{c} dw_1\\
dw_2\end{array}\right)
&=& \left(\begin{array}{cc}\dbar \varphi_1 & \dbar \varphi_2\end{array}\right) \wedge\left(\begin{array}{c} dz_1\\
dz_2\end{array}\right).
\end{eqnarray*}
So
\begin{eqnarray}\label{302}
\left(\begin{array}{cc}\dbar \varphi_1 & \dbar
\varphi_2\end{array}\right)=\left(\begin{array}{cc}\dbar \gamma_1
& \dbar \gamma_2\end{array}\right)J.
\end{eqnarray}
On the other hand, we have
\begin{equation}\label{303}
g(z)=J^t g(w)\overline{J},
\end{equation}
where $g(z)=(g_{i\bar j}(z))$ and $g(w)=(g_{i\bar j}(w))$. Then on
$U\cap W$, using (\ref{302}), (\ref{303}), we have
\begin{eqnarray*}
&&\text{tr}\left(\begin{array}{c} \dbar \gamma_1\\ \dbar
\gamma_2\end{array}\right)\wedge\left(\begin{array}{cc}
\partial \bar \gamma_1& \partial\bar \gamma_2\end{array}\right)\cdot
g^{-1}(w)\\
&=&\text{tr}\left(\begin{array}{c} \dbar \gamma_1\\ \dbar
\gamma_2\end{array}\right)\wedge\left(\begin{array}{cc}
\overline{\overline\partial \gamma_1}& \overline{\overline\partial
\gamma_2}\end{array}\right)\cdot g^{-1}(w)\\
&=&\text{tr}(J^t)^{-1}\left(\begin{array}{c} \dbar \varphi_1\\
\dbar \varphi_2\end{array}\right)\wedge\left(\begin{array}{cc}
\overline{\overline\partial \varphi_1}&
\overline{\overline\partial
\varphi_2}\end{array}\right)\bar J^{-1}\cdot \bar J\cdot g^{-1}(z)\cdot J^t\\
&=&\text{tr}J^t\cdot (J^t)^{-1}\left(\begin{array}{c} \dbar \varphi_1\\
\dbar \varphi_2\end{array}\right)\wedge\left(\begin{array}{cc}
\overline{\overline\partial \varphi_1}&
\overline{\overline\partial
\varphi_2}\end{array}\right)\cdot g^{-1}(z)\\
&=&\text{tr}\left(\begin{array}{c} \dbar \varphi_1\\ \dbar
\varphi_2\end{array}\right)\wedge\left(\begin{array}{cc}
\partial\bar \varphi_1& \partial
\bar \varphi_2\end{array}\right)\cdot g^{-1}(z),\\
\end{eqnarray*}
which proves that  $\text{tr}(\dbar B\wedge \partial B^*\cdot
g^{-1})$ is a well-defined $(1,1)$-form on $S$.
\end{proof}
Although $\text{tr}(\dbar B\wedge \partial B^*\cdot g^{-1})$ is a
well-defined (1,1)-form on $S$, we can not express it by
$\omega_1$ and $\omega_2$. But in some  particular case, we can.
\begin{prop}\lab{prop 9}
When $\omega_2=n\omega_1$, $n\in \mathbb{Z}$,
\begin{equation}\label{119}
\textup{tr}(\dbar B\wedge \partial B^*\cdot
g^{-1})=\frac{\sqrt{-1}}{4}(1+n^2)\parallel
\omega_1\parallel^2_{\omega_S}\omega_S,
\end{equation}
where $g$ is the  given Calabi-Yau metric on $S$ and $\omega_S$ is
the corresponding K\"{a}hler form.
\end{prop}
\begin{proof}
We recall that locally,
\begin{eqnarray*}
\omega_1&=&\dbar \xi,\ \ \ \ \xi=\xi_1dz_1+\xi_2dz_2,\\
\omega_2&=&\dbar \zeta, \ \ \ \ \zeta=\zeta_1dz_1+\zeta_2dz_2,\\
\varphi_j&=&\xi_j+\sqrt{-1}\zeta_j,\ \ \ \ \text{for}\ \ j=1,2,\\
B&=&\left(\begin{array}{c}\varphi_1\\ \varphi_2\end{array}\right),
\, \ B^*=\left(\begin{array}{cc}\bar \varphi_1&\bar
\varphi_2\end{array}\right).
\end{eqnarray*}
When $\omega_2=n\omega_1$,we take $\zeta=n\xi$. Then
$\dbar\zeta_j=n\dbar\xi_j$,
\begin{eqnarray*}
\dbar B=\left(\begin{array}{c}\dbar \varphi_1\\ \dbar
\varphi_2\end{array}\right)=(1+n\sqrt{-1})\left(\begin{array}{c}\dbar
\xi_1\\ \dbar \xi_2\end{array}\right)
\end{eqnarray*}
and
\begin{equation*}
\partial B^*=\left(\begin{array}{cc}\partial \bar \varphi_1& \partial\bar
\varphi_2\end{array}\right)=
(1-n\sqrt{-1})\left(\begin{array}{cc}\partial\bar\xi_1&
\partial\bar \xi_2\end{array}\right).
\end{equation*}
Using above equalities, we find
\begin{equation}\label{0301}
\begin{aligned}
&\text{tr}(\dbar B\wedge \partial B^*\cdot
g^{-1})\\
=&(1+n^2)\text{tr}\left(\begin{array}{c}\dbar \xi_1\\ \dbar
\xi_2\end{array}\right)\wedge\left(\begin{array}{cc}\partial\bar\xi_1&
\partial\bar \xi_2\end{array}\right)\cdot g^{-1}\\
=&\frac{1+n^2}{\det g}\text{tr}
\left(\begin{array}{c}\frac{\partial \xi_1}{\partial\bar z_i}d\bar z_i\\
\frac{\partial \xi_2}{\partial\bar z_i}d\bar
z_i\end{array}\right)\wedge\left(\begin{array}{cc}\overline{\frac{\partial\xi_1}{\partial\bar
z_j}}dz_j &\overline{\frac{\partial\xi_2}{\partial\bar z_j}}dz_j
\end{array}\right)\cdot
\left(\begin{array}{cc}g_{2\bar2}&  -g_{1\bar2}\\
-g_{2\bar1}& g_{1\bar1}\end{array}\right)\\
=&\frac{1+n^2}{\det g}\text{tr}
\left(\begin{array}{c}\frac{\partial \xi_1}{\partial\bar z_i}\\
\frac{\partial \xi_2}{\partial\bar
z_i}\end{array}\right)\wedge\left(\begin{array}{cc}\overline{\frac{\partial\xi_1}{\partial\bar
z_j}} &\overline{\frac{\partial\xi_2}{\partial\bar z_j}}
\end{array}\right)\cdot
\left(\begin{array}{cc}g_{2\bar2}&  -g_{1\bar2}\\
-g_{2\bar1}& g_{1\bar1}\end{array}\right)d\bar z_i\wedge dz_j.
\end{aligned}
\end{equation}
In order to get the global formula, we need to calculate
$\omega_1$. As $\omega_1$ is real,
\begin{equation}\label{305}
\overline{\frac{\partial\xi_i}{\partial\bar z_j}}=-
\frac{\partial\xi_j}{\partial\bar z_i}\ \ \ \ \text{for}\ \
i,j=1,2.
\end{equation}
Since $\omega_1$ is anti-self-dual, i.e.,  $\omega_1\wedge
\omega_S=0$,  we have
\begin{equation}\label{306}
g_{1\bar1}\frac{\partial\xi_2}{\partial\bar z_2}
+g_{2\bar2}\frac{\partial\xi_1}{\partial\bar
z_1}-g_{1\bar2}\frac{\partial\xi_2}{\partial\bar z_1}
-g_{2\bar1}\frac{\partial\xi_1}{\partial\bar z_2}=0.
\end{equation}
Because
\begin{equation}\label{307}
\omega_1\wedge\omega_1=-\omega_1\wedge\ast\omega_1=
-\omega_1\ast\bar\omega_1=-\parallel\omega_1\parallel_{\omega_S}^2\frac{\omega^2_S}{2!},
\end{equation}
locally we also have
 \begin{equation}\label{308}
\frac{1}{\det(g)}\left(\frac{\partial\xi_1}{\partial\bar
z_1}\frac{\partial\xi_2}{\partial\bar
z_2}-\frac{\partial\xi_1}{\partial\bar
z_2}\frac{\partial\xi_2}{\partial\bar
z_1}\right)=\frac{1}{8}\parallel\omega_1\parallel_{\omega_S}^2.
\end{equation}
Now using above (\ref{305}), (\ref{306}) and (\ref{308}), we
calculate the component of $d\bar z_1\wedge d z_1$ in (\ref{0301})
to be
\begin{equation}\label{309}
\begin{aligned}
&\frac{1+n^2}{\det(g)}\left(g_{2\bar2}\frac{\partial\xi_1}{\partial\bar
z_1}\overline{\frac{\partial\xi_1}{\partial\bar
z_1}}-g_{2\bar1}\frac{\partial\xi_1}{\partial\bar
z_1}\overline{\frac{\partial\xi_2}{\partial\bar
z_1}}-g_{1\bar2}\frac{\partial\xi_2}{\partial\bar
z_1}\overline{\frac{\partial\xi_1}{\partial\bar z_1}}-g_{1\bar1}
\frac{\partial\xi_2}{\partial\bar
z_1}\overline{\frac{\partial\xi_2}{\partial\bar z_1}}\right)\\
=&\frac{1+n^2}{\det(g)}\left(g_{2\bar1}\frac{\partial\xi_1}{\partial\bar
z_1}\frac{\partial\xi_1}{\partial\bar
z_2}+g_{1\bar2}\frac{\partial\xi_2}{\partial\bar
z_1}\frac{\partial\xi_1}{\partial\bar
z_1}-g_{2\bar2}\left(\frac{\partial\xi_1}{\partial\bar
z_1}\right)^2-g_{1\bar1} \frac{\partial\xi_2}{\partial\bar
z_1}\frac{\partial\xi_1}{\partial\bar z_2}\right)\\
=&\frac{1+n^2}{\det(g)}\left(\frac{\partial\xi_1}{\partial\bar
z_1}\left(g_{2\bar1} \frac{\partial\xi_1}{\partial\bar
z_2}+g_{1\bar2}\frac{\partial\xi_2}{\partial\bar
z_1}\right)-g_{2\bar2}\left(\frac{\partial\xi_1}{\partial\bar
z_1}\right)^2-g_{1\bar1} \frac{\partial\xi_2}{\partial\bar
z_1}\frac{\partial\xi_1}{\partial\bar z_2}\right)\\
=&\frac{1+n^2}{\det(g)}\left(\frac{\partial\xi_1}{\partial\bar
z_1}\left(g_{1\bar1} \frac{\partial\xi_2}{\partial\bar
z_2}+g_{2\bar2}\frac{\partial\xi_1}{\partial\bar
z_1}\right)-g_{2\bar2}\left(\frac{\partial\xi_1}{\partial\bar
z_1}\right)^2-g_{1\bar1} \frac{\partial\xi_2}{\partial\bar
z_1}\frac{\partial\xi_1}{\partial\bar z_2}\right)\\
=&\frac{1+n^2}{\det(g)}g_{1\bar1}\left(\frac{\partial\xi_1}{\partial\bar
z_1} \frac{\partial\xi_2}{\partial\bar z_2}-
\frac{\partial\xi_2}{\partial\bar
z_1}\frac{\partial\xi_1}{\partial\bar z_2}\right)\\
=&\frac{1+n^2}{8}\parallel\omega_1\parallel_{\omega_S}^2g_{1\bar1}.
\end{aligned}
\end{equation}
Similarly, the components of $d\bar z_2\wedge dz_1$, $d\bar
z_1\wedge dz_2$ and $d\bar z_2\wedge dz_2$ in (\ref{0301}) are
$\frac{1+n^2}{8}\parallel\omega_1\parallel_{\omega_S}^2g_{1\bar2}$,
$\frac{1+n^2}{8}\parallel\omega_1\parallel_{\omega_S}^2g_{2\bar1}$
and
$\frac{1+n^2}{8}\parallel\omega_1\parallel_{\omega_S}^2g_{2\bar2}$
respectively. So  we obtain
\begin{eqnarray*}
&&\text{tr}(\dbar A\wedge \partial A^*\cdot
g^{-1})\\
&=&\frac{1+n^2}{8}\parallel\omega_1\parallel_{\omega_S}^2(g_{1\bar1}d\bar
z_1\wedge dz_1+g_{1\bar2}d\bar z_2\wedge dz_1+g_{2\bar1}d\bar
z_1\wedge dz_2+g_{2\bar2}d\bar z_2\wedge dz_2)\\
&=&\sqrt{-1}\frac{1+n^2}{4}\parallel\omega_1\parallel_{\omega_S}^2\omega_S.
\end{eqnarray*}
\end{proof}

 \section{Reduction of  the  Strominger's system}
Consider a  3-dimensional Hermitian manifold $(X,\omega_0,\Omega)$
as described in the section 2. Let $\omega_S$ be the Calabi-Yau
metric on $S$. Let
$$\theta=dx+\alpha_1+\sqrt{-1}(dy+\alpha_2),$$
then the Hermitian  form $\omega_0$ in (\ref{2011}) is
$$
\omega_0=\pi^*\omega_S+\frac{\sqrt{-1}}{2}\theta\wedge\bar\theta.$$
 Because  $\parallel\Omega\parallel=1$, and $\omega_1$ and $\omega_2$ are anti-self-dual,we use
 (\ref{2004}) to compute
\begin{equation}\label{0401}
\begin{aligned}
&\ d(\parallel\Omega\parallel_{\omega_0}\omega_0^2)\\
=&\ d\omega^2_0=d(\pi^*\omega_S^2+
\sqrt{-1}\pi^*\omega_S\wedge\theta\wedge\bar\theta)\\
=&\ \sqrt{-1}\pi^*\omega_S\wedge
d\theta\wedge\bar\theta-\sqrt{-1}\pi^*\omega_S\wedge
\theta\wedge d\bar\theta\\
=&\ \sqrt{-1}\pi^*\omega_S\wedge
(\omega_1+\sqrt{-1}\omega_2)\wedge\bar\theta-\sqrt{-1}\pi^*\omega_S\wedge
(\omega_1-\sqrt{-1}\omega_2)
\wedge\theta\\
=&\ 0.
\end{aligned}
\end{equation}
According to Lemma 1, $(\omega_0,\Omega)$ is the solution of
equation (\ref{104}).
 Let $u$ be any smooth function on $S$ and let
\begin{equation}\lab{401}
\omega_u=\pi^*(e^u\omega_S)+\frac{\sqrt{-1}}{2}\theta\wedge\bar\theta.
\end{equation}
Then
$$\parallel\Omega\parallel^2_{\omega_u}=\frac{\omega_0^3}{\omega^3_u}=\frac{1}{e^{2u}}$$
and
\begin{eqnarray*}
\parallel\Omega\parallel_{\omega_u}\omega^2_u&=&e^{-u}(e^{2u}\omega_S^2+
\sqrt{-1}e^u\omega_S\wedge\theta\wedge\bar\theta)\\
&=&\omega_0^2+(e^u-1)\omega_S^2.
\end{eqnarray*}
Using (\ref{0401}), we obtain
$$d(\parallel\Omega\parallel_{\omega_u}\omega^2_u)=d\omega^2_0+d(e^u-1)\wedge\omega_S^2=0$$
because $e^u$ is a function on $S$. Hence  we have proven the
following
\begin{lemm}\cite{GP}\lab{lemm 10}
The metric (\ref{401}) defined on $X$ satisfies  equation
(\ref{105}) and so satisfies  equation (\ref{104}).
\end{lemm}

Let $V$ be a stable vector bundle over $X$ with degree $0$ with
respective to the metric $\omega_u$. According to Li-Yau's theorem
\cite{LY1},  there is a Hermitian-Yang-Mills  metric $H$ on $V$,
which is unique up to constant. Then $(V, H, X, \omega_u)$
satisfies equation (\ref{101}), (\ref{102}) and (\ref{104}) of the
Strominge's system. Hence to look for  a solution to Strominger's
system, we  need only to consider equation (\ref{103}):
\begin{equation}\lab{403}
\sqrt{-1}\partial\bar\partial
\omega_u=\frac{\alpha'}{4}(\text{tr}R_u\wedge
R_u-\text{tr}F_H\wedge F_H),
\end{equation}
where $R_u$ is the curvature of Hermitian connection of metric
$\omega_u$ on the holomorphic tangent bundle $T'X$.
 Define the
Laplacian operator $\bigtriangleup$ with respective to the metric
$\omega_S$ as
\begin{equation*}
\bigtriangleup
\psi\frac{\omega_S^2}{2!}=\sqrt{-1}\partial\bar\partial \psi\wedge
\omega_S.
\end{equation*}
\begin{lemm}\lab{lemm 11}
$\sqrt{-1}\partial\dbar\omega_u=\bigtriangleup
e^u\cdot\frac{\omega_S^2}{2!}+\frac{1}{2}(\parallel\omega_1\parallel_{\omega_S}^2+
\parallel\omega_2\parallel_{\omega_S}^2)\frac{\omega^2_S}{2!}$.
\end{lemm}
\begin{proof}
Using (\ref{2004}) and (\ref{307}), we compute
\begin{eqnarray*} \sqrt{-1}\partial\overline{\partial}\omega_u
&=&\sqrt{-1}\partial\overline{\partial}(e^u\omega_S+\frac{\sqrt{-1}}{2}\theta\wedge\overline\theta)\\
&=&\sqrt{-1}\partial\dbar
e^u\wedge \omega_S-\frac{1}{2}\dbar\theta\wedge\partial\bar\theta\\
&=&\bigtriangleup
e^u\cdot\frac{\omega^2_S}{2!}-\frac{1}{2}(\omega_1+\sqrt{-1}\omega_2)
\wedge(\omega_1-\sqrt{-1}\omega_2)\\
&=&\bigtriangleup
e^u\cdot\frac{\omega^2_S}{2!}-\frac{1}{2}(\omega_1\wedge\omega_1+\omega_2\wedge\omega_2)
\\
&=&\bigtriangleup
e^u\cdot\frac{\omega^2_S}{2!}+\frac{1}{2}(\parallel\omega_1\parallel_{\omega_S}^2+
\parallel\omega_2\parallel_{\omega_S}^2)\frac{\omega^2_S}{2!}.
\end{eqnarray*}
\end{proof}
\begin{lemm}\lab{lemm 12}
 $\textup{tr}R_u\wedge R_u=\pi^*\textup{tr}R_S\wedge
R_S+2\pi^*(\partial\dbar u\wedge
\partial\dbar u)+2\pi^*(\partial\dbar(e^{-u}\textup{tr}(\dbar
B\wedge \partial B^*\cdot g^{-1}))).$
\end{lemm}
\begin{proof}
In the proof of the Proposition \ref{prop 6} we don't use the
condition that $\omega_S$ is K\"ahler. So if we replace metric $g$
by $e^ug$,  we can still obtain:
\begin{equation}\label{0402}
\begin{aligned}
\text{tr}R_u\wedge R_u=&\pi^*(\text{tr}R_S^u\wedge
R_S^u+2\text{tr}\partial\dbar(\dbar B\wedge \partial B^*\cdot
(e^{u}g)^{-1}))\\
=&\pi^*(\text{tr}R_S^u\wedge
R_S^u+2\partial\dbar(e^{-u}\text{tr}(\dbar B\wedge
\partial B^*\cdot g^{-1}))),
\end{aligned}
\end{equation}
here $R_S^u$ denotes the curvature of Hermitian connection of the
metric $e^ug$ on holomorphic tangent bundle $T'S$. So
\begin{eqnarray*}
R^u_S&=&\dbar(\partial(e^ug)\cdot(e^ug)^{-1})\\
&=&\dbar(\partial u\cdot I+\partial g\cdot g^{-1})\\
&=&\dbar\partial u\cdot I+R_S
\end{eqnarray*}
and
\begin{equation}\label{0403}
\begin{aligned}
\text{tr}R_S^u\wedge R_S^u=&\text{tr}R_S\wedge R_S+2\partial\dbar
u\wedge \partial\dbar u+2\partial\dbar u\wedge \text{tr}R_S\\
=&\text{tr}R_S\wedge R_S+2\partial\dbar u\wedge \partial\dbar u,
\end{aligned}
\end{equation}
here we use the fact that $\text{tr}R_S=0$ because the Hermitian
metric $g$ is the Calabi-Yau metric on $S$. Inserting (\ref{0403})
into (\ref{0402}), we have proven the lemma.
\end{proof}

From Lemma \ref{lemm 11} and \ref{lemm 12},  we can rewrite
equation (\ref{403}) as
\begin{equation}\label{406}
\begin{aligned}
&\ \ \sqrt{-1}\partial\bar\partial e^u\wedge
\omega_S-\frac{\alpha'}{2}\partial\dbar(e^{-u}\text{tr}(\dbar
B\wedge
\partial B^*\cdot g^{-1}))-\frac{\alpha'}{2}\partial\dbar u\wedge
\partial\dbar u\\
=&\ \ \frac{\alpha'}{4}\text{tr} R_S\wedge
R_S-\frac{\alpha'}{4}\text{tr}F_{ H}\wedge F_{ H}- 1/2 (\parallel
\omega_1\parallel^2+\parallel\omega_2\parallel_{\omega_S}^2)\omega_S^2/{2!}.
\end{aligned}
\end{equation}

\begin{prop}\lab{prop 13}
 There is no solution of Strominger's
system on the torus bundle $X$ over $T^{4}$ if the metric is
$e^{u}\omega_{S}+\frac{\sqrt{-1}}{2}\theta\wedge\bar{\theta}$.
\end{prop}
\begin{proof}
Wedging left-hand side of equation (\ref{406}) by $\omega_u$ and
integrating over $X$, we get
\begin{equation}\lab{407}
\begin{aligned}
&\int_X \{\sqrt{-1}\partial\bar\partial e^u\wedge
\omega_S-\frac{\alpha'}{2}\partial\dbar(e^{-u}\text{tr}(\dbar
B\wedge
\partial B^*\cdot g^{-1}))-\frac{\alpha'}{2}\partial\dbar u\wedge
\partial\dbar u\}\wedge \omega'_u\\
=&\int_X\{\sqrt{-1}\bar\partial e^u\wedge
\omega_S-\frac{\alpha'}{2}\dbar(e^{-u}\text{tr}(\dbar B\wedge
\partial B^*\cdot g^{-1}))-\frac{\alpha'}{2}\dbar u\wedge
\partial\dbar u\}\wedge\partial\omega'_u=0
\end{aligned}
\end{equation}
because $\partial \omega_u=\partial(e^u)\wedge
\omega_S+2\theta\wedge(\omega_1-\sqrt{-1}\omega_2)$. When $S=T^4$,
$R_{T^4}=0$. Integrating both sides of (\ref{406}) and applying
(\ref{407}), we get
\begin{equation}\lab{409}
\alpha'\int_X\text{tr}F_{H}\wedge F_{H}\wedge \omega_u+\frac 12
\int_X(\parallel\omega_1\parallel_{\omega_S}^2+\parallel\omega_2\parallel_{\omega_S}^2)\frac
{\omega^2}{2!}\wedge \omega_u=0.
\end{equation}
Certainly
\begin{equation}\lab{410}
2\int_X(\parallel\omega_1\parallel_{\omega_S}^2+\parallel\omega_2\parallel_{\omega_S}^2)\frac
{\omega_S^2}{2!}\wedge
\omega_u=2\int_Xe^{-2u}(\parallel\omega_1\parallel_{\omega_S}^2+\parallel\omega_2\parallel_{\omega_S}^2)\frac
{\omega_u^3}{3!}>0.
\end{equation}
On the other hand, it is well-known that
\begin{equation*}
\frac{\text{tr}F_{H}^2}{8\pi^2}=\frac {1} {2}
c_1^2(V)-c_2(V)=\frac 1 {2r}c_1^2(V)-\frac 1
{2r}(2rc_2(V)-(r-1)c_1^2(V)),
\end{equation*}
where $r$ is a rank of the bundle $V$ and that
\begin{equation*}
(2r(c_2(V)-(r-1)c_1^2(V))\wedge \omega_u=\frac {r}{4\pi^2}\mid
F_0\mid^2\frac {\omega_u^3}{3!},
\end{equation*}
where $F_0=F_{ H}-\frac 1 r \text{tr} F_{H}\cdot {\text id}_V$. So
\begin{equation*}
\text{tr}F_{H}^2\wedge \omega_u=\frac{8\pi^2}{2r}c_1^2(V)-\mid
F_0\mid^2\frac {\omega_u^3}{3!}.
\end{equation*}
Now according to equation (\ref{102}), $F_H\wedge \omega_u^2=0$
and so  $c_1(V)\wedge \omega_u^2=0$. Therefore  $c_1(V)$ is an
anti-self-dual $(1,1)$-form on $X$. Thus
\begin{equation*}
c_1^2(V)\wedge \omega_u=-\mid c_1(V)\mid^2\frac {\omega_u^3}{3!}
\end{equation*}
and \begin{equation}\lab{415}
 \int_X\text{tr}F_{H}^2\wedge \omega_u=-\frac {4\pi^2}{r}\int_X \mid
c_1(V)\mid^2\frac{\omega_u^3}{3!}-\int_X \mid F_0\mid^2\frac
{\omega_u^3}{3!}\leq 0.
\end{equation}
Inserting  (\ref{410}) and (\ref{415}) into (\ref{409}),  we get a
contradiction.
\end{proof}
This situation is different if the base is  $K3$ surface. At first
we observe
\begin{lemm}\lab{lemm 15}
Let $E$ be   a stable vector bundle over $S$ with degree $0$ with
respective to the Calabi-Yau metric $\omega_S$. Then $V=\pi^*E$ is
also a stable vector bundle over $X$ with degree $0$ with
respective to Hermitian metric $\omega_u$ for any smooth function
$u$ on $S$.
\end{lemm}
\begin{proof}
According to the Donaldson-Uhlenbeck-Yau theorem, there is an
unique Hermitian-Yang-Mills metric $H$ on $E$ up to constant.
Since  we assume that the degree of $E$ is zero,  the curvature
$F_H$ of $H$ satisfies the equation
\begin{equation*}
F_H\wedge \omega_S=0.
\end{equation*}
For the metric $\pi^*H$ on $V=\pi^*E$, the curvature $\pi^*(F_H)$
satisfies
\begin{equation*}
\pi^*F_H\wedge \omega_u^2=\pi^*(F_H\wedge \omega_S)\wedge
(\pi^*(e^{2u}\omega_S)+\pi^*(e^u)\theta\wedge\bar\theta)=0.
\end{equation*}
So $\pi^*H$ is also the Hermitian-Yang-Mills metric on $V=\pi^*E$
with degree $0$. Thus $V$ is a stable vector bundle over $X$ with
respective to the Hermitian metric $\omega_u$ for any smooth
function $u$.
\end{proof}
When we restrict ourselves to consider such a vector bundle
$(V=\pi^*E, \pi^*F_H)$ over $X$, we see that equation (\ref{406})
on $X$ can be considered as an equation on $S$. Integrating
equation (\ref{406}) over $S$, we get
\begin{equation}\lab{4018}
\alpha'\int_S\{\text{tr} R_S\wedge R_S-\text{tr}F_{H}\wedge
F_{H}\}=2\int_S(\parallel
\omega_1\parallel_{\omega_S}^2+\parallel\omega_2\parallel_{\omega_S}^2)\frac
{\omega_S^2}{2!}.
\end{equation}
 As $\int_S\text{tr} R_S\wedge
R_S=8\pi^2c_2(V)=8\pi^2\times 24$, and $\int_S\text{tr} F_H\wedge
F_H=8\pi^2\times (c_2(E)-\frac 1 2 c_1^2(E))\geq 0$,  we can
rewrite equation (\ref{4018}) as
\begin{equation}\lab{1110}
\alpha'(24-(c_2(E)-\frac 1 2 c_1^2(E)))=\int_S(\parallel\frac
{\omega_1}{2\pi}\parallel_{\omega_S}^2+\parallel\frac
{\omega_2}{2\pi}\parallel_{\omega_S}^2)\frac {\omega_S^2}{2!}.
\end{equation}
Using notations of section 1, above equation implies:
\begin{equation}\lab{425}
\alpha'(24-\kappa(E))+\left(Q\left(\frac
{\omega_1}{2\pi}\right)+Q\left(\frac{\omega_2}{2\pi}\right)\right)=0.
\end{equation}
This equation implies that  there is a smooth function $\mu$ such
that
\begin{equation}\lab{426}
\frac{\alpha'}{4}\text{tr} R_S\wedge
R_S-\alpha'\text{tr}F_{H}\wedge F_{H}-\frac 1 2 (\parallel
\omega_1\parallel^2+\parallel\omega_2\parallel_{\omega_S}^2)\frac{\omega_S^2}{2!}=-\mu\frac
{\omega_S^2}{2!}
\end{equation}
and $\int_S\mu\frac {\omega_S^2}{2!}=0$. Inserting (\ref{426})
into (\ref{406}), we obtain the following equation:
\begin{equation}\lab{427}
\sqrt{-1}\partial\bar\partial
e^u\wedge\omega_S-\frac{\alpha'}{2}\partial\dbar(e^{-u}\text{tr}(\dbar
B\wedge
\partial B^*\cdot g^{-1}))-\frac{\alpha'}{2}\partial\dbar u\wedge
\partial\dbar u+\mu\frac {\omega_S^2}{2!}=0
\end{equation}
where $\mu$ is a smooth function satisfying the integrable
condition $\int_S \mu=0$ and $\text{tr}(\dbar B\wedge
\partial B^*\cdot g^{-1})$ is a smooth well-defined real $(1,1)$-form on $S$.
 In the next section we will  use the continuity method to solve
equation (\ref{427}). We will prove that equation (\ref{427}) has
a smooth solution $u$.
\begin{theo}\lab{theo 2}
Let $S$ be a $K3$ surface with a Calabi-Yau metric $\omega_S$. Let
$\omega_1$ and $\omega_2$ be anti-self-dual $(1,1)$-forms on $S$
such that $\frac{\omega_1}{2\pi}\in H^2(S,\mathbb{Z})$ and
$\frac{\omega_2}{2\pi}\in H^2(S,\mathbb{Z})$. Let $X$ be
 a $T^2$-bundle over $S$ constructed by $\omega_1$ and $\omega_2$.
 Let $E$ be a stable bundle over $S$ with degree 0.
 Suppose that $\omega_1$, $\omega_2$ and $\kappa(E)$ satisfy the
 condition (\ref{425}).
 Then there exist a smooth function $u$ on $S$ and a Hermitian-Yang-Mills metric $H$ on $E$
  such that
 $(V=\pi^*E,\pi^*F_H, X, \omega_u)$ is a solution of Strominger's
 system.
 \end{theo}
 \begin{proof}
 Because we assume that $E$ is a stable bundle over $S$ with
 degree $0$ with respective to the Calabi-Yau metric $\omega_S$,
 according to the Donaldson-Uhlenbeck-Yau theorem, there is an unique
 Hermitian-Yang-Mills metric $H$ on $E$ up to constant such that
 the curvature $F_H$ of metric $H$ satisfies
 \begin{equation*}
 F_H^{2,0}=F_H^{0,2}=0,\ \ \ \ F_H\wedge \omega_S=0.
 \end{equation*}
So we have $\pi^*F_H^{2,0}=\pi^*F_H^{0,2}=0$ and according to
Lemma \ref{lemm 15}, we also have $\pi^*F_H\wedge\omega_u^2=0$.
Now according to our assumption, $(\omega_1,\omega_2,E)$ satisfies
the condition (\ref{425}), and hence  there is a function $\mu$
satisfying equation (\ref{426}). Then we solve equation
(\ref{427}). According to Theorem 18 in the next section, there
exists a smooth solution $u$ of equation (\ref{427}). Combining
equation (\ref{427}) with (\ref{426}), we know that $u$ is the
solution of equation (\ref{406}). So $(\pi^*F_H, \omega_u)$
satisfies equation (\ref{103}). On the other hand, according to
Lemma \ref{lemm 10}, the metric
 $\omega_u=e^u\omega_S+\frac{\sqrt{-1}}{2}\theta\wedge\bar\theta$
 on $X$ satisfies equation (\ref{104}).
 Thus we have proven that
$(V=\pi^*E,\pi^*F_H,X,\omega_u)$ satisfy all equations of
Strominger's system.
 \end{proof}

\section{Solving the equation}
In this section, we want  to prove
\begin{theo}
The  equation
\begin{equation}\lab{5001}
\sqrt{-1}\partial\bar\partial
e^u\wedge\omega_S-\frac{\alpha'}{2}\partial\dbar(e^{-u}\textup{tr}(\dbar
B\wedge
\partial B^*\cdot g^{-1}))-\frac{\alpha'}{2}\partial\dbar u\wedge
\partial\dbar u+\mu\frac {\omega_S^2}{2!}=0
\end{equation}
has a smooth solution $u$ such that
$\omega'=e^u\omega_S-\frac{\sqrt{-1}}{2}t\alpha'
e^{-u}\textup{tr}(\bar\partial B\wedge\partial B^*\cdot
g^{-1})+\alpha'\sqrt{-1}\partial\bar\partial u$ defines a
Hermitian metric on $S$.
\end{theo}
\begin{proof}
We solve equation (\ref{5001}) by the continuity method. More
precisely we introduce a parameter $t\in [0,1]$  and consider the
following equation
\begin{equation}\label{5004}
\sqrt{-1}\partial\bar\partial e^u\wedge\omega_S-
t\alpha\partial\bar\partial(e^{-u}\text{tr}(\bar\partial
B\wedge\partial B^*\cdot g^{-1}))-\alpha\partial\bar\partial
u\wedge\partial\bar\partial u+t\mu\omega_S^2/2!=0,
\end{equation}
where we have replace $\frac{\alpha'}{2}$ by $\alpha$. Let
\begin{equation*}
\rho=-\sqrt{-1}\text{tr}(\bar\partial B\wedge
\partial B^*\cdot g^{-1}),
\end{equation*}
then according to Lemma \ref{lemm 8}, $\rho$ is a  well-defined
real $(1,1)$-form on $S$. We can rewrite the equation as
\begin{equation}\lab{5007}
\sqrt{-1}\partial\bar\partial e^u\wedge
\omega_S-t\alpha\sqrt{-1}\partial\bar\partial(e^{-u}\rho)-\alpha\partial\bar\partial
u\wedge\partial\bar\partial u +t\mu\frac {\omega_S^2}{2!}=0.
\end{equation}
We shall impose the following:
\begin{equation}\lab{5008}
\text{Elliptic condition}:\ \ \omega'=e^u\omega_S+t\alpha
e^{-u}\rho+2\alpha\sqrt{-1}\partial\bar\partial u>0
\end{equation}
and
\begin{equation}\lab{50009}
 \text{Normalization}:\ \ \ \ \left(\int_S e^{-4u}\frac {\omega_S^2}{2!}\right)^{\frac
14}=A, \ \ \ \ \int_S 1\frac {\omega_S^2}{2!}=1.
\end{equation}
Let $C^{k,\alpha_0}(S)$ be the space of functions whose
$k$-derivatives are H\"{o}lder continuous with exponent
$0<\alpha_0<1$. We consider the solution in the following space
\begin{equation}\lab{5010}
B_A=\{u\in C^{2,\alpha_0}(S)\mid u \ \ \text{satisfies the
normalization  (\ref{50009})}\}
\end{equation}
and
\begin{equation}\lab{5011}
 B_{A,t}=\{u\in B_A\mid u\ \
\text{also satisfies the elliptic condition (\ref{5008})}\}.
\end{equation}
  Let
\begin{equation}\lab{1506}
{\bf T}=\{s\in [0,1]\mid \text{for}\ \ t\in [0,s] \ \
\text{equation (\ref{5007}) admits a solution in} \  B_{A,t}\}.
\end{equation}
Obviously $0\in {\bf T}$ with a solution $u=-\ln A$. Hence we need
only to show that ${\bf T}$ is both closed and open in $[0,1]$.
This will imply that $1\in {\bf T}$ and that our original equation
has a solution in $C^{2,\alpha_0}$. To see that the set ${\bf T}$
is open, we use the standard implicity function theorem.

Let $t_0\in {\bf T}$ and $u_{t_0}$ be a solution of equation
(\ref{5007}). Let $B_{[0,1]}=\{(t,u)\in[0,1]\times B_A\mid u\in
B_{A,t}\}$. Then $B_{[0,1]}$ is an open set of $[0,1]\times B_A$.
Let $C_0^{0,\alpha_0}(S)=\{\psi\in C^{0,\alpha_0}\mid
\int_S\psi\frac {\omega_S^2}{2!}=0\}$. We have a map:
$\tilde{L}:B_{[0,1]}\rightarrow C_0^{0,\alpha_0}(S)$,
\begin{equation}
\tilde L(t,u)=\ast_{\omega_S} (\sqrt{-1}\partial\bar\partial
e^u\wedge\omega_S-
\sqrt{-1}t\alpha\partial\bar\partial(e^{-u}\rho)-\alpha\partial\bar\partial
u\wedge\partial\bar\partial{ u}+t\mu\omega_S^2/2!).
\end{equation}
According to  the definition of $t_0$,  $\tilde L(t_0,u_{t_0})=0$.
The differential $d\tilde L$ of $\tilde L$ at $u_{t_0}$ evaluated
at $\varphi$ is $L(\varphi)$, where the linear operator $L$ from
$C^{2,\alpha_0}(S)$ to $C^{0,\alpha_0}(S)$ is defined as:
\begin{equation}\lab{1507} L(\varphi)=\ast_{\omega_S}
(\sqrt{-1}\partial\bar\partial
(e^{u_{t_0}}\varphi)\wedge\omega_S+\sqrt{-1}
t_0\alpha\partial\bar\partial(e^{-u_{t_0}}\varphi\rho)-2\alpha\partial\bar\partial
u_{t_0} \wedge\partial\bar\partial\varphi).
\end{equation}
So $d\tilde L=L\mid_{T_{u_{t_0}}B_A}$, where
$T_{u_{t_0}}B_A=\{\varphi\in C^{2,\alpha_0}(S)\mid\int
e^{-4u_{t_0}}\varphi=0\}$ is the tangent space of $B_A$ at
$u_{t_0}$. The principle part of the operator $\ast_{\omega_S} L$
is
\begin{equation}\lab{1508}
\sqrt{-1}\partial\bar\partial
\varphi\wedge(e^{u_{t_0}}\omega_S+t_0\alpha
e^{-u_{t_0}}\rho+2\alpha\sqrt{-1}\partial\bar\partial u_{t_0}).
\end{equation}
From the elliptic condition (\ref{5008}), we get:
\begin{equation}\lab{1509}
\omega'_{t_0}=e^{u_{t_0}}\omega+t_0\alpha
e^{-u_{t_0}}\rho+2\alpha\sqrt{-1}\partial\bar\partial u_{t_0}>0.
\end{equation}
$\omega_{t_0}'$ can be
 taken as a Hermitian (not K\"ahler !) metric on
$S$. Let
\begin{equation}\label{15011}
P=\sqrt{-1}\Lambda_{\omega_{t_0}'}\partial\dbar.
\end{equation}
Then $P$ is an elliptic operator on $S$. Because $u_{t_0}$ is a
solution in $C^{2,\alpha_0}$ and our $\mu$ and $\rho$ are smooth,
according to Schauder theory, $u_{t_0}$ is smooth. So the operator
$P$ is smooth and can be defined by
\begin{equation}\label{1512}
\sqrt{-1}\partial\dbar
\psi\wedge\omega_{t_0}'=P(\psi)\omega'^2_{t_0}/2!
\end{equation}
for any $C^2(S)$ function $\psi$ on $S$.
 For any $\phi,\psi\in
C^{2,\alpha_0}(S,\mathbb{R})$, we compute
\begin{eqnarray*}
&&\int L^*(\psi)\varphi\frac{\omega_S^2}{2!}=\int \psi\cdot L(\varphi)\frac{\omega_S^2}{2!}\\
&=&\int \psi\cdot\{\sqrt{-1}\partial\bar\partial
(e^{u_{t_0}}\varphi)\wedge\omega_S+\sqrt{-1}t_0\alpha\partial\bar\partial(e^{-u_{t_0}}\varphi
\rho)-2\alpha\partial\bar\partial u_{t_0}
\wedge\partial\bar\partial\varphi\}
\\
&=&\int \varphi\sqrt{-1}\bar\partial\partial
\psi\wedge(e^{u_{t_0}}\omega_S+ t_0\alpha
e^{-u_{t_0}}\rho+2\alpha\sqrt{-1}\partial\bar\partial u_{t_0})\\
 &=&\sqrt{-1}\int \varphi\partial\dbar \psi\wedge \omega'_{t_0}\\
&=&\int\varphi\cdot P(\psi)\frac{\omega'^2_{t_0}}{2!}=\int
P^*(\varphi)\psi\frac{\omega'^2_{t_0}}{2!}.
\end{eqnarray*}
 Thus using the Corollary in page 227 of \cite{LT}, we obtain
 \begin{equation*}
 \ker L^*=\ker P=\mathbb{R}
 \end{equation*}
  and
  \begin{equation*}
\ker L=\ker P^*=\{\mathbb{R}\varphi_0\mid \varphi_0 \ \ \textup{is
a nonzero  function that has constant sign}\}.
\end{equation*}

Now we are ready to prove $d\tilde{L}$ is invertible. Because
$d\tilde L=L\mid_{T_{u_{t_0}}B_A}$, we only need to prove
$L\mid_{T_{u_{t_0}}B_A}:T_{u_{t_0}}B_A\rightarrow
C_0^{0,\alpha_0}(S)$ is invertible. It is clearly that $\ker L\cap
T_{u_{t_0}}B_A=0$. So $d\tilde{L}=L\mid_{T_{u_{t_0}}B_A}$ is
injective. Next we prove that $d\tilde L=L\mid_{T_{u_{t_0}}B_A}$
is surjective. For any $\psi\in C_0^{0,\alpha_0}(S)$, we have
 $\psi\perp\ker L^*$.  It
 is well known that there is a weak solution  $\varphi_1$ of linear elliptic equation
 $L(\varphi_1)=\psi$.  The Schauder theory shows  that
 $\varphi\in C^{2,\alpha_0}(S)$ when $\psi\in C^{0,\alpha_0}(S)$.
 Take $c_0=-\frac{\int e^{-4u_{t_0}}\varphi_1}{\int e^{-4u_{t_0}}\varphi_0}$,
then $\varphi_1+c_0\varphi_0\in T_{u_{t_0}}B_A$ and
$L(\varphi_1+c_0\varphi_0)=\psi$. So $d\tilde
L=L\mid_{T_{u_{t_0}}B_A}$ is surjective. Hence $d\tilde L$ of
$\tilde L$ at $u_{t_0}$ is invertible and $\tilde L$ maps an open
neighborhood of $(t_0,u_{t_0})$ in $B_{[0,1]}$ to an open
neighborhood of $\tilde L(t_0,u_{t_0})$ in $C_0^{0,\alpha_0}(S)$.
This proves the set ${\bf T}$ is open.

 It remains to prove that
${\bf T}$ is closed. Let $\rho=\frac{\sqrt{-1}}{2}\rho_{i\bar
j}dz_i\wedge d\bar z_j$, then we can write $g'_{i\bar j}$ as
\begin{equation*}
g'_{i\bar j}=e^ug_{i\bar j}+t\alpha e^{-u}\rho_{i\bar j}+4\alpha
u_{i\bar j}.
\end{equation*}
By directly computation, we get
\begin{equation}\lab{5019}
\begin{aligned}
\frac{\det g'_{i\bar j}}{\det g_{i\bar j}}=&e^{2u}+ 2\alpha
e^u\bigtriangleup  u+t\alpha g^{i\bar j}\rho_{i\bar
j}+2t\alpha^2e^{-u}(\sqrt{-1}\partial\bar\partial u\wedge
\rho,\frac{\omega_S^2}{2!})\\
& +t^2\alpha^2e^{-2u}\frac {\det \rho_{i\bar j}}{\det g_{i\bar
j}}+16\alpha^2\frac {\det u_{i\bar j}}{\det g_{i\bar j}}.
\end{aligned}
\end{equation}
We can rewrite equation (\ref{5007}) as
\begin{equation}\lab{5020}
\begin{aligned}
8\alpha\frac{\det u_{i\bar j}}{\det g_{i\bar
j}}=&-e^u\bigtriangleup u-2e^u\mid\bigtriangledown
u\mid^2-t\mu-t\alpha e^{-u}(\sqrt{-1}\partial\bar\partial
u\wedge\rho,\frac{\omega_S^2}{2!})\\
&+t\alpha e^{-u}(\sqrt{-1}\partial u\wedge \bar\partial u\wedge
\rho,\frac{\omega_S^2}{2!})-t\alpha e^{-u}(\sqrt{-1}\partial
u\wedge \bar\partial \rho,\frac{\omega_S^2}{2!})\\
&+t\alpha e^{-u}(\sqrt{-1}\bar\partial u\wedge \partial
\rho,\frac{\omega_S^2}{2!})+t\alpha
e^{-u}(\sqrt{-1}\partial\bar\partial \rho,\frac{\omega_2^2}{2!}).
\end{aligned}
\end{equation}
Then inserting (\ref{5020}) into (\ref{5019}), we find the
Monge-Amp\`ere-type equation: \begin{equation}\lab{5021}
\begin{aligned}
&\frac{\det (e^ug_{i\bar j}+t\alpha e^{-u}\rho_{i\bar j}+4\alpha
u_{i\bar j})}{\det g_{i\bar j}}=F_{t,u_t}
\end{aligned}
\end{equation}
where
\begin{equation*}
\begin{aligned}
F_{t,u_t}=&e^{2u}+t\alpha g^{i\bar j}\rho_{i\bar j}
+t^2\alpha^2e^{-2u}\frac {\det \rho_{i\bar j}}{\det g_{i\bar
j}}-2e^u\mid\bigtriangledown
u\mid^2\\
&+2t\alpha^2 e^{-u}(\sqrt{-1}\partial u\wedge \bar\partial u\wedge
\rho,\frac{\omega_S^2}{2!})-2t\alpha^2 e^{-u}(\sqrt{-1}\partial
u\wedge \bar\partial \rho,\frac{\omega_S^2}{2!})\\
&+2t\alpha^2 e^{-u}(\sqrt{-1}\bar\partial u\wedge \partial
\rho,\frac{\omega_S^2}{2!})+2t\alpha^2
e^{-u}(\sqrt{-1}\partial\bar\partial
\rho,\frac{\omega_2^2}{2!})-2t\alpha\mu.
\end{aligned}
\end{equation*}
In particular, when $\omega_2=n\omega_1$,
\begin{equation*}
\begin{aligned}
F_{t,u_t}=&(e^u+t\alpha fe^{-u})^2 -2\alpha(e^u-t\alpha
fe^{-u})\mid\bigtriangledown u\mid^2\\
&-4 t\alpha^2 e^{-u}\bigtriangledown u\cdot \bigtriangledown f
+2t\alpha^2 e^{-u}\bigtriangleup f-2t\alpha\mu.
\end{aligned}
\end{equation*}

If ${t_q}$ is a sequence in ${\bf T}$, then we have a sequence
$u_q\in C^{2,\alpha_0}(S)$ such that
\begin{equation}\lab{5022}
\begin{aligned}
\frac{\det (e^{u_q}g_{i\bar j}+t_q\alpha e^{-u_q}\rho_{i\bar
j}+4\alpha \frac{\partial^2 u_q}{\partial z_i\partial \bar
z_j})}{\det g_{i\bar j}}=F_{t_q,u_{t_q}}.
\end{aligned}
\end{equation}

Differentiating equation (\ref{5022}), we have
\begin{equation}\lab{5024}
\begin{aligned}
&4\alpha\det\left(e^{u_q}g_{i\bar j}+t_q\alpha e^{-u_q}\rho_{i\bar
j}+4\alpha\frac{\partial^2 u_q}{\partial z_i\partial \bar
z_j}\right)\cdot \sum_{i\bar j}g_q'^{i\bar
j}\frac{\partial^2}{\partial z_i\partial\bar z_j}\left(\frac
{\partial u_q}{\partial
z_k}\right)\\
=&-\det\left(e^{u_q}g_{i\bar j}+t_q\alpha e^{-u_q}\rho_{i\bar
j}+4\alpha\frac{\partial^2 u_q}{\partial z_i\partial \bar
z_j}\right)\cdot \sum_{i\bar j}g_q'^{i\bar
j}\frac{\partial}{\partial
z_k}(e^{u_q}g_{i\bar j}+t_q\alpha e^{-u_q}\rho_{i\bar j})\\
&+\frac {\partial}{\partial z_k}\{\det g_{i\bar j}\cdot
F_{t_q,u_{t_q}}\}.
\end{aligned}
\end{equation}
 Proposition 24 ( and  Proposition \ref{prop
 2}-\ref{prop 4} for a special case $\omega_2=n\omega_1$)
 shows that the operator on the left-hand side of
(\ref{5024}) is uniformly elliptic.  Proposition 25 (and
Proposition \ref{prop 5} for the special case)  shows that the
coefficients are H\"older continuous with exponent $\alpha$ for
any $0\leq \alpha_0\leq 1$. The Schauder estimate then gives an
estimate for the $C^{2,\alpha_0}$-estimates of $\partial
u_q/\partial z_k$. Similarly we can find  $C^{2,\alpha_0}$-norm of
$\partial u_q/\partial \bar z_k$. Therefore the sequence $\{u_q\}$
converges in the $C^{2,\alpha_0}$-norm to a solution of the
equation
\begin{equation*}
\begin{aligned}
\frac{\det (e^{u}g_{i\bar j}+t_0\alpha e^{-u}\rho_{i\bar
j}+4\alpha \frac{\partial^2 u}{\partial z_i\partial \bar
z_j})}{\det g_{i\bar j}}=F_{t_0},
\end{aligned}
\end{equation*}
where $t_0=\lim_{q\rightarrow \infty}t_q$. Thus we find a
$C^{2,\alpha_0}(S)$ solution $u$ of equation (\ref{5021}). But
equation (\ref{5021}) is equivalent to equation (5.3). Hence ${\bf
T}$ is closed. So there is a solution $u$ of equation (\ref{5001})
in $C^{2,\alpha_0}(S)$. Because our function $\mu$ and
$(1,1)$-form $-\sqrt{-1}\text{tr}(\bar\partial B\wedge
\partial B^*\cdot g^{-1})$ is smooth, again by the Schauder
theory, we get the smooth solution of equation (\ref{5001}).
\end{proof}

\section{Zeroth order estimate}\lab{sec: 6}
From this section to the section \ref{sec: 9}, we do a priori
estimates of $u$ up to the third order. We deal with the simpler
case $\omega_2=n\omega_1$, where $\omega_1$ is an anti-self-dual
$(1,1)$-form on $S$. We let
$f=\frac{1+n^2}{4}\parallel\omega_1\parallel_{\omega_S}^2$.  Then
the equation is
\begin{equation*}
\bigtriangleup(e^u-t\alpha fe^{-u})+8\alpha\frac{\det u_{i\bar
j}}{\det g_{i\bar j}}+t \mu=0,
\end{equation*}
where $f$ and $\mu$ are smooth functions on $S$ such that $f\geq
0$ and $\int_S \mu\frac {\omega_S^2}{2!}=0$. According to our
assumption, $u\in C^{2,\alpha_0}(S)$. So by the Schauder theory,
the solution $u$ is smooth. We  denote partial derivatives by
$u_{i\bar j}=\partial_{i\bar j}u=\frac{\partial^2u}{\partial
z_i\partial_{\bar j}}$. If we replace $t\alpha f$ by  $f$ and
$t\mu$ by $\mu$,  then the equation can be written as
\begin{equation}\lab{2801}
\bigtriangleup(e^u-fe^{-u})+8\alpha \frac {\det u_{i\bar j}}{\det
g_{i\bar j}}+\mu=0.
\end{equation}
We impose the elliptic condition
\begin{equation*}
\omega'=(e^u+fe^{-u})\omega_S+2\alpha\sqrt{-1}\partial\bar\partial
u>0
\end{equation*}
and the normalization condition
\begin{equation}\lab{2804}
\left(\int_S e^{-4u}\frac{\omega_S^2}{2!}\right)^{\frac 14}=A,\ \
\ \ \int_S 1\frac{\omega_S^2}{2!}=1.
\end{equation}
In this section we prove that if $A$ is small enough, then the
solution $u$ has an  upper bound and a  lower bound  depending
only on $\alpha$, $f$, $\mu$, Sobolev constant of metric
$\omega_S$ and $A$. In the next section, we shall prove that if
$A$ is small enough, then the determinant of   $\omega'$ has a
lower bound greater than 0 and  the metric $\omega'$ is uniformly
positive. Let $g'=\frac{\sqrt{-1}}{2}g'_{i\bar j}dz_i\wedge
dz_{\bar j}$, where
\begin{equation*}
g'_{i\bar j}=(e^u+fe^{-u})g_{i\bar j}+4\alpha u_{i\bar j}.
\end{equation*}
We note that
$$\frac{\omega'^2}{2!}=\frac{\det g'_{i\bar j}}{\det g_{i\bar
j}}\frac{\omega_S^2}{2!}.$$ The matrix $(g'^{i\bar j})$ satisfies
the equation
\begin{equation*}
\sum g'^{i\bar j}g_{i\bar j}=\delta^{\bar j}_{\bar k}.
\end{equation*}
So
\begin{equation*}
g'^{1\bar 1}=\frac {g'_{2\bar 2}}{\det g'_{i\bar j}},\ \ g'^{1\bar
2}=-\frac{g'_{2\bar 1}}{\det g'_{i\bar j}}, \ \ g'^{2\bar
1}=-\frac{g'_{1\bar 2}}{\det g'_{i\bar j}},\ \ g'_{2\bar
2}=\frac{g'_{1\bar 1}}{\det g'_{i\bar j}}.
\end{equation*}
Hence from the definition (\ref{1512}) of the operator $P$, we
 have $P(\varphi)=2g'^{i\bar j}\varphi_{i\bar j}$. We apply equation (\ref{2801}) to compute
\begin{equation}\lab{503}
\begin{aligned}
&P(u)\frac {\det g'_{i\bar j}}{\det g_{i\bar j}}=(2g'^{i\bar
j}u_{i\bar j})
\frac {\det g'_{i\bar j}}{\det g_{i\bar j}}\\
=&2(g'_{2\bar 2}\partial_{1\bar 1}u+g'_{1\bar 1}\partial_{2\bar
2}u-g'_{1\bar 2}\partial_{2\bar 1}u-g'_{2\bar 1}\partial_{1\bar
2}u)\cdot (\det g_{i\bar j})^{-1}\\
=&(e^u+fe^{-u})\bigtriangleup u+16\alpha\frac{\det u_{i\bar
j}}{\det g_{i\bar j}}\\
=&(e^u+fe^{-u})\bigtriangleup u-2\bigtriangleup(e^u-fe^{-u})-2\mu.
\end{aligned}
\end{equation}
 In the following,  the volume form will be  $\frac {\omega_S^2}{2!}$ unless it is clear
 from the context.
 We can use (\ref{503}) to compute
\begin{equation}\lab{508}
\begin{aligned}
&\ \ \int P(e^{-ku})\frac{\omega'^2}{2!}
=2\int g'^{i\bar j}\partial_{i\bar j}(e^{-ku})\frac{\omega'^2}{2!}\\
&=k^2\int e^{-ku}(2g'^{i\bar j}\partial_iu\partial_{\bar
j}u)\frac{\omega'^2}{2!}-k\int e^{-ku}(2g'^{i\bar
j}\partial_{i\bar
j}u)\frac{\omega'^2}{2!}\\
&\geq -k\int e^{-ku}P(u)\frac{\omega'^2}{2!}=-k\int e^{-ku}P(u)
\frac{\det g'_{i\bar j}}{\det g_{i\bar j}}\frac{\omega^2}{2!}\\
&= -k\int e^{-ku}(e^u+fe^{-u})\bigtriangleup u+2k\int
e^{-ku}\bigtriangleup (e^u-fe^{-u})+2k\int e^{-ku}\mu.
\end{aligned}
\end{equation}
On the other hand, we can also use (\ref{1512}) to compute
\begin{equation}\lab{509}
\begin{aligned}
&\int P(e^{-ku})\frac{\omega'^2}{2!}=\sqrt{-1}\int
\partial\bar\partial
(e^{-ku})\wedge\omega'\\
=&\sqrt{-1}\int
\partial\bar\partial(e^{-ku})\wedge((e^u+fe^{-u})\omega_S+2\alpha\sqrt{-1}\partial\bar\partial
u)\\
=&\int (e^u+fe^{-u})\bigtriangleup (e^{-ku})\\
=&-k\int e^{-ku}(e^u+fe^{-u})\bigtriangleup u+k^2\int
e^{-ku}(e^u+fe^{-u})\mid\bigtriangledown u\mid^2,
\end{aligned}
\end{equation}
where $\mid\bigtriangledown u\mid^2=2g^{i\bar j}u_iu_{\bar j}$.
Combing (\ref{508}) and (\ref{509}),
\begin{equation}\lab{506}
\begin{aligned}
&k\int (e^u+fe^{-u})e^{-ku}\mid\bigtriangledown
u\mid^2\\
\geq &2\int e^{-ku}\bigtriangleup (e^u-fe^{-u})+2\int
e^{-ku}\mu\\
=&2\int e^{-ku}(e^u+fe^{-u})\bigtriangleup u+2\int
e^{-ku}(e^u-fe^{-u})\mid\bigtriangledown u\mid^2\\
&-2\int e^{-(k+1)u}\bigtriangleup f+4\int
e^{-(k+1)u}\bigtriangledown u\cdot \bigtriangledown f+2\int
e^{-ku}\mu,
\end{aligned}
\end{equation}
where $\bigtriangledown u\cdot \bigtriangledown f=g^{i\bar
j}(u_if_{\bar j}+u_{\bar j}f_i)$. When $k\geq 2$, we integrate by
part and obtain
\begin{equation}\lab{507}
\begin{aligned}
&2\int e^{-ku}(e^u+fe^{-u})\bigtriangleup u\\
=&2(k-1)\int e^{-(k-1)u}\mid\bigtriangledown u\mid^2+2(k+1)\int
fe^{-(k+1)u}\mid\bigtriangledown u\mid^2\\
&+\frac{2}{k+1}\int e^{-(k+1)u}\bigtriangleup
f\frac{\omega_S^2}{2!}-4\int e^{-(k+1)u}\bigtriangledown u\cdot
\bigtriangledown f.
\end{aligned}
\end{equation}
Inserting (\ref{507}) into (\ref{506}),
\begin{equation*}
\begin{aligned}
&k\int e^{-(k-1)u}\mid \bigtriangledown u\mid^2+k\int
fe^{-(k+1)u}\mid\bigtriangledown u\mid^2\\
\leq &2(1-\frac{1}{k+1})\int e^{-(k+1)u}\bigtriangleup f-2\int
e^{-ku}\mu.
\end{aligned}
\end{equation*}
Because $f\geq 0$,  above inequality implies
\begin{equation}\lab{5009}
k\int e^{-(k-1)u}\mid\bigtriangledown u\mid^2\leq C_0\int
e^{-(k+1)u}+C_0\int e^{-ku},
\end{equation}
where $C_0$  depends only on $f$ (so also depends on $\alpha$) and
$\mu$. In the following, $C_0$ may depend on $\alpha$, $f$, $\mu$
and the Sobolev constant of $S$ about the metric $\omega_S$. We
use the constant $C_0$ in the generic sense. So $C_0$ may mean
different constants in different equations. Now from above
inequality, if we replace $k-1$ by $k$, then when $k\geq 1$,
\begin{equation}\lab{510}
\int \mid \bigtriangledown(e^{-u})^{\frac{k}{2}}\mid^2\leq C_0
k\int e^{-(k+2)u}+C_0k\int e^{-(k+1)u}.
\end{equation}
We  apply the Sobolev inequality
$$\parallel e^{-\frac k 2  u}\parallel_{L^r}\leq C_0(\parallel
e^{-\frac k 2 u}\parallel_{L^p}+\parallel \bigtriangledown
e^{-\frac k 2 u}\parallel_{L^p}$$ with $r=\frac {4p}{4-p}=4$. In
the case  $p=2$, we have
\begin{equation*}
\left(\int (e^{-u})^{2k}\right)^{\frac{1}{2}}\leq C_0\int
(e^{-u})^{k}+C_0\int \mid
\bigtriangledown(e^{-u})^{\frac{k}{2}}\mid^2.
\end{equation*}
Inserting (\ref{510}) into above inequality, we get
\begin{equation*}
\left(\int (e^{-u})^{2k}\right)^{\frac{1}{2}}\leq C_0\int
(e^{-u})^k+C_0k\int (e^{-u})^{k+2}+C_0k\int (e^{-u})^{k+1}.
\end{equation*} Because we have normalized the metric $\omega_S$ such
that  $\int_S 1\frac{\omega_S^2}{2!}=1$, we apply the H\"older
inequality to above inequality to get
\begin{equation}\lab{511}
\left(\int (e^{-u})^{2k}\right)^{\frac{1}{2}}\leq C_0\left(\int
(e^{-u})^{k+2}\right)^{\frac{k}{k+2}}+C_0k\left(\int
(e^{-u})^{k+2}\right)^{\frac{k+1}{k+2}}+C_0k\int (e^{-u})^{k+2}.
\end{equation}
 Note that when $k=2$, above inequality has no use. This
explains why we need the normalization
 (\ref{2804}). In the following we assume that
\begin{equation}\lab{512}
\left(\int (e^{-u})^4\right)^{\frac{1}{4}}=A<1, \ \ \ \ \int_S
1\frac{\omega_S^2}{2!}=1.
\end{equation}
There are two cases:

Case (1): For any $k\geq 4$, $\int (e^{-u})^k\leq1$. Then
(\ref{511}) implies
\begin{equation}\lab{513}
\left(\int (e^{-u})^{2k}\right)^{\frac{1}{2}}\leq C_0k\left(\int
(e^{-u})^{k+2}\right)^{\frac{k}{k+2}}.
\end{equation}
Applying the H\"{o}lder inequality,
\begin{equation}\lab{514}
\begin{aligned}
\int (e^{-u})^{k+2}&=\int (e^{-u})^{k-2}(e^{-u})^4\\
&\leq \left(\int((e^{-u})^{k-2})^{\frac{k}{k-2}}\right)^{\frac
{k-2}{k}}\left(\int((e^{-u})^{4})^{\frac{k}{2}}\right)^{\frac
{2}{k}}\\
&=\left(\int (e^{-u})^k\right)^{\frac{k-2}{k}}\left(\int
(e^{-u})^{2k}\right)^{\frac{2}{k}}.
\end{aligned}
\end{equation}
Inserting above inequality into (\ref{513}), we see
\begin{equation}\lab{515}
\int (e^{-u})^{2k}\leq C_0 k^{2\frac{k+2}{k-2}}\left(\int
(e^{-u})^k\right)^{2}\leq C_0 k^2\left(\int (e^{-u})^k\right)^2.
\end{equation}
Take $k=2^{\beta}$ for $\beta\geq 2$. Then $\beta\geq 2$ and
rewrite (\ref{515}) as
\begin{equation*}
\int (e^{-u})^{2^{\beta+1}}\leq C_0 2^{2\beta}\left(\int
(e^{-u})^{2^\beta}\right)^2.
\end{equation*}
Iterating above inequality, we  get
\begin{equation}\lab{517}
\left(\int
(e^{-u})^{2^{\beta+1}}\right)^{\frac{1}{2^{\beta+1}}}\leq C_0
\left(\int (e^{-u})^{4}\right)^{\frac 14}.
\end{equation}
We fix the constant $C_0$ and denote it by $C_1$, which depends
only on $f$, $\mu$, $\alpha$ and the Sobolev constant of $S$ with
respective to the metric $\omega_S$. Letting $\beta\rightarrow
\infty$, we find
\begin{equation}\lab{518}
\exp(-\inf u)=\parallel e^{-u}\parallel_\infty\leq C_1A.
\end{equation}

Case(2). There is an integer $k$ such that $\int (e^{-u})^k>1$.
Let $k_0$ be the first such an integer. According to the
assumption (\ref{512}), $k_0>4$.  Then for any $k\geq k_0$, by the
H\"older inequality, we have $\int (e^{-u})^k>1$. For any $k\geq
k_0>4$, inequality  (\ref{511}) and (\ref{514}) imply
\begin{equation*}
\begin{aligned}
\left(\int (e^{-u})^{2k}\right)^{\frac{1}{2}}& \leq
C_0k\int(e^{-u})^{k+2}\\
\leq &C_0k\left(\int (e^{-u})^k\right)^{\frac{k-2}{k}}\left(\int
(e^{-u})^{2k}\right)^{\frac{2}{k}}.
\end{aligned}
\end{equation*}
We can see from above inequality:
\begin{equation*}
\int (e^{-u})^{2k}\leq C_0
k^2\left(\int(e^{-u})^k\right)^{2\frac{k-2}{k-4}}\ \ \ \text{for}\
\ \ k\geq k_0>4.
\end{equation*}
Using above inequality for $k\geq k_0$ and the inequality
(\ref{515}) for $k<k_0$, we can still get the estimate (\ref{518})
of $\inf u$, because $A^a<A$ when $A<1$ and $a>1$.

Next we  estimate  $\sup_S u$. Similar to the way we  estimate
$\inf u$, we compute $\int_SP(e^{pu})\frac {\omega'^2}{2!}$ by two
methods and get
\begin{equation}\lab{522}
p\int (e^u+fe^{-u})e^{pu}\mid\bigtriangledown u\mid^2\geq -2\int
e^{pu}\bigtriangleup(e^u-fe^{-u})-2\int
 e^{pu}\mu.
 \end{equation}
Integrating by part, when $p\geq 2$,
\begin{equation}\lab{523}
\begin{aligned}
&\int
e^{pu}\bigtriangleup(e^u-fe^{-u})\\
=&-p\int e^{(p+1)u}\mid\bigtriangledown u\mid^2-p\int
e^{(p-1)u}f\mid\bigtriangledown
u\mid^2-\left(1+\frac{1}{p-1}\right)\int e^{(p-1)u}\bigtriangleup
f
\end{aligned}
\end{equation}
and when $p=1$,
\begin{equation}\lab{524}
\int e^u\bigtriangleup (e^u-fe^{-u})=-\int
e^{2u}\mid\bigtriangledown u\mid^2-\int f\mid\bigtriangledown
u\mid^2-\int u\bigtriangleup f.
\end{equation}
Inserting (\ref{523}) or (\ref{524}) into (\ref{522}), because
$f>0$, we get
\begin{equation}\lab{525}
p\int e^{(p+1)u}\mid\bigtriangledown u\mid^2\leq C_0\int
e^{pu}+C_0\int e^{(p-1)u}\  \ \ \text{for} \ \ \ p\geq 2.
\end{equation}
 When $p=1$,
\begin{equation}\lab{526}
\int e^{2u}\mid\bigtriangledown u\mid^2\leq 2\int e^u\mu-2\int
u\bigtriangleup f \leq C_0\int e^u+C_0\int \mid u\mid.
\end{equation}
\begin{rema}When $t=0$, $f$  and $\mu$
(actually actually $t\alpha f$ and $t\mu$) are equal to zero. From
above inequality we have
\begin{equation*}
\int e^{2u}\mid\bigtriangledown u\mid^2\leq 2\int e^u\mu-2\int
u\bigtriangleup f=0,
\end{equation*}
which implies $\mid\bigtriangledown u\mid^2\equiv 0$. So when
$t=0$, there is an unique constant solution under the
normalization and the elliptic condition.
\end{rema}
We choose $A$ small enough such that
\begin{equation}\lab{527}
A<C_1^{-1}.
\end{equation}
Then from $e^{-\inf u}\leq C_1A<1$, $u>0$. (\ref{526}) implies
\begin{equation}\lab{528}
\int \mid\bigtriangledown e^u\mid^2\leq C_0\int e^u
\end{equation}
and  (\ref{525}) implies
\begin{equation}\lab{529}
\int \mid \bigtriangledown e^{\frac p2 u}\mid^2\leq C_0 p\int
e^{pu}\ \ \ \text{when} \ \ \ p\geq 3.
\end{equation}
Applying the Sobolov inequality and using (\ref{528}),
(\ref{529}), we obtain
\begin{equation*}
\left(\int (e^u)^{2p}\right)^{\frac{1}{2}}\leq C_0p\int e^{pu},\ \
\  \text{for} \ \ \ p\geq 2.
\end{equation*}
Take $p=2^\beta$ for $\beta\geq 1$. Then
\begin{equation*}
\int (e^u)^{2^{\beta+1}}\leq C_0 2^{2\beta}\left(\int
(e^u)^{2^\beta}\right)^2.
\end{equation*}
Iterating above inequality and take the limit $\beta\rightarrow
\infty$,  we get
\begin{equation}\lab{532}
\exp(\sup u)\leq C_0\left( \int e^{2u}\right)^{\frac{1}{2}}.
\end{equation}
Let $\int e^u=M_u$, then $\int (e^u-M_u)=0$.  The Poincar\'e
inequality and (\ref{528}) imply
\begin{equation}\lab{533}
\begin{aligned}
&\int (e^{u})^2-\left(\int e^u\right)^2=\int (e^u-M_u)^2\\
\leq & C_0\int \mid \bigtriangledown(e^u-M_u)\mid^2 \leq C_0\int
\mid \bigtriangledown e^u\mid^2\leq C_0\int e^u.
\end{aligned}
\end{equation}
Let $U_1=\{x\in S\mid e^{-u(x)}\geq \frac{A}{2}\}$ and $U_2=\{x\in
S\mid e^{-u(x)}<\frac{A}{2}\}$. Then
\begin{equation*}
\begin{aligned}
A^4&=\int_S e^{-4u}=\int_{U_1}e^{-4u}+\int_{U_2}e^{-4u}\\
&\leq \int_{U_1}e^{-4\inf
u}+\int_{U_2}(A/2)^4\\
&=e^{-4\inf
u}\text{Vol}(U_1)+\left(A/2\right)^4\text{Vol}(U_2)\\
&=\left[(e^{-\inf
u})^4-(A/2)^4\right]\text{Vol}(U_1)+\left(A/2\right)^4.
\end{aligned}
\end{equation*}
So
\begin{equation*}
\text{Vol}(U_1)\geq\frac{A^4-(A/2)^4}{(e^{-\inf
u})^4-(A/2)^4}\geq\frac
{A^4-(A/2)^4}{(C_1A)^4-(A/2)^4}=\frac{2^4-1}{(2C_1)^4-1}=m_0>0.
\end{equation*}
Thus
\begin{equation*}
\text{Vol}(U_2)=1-\text{Vol}(U_1)\leq 1-m_0<1.
\end{equation*}
Applying the Young inequality, the H\"older inequality, and then
using (\ref{533}), we find
\begin{equation}\lab{537}
\begin{aligned}
&\left(\int
e^u\right)^2=\left(\int_{U_1}e^u+\int_{U_2}e^u\right)^2\\
\leq& \left(1+\frac
1\epsilon_0\right)\left(\int_{U_1}e^u\right)^2+(1+\epsilon_0)\left(\int_{U_2}e^u\right)^2\\
\leq& \left(1+\frac
1\epsilon_0\right)\left(\int_{U_1}e^{2u}\right)\text{Vol}(U_1)+(1+\epsilon_0)\text{Vol}(U_2)
\int_{U_2}e^{2u}\\
\leq &\left(1+\frac
1\epsilon_0\right)\left(\frac 2A\right)^2+(1+\epsilon_0)\text{Vol}(U_2)\int_S e^{2u}\\
\leq &\left(1+\frac{1}{\epsilon_0}\right)\left(\frac
2A\right)^2+(1+\epsilon_0)(1-m_0)\left(\left(\int
e^u\right)^2+C_0\int e^u\right).
\end{aligned}
\end{equation}
Take $\epsilon_0$ small enough such that
\begin{equation*}
(1+\epsilon_0)(1-m_0)<1.
\end{equation*}
Then from (\ref{537}),
\begin{equation}\lab{521}
\left(\int_S
e^u\right)^2-\frac{(1+\epsilon_0)(1-m_0)C_0}{1-(1+\epsilon_0)(1-m_0)}\int
e^u+\frac{\left(1+\frac 1{\epsilon_0}\right)\left(\frac
2A\right)^2}{1-(1+\epsilon_0)(1-m_0)}\leq 0,
\end{equation}
which implies an  upper bound of $\int e^u$. Now the estimate of
$\int e^{2u}$ follows from (\ref{533}) and  the estimate of $\sup
u$ then follows from (\ref{532}). We summarize above discussion in
the following
\begin{prop}\lab{prop 2}
Let  $t\in {\bf T}$ and $u$ is a solution of equation (\ref{2801})
under the elliptic condition $\omega'=(e^u+t\alpha
fe^{-u})\omega_S+2\alpha\sqrt{-1}\partial\bar\partial u>0$ and and
normalization $\left(\int e^{-4u}\right)^{\frac 14}=A$ and $\int
1\frac{\omega_S^2}{2!}=1$. If $A<1$, then there is a constant
$C_1$ which depends on $\alpha$, $f$, $\mu$ and the Sobolev
constant of $\omega_S$ such that
\begin{equation*}
\inf_S u\geq -\ln (C_1A).
\end{equation*}
Moreover, if $A$ is small enough such that $A<(C_1)^{-1}$, then
there is an upper bound of $\sup_S u$ which depends on $\alpha$,
$f$, $\mu$, the Sobolev constant of $\omega_S$ and $A$.
\end{prop}

\section{An estimate of the determinant}\lab{sec:7}
In this section, we want to obtain a lower bound of the
determinant, which is equal to
\begin{equation}\lab{601}
\begin{aligned}
 F=\frac{\det g'_{i\bar j}}{\det g_{i\bar j}}=&(e^u+t\alpha
 fe^{-u})^2+2\alpha(e^u+t\alpha fe^{-u})\bigtriangleup
 u+16\alpha^2\frac{\det u_{i\bar j}}{\det g_{i\bar j}}\\
 =&(e^u+t\alpha fe^{-u})^2+2\alpha(e^u+t\alpha f
 e^{-u})\bigtriangleup u-2\alpha(\bigtriangleup(e^u-t\alpha f
 e^{-u})+t \mu)\\
 = &(e^u+t\alpha
 fe^{-u})^2-2\alpha(e^u-t\alpha fe^{-u})\mid\bigtriangledown
 u\mid^2\\
 &-4t\alpha^2e^{-u}\bigtriangledown u\cdot\bigtriangledown
 f+2t\alpha^2 e^{-u}\bigtriangleup f-2t\alpha \mu.
\end{aligned}
\end{equation}
From (\ref{601}), we see
\begin{equation}\lab{602}
e^{-2u}F=1-2\alpha e^{-u}\mid\bigtriangledown u\mid^2+e^{-2u}O(1),
\end{equation}
where
\begin{equation}\lab{603}
O(1)=2t\alpha f+t^2\alpha^2
f^2e^{-2u}+2t\alpha^2fe^{-u}\mid\bigtriangledown
u\mid^2-4t\alpha^2e^{-u}\bigtriangledown u\cdot \bigtriangledown
f+2t\alpha^2e^{-u}\bigtriangleup f-2t\alpha\mu.
\end{equation}
Our elliptic condition is $\omega'>0$, which is equivalent to
$F>0$. The first step is to  derive an upper bound of
$\mid\bigtriangledown u\mid^2$.  In the above section we have
proven that $e^{-\inf u}\leq C_1A$ and have assumed that $C_1A<1$.
Applying this assumption, we estimate
\begin{equation}\lab{604}
\begin{aligned}
&e^{-2u}F=1-2\alpha e^{-u}\mid\bigtriangledown u\mid^2+e^{-2u}O(1)\\
\leq &1-2\alpha e^{-u}\mid\bigtriangledown
u\mid^2+(2t\alpha^2fe^{-3u}+2t\alpha^2
e^{-3u})\mid\bigtriangledown u\mid^2\\
&+e^{-2u}\{2t\alpha f+t^2\alpha^2
f^2e^{-2u}+2t\alpha^2e^{-u}\mid\bigtriangledown
f\mid^2+2t\alpha^2e^{-u}\bigtriangleup f-2t\alpha \mu\}\\
\leq&1-2\alpha \{1-t\alpha(1+\sup f)e^{-2\inf
u}\}e^{-u}\mid\bigtriangledown u\mid^2\\
&+e^{-2\inf u}\{2\alpha \sup f+\alpha^2(\sup
f)^2+2\alpha^2\sup\mid\bigtriangledown
f\mid^2+2\alpha^2\sup\mid\bigtriangleup f\mid+2\alpha \sup
\mid\mu\mid\}\\
\leq& 1-2\alpha \{1-\alpha(1+\sup
f)(C_1A)^2\}e^{-u}\mid\bigtriangledown u\mid^2+C_2(C_1A)^2,
\end{aligned}
\end{equation}
where
\begin{equation}\lab{605}
C_2=2\alpha \sup f+\alpha^2(\sup
f)^2+2\alpha^2\sup\mid\bigtriangledown
f\mid^2+2\alpha^2\sup\mid\bigtriangleup f\mid+2\alpha \sup
\mid\mu\mid.
\end{equation}
Applying $F>0$ to (\ref{604}), we get
\begin{equation}\lab{606}
1-2\alpha \{1-\alpha(1+\sup f)(C_1A)^2\}e^{-u}\mid\bigtriangledown
u\mid^2+C_2(C_1A)^2>0.
\end{equation}
If we take
\begin{equation*}
A\leq \{2\alpha(1+\sup f)\}^{-\frac 1 2} C_1^{-1},
\end{equation*}
then
\begin{equation}\lab{608}
1-\alpha (1+\sup f)(C_1A)^2\geq\frac 12 >0.
\end{equation}
 Then from (\ref{606}) and (\ref{608}), we can get
 \begin{equation}\lab{609}
 \begin{aligned}
  \mid \bigtriangledown u\mid^2\leq \frac{1+C_2(C_1A)^2}{2\alpha
  \cdot\frac 12}e^u\leq \frac{1+C_2}{\alpha }e^u.
 \end{aligned}
 \end{equation}
 So $\mid\bigtriangledown u\mid^2$ has an upper bound. In the following we
 want to prove  that for any given  constant $\kappa$ satisfying $0<\kappa<1$,
  we can choose $A$ small enough (depending on $\kappa$) so that  $e^{-2u}F(t,\cdot)>\kappa$.
 In the above section, we have
seen that when $t=0$, the equation has an unique solution $u=-\ln
A$. So $e^{-2u}F(0,\cdot)\equiv 1$. By the continuity assumption
(\ref{1506}), we only need to prove that there is not $t=t_0\in
{\bf T}$ such that $\inf (e^{-2u}F(t_0,\cdot))=\kappa$. If not,
there is a $t_0\in {\bf T}$ and $q_1$ such that $F(t_0,q_1)=\inf
(e^{-2u}F(t_0,\cdot))=\kappa$. We fix this $t_0$ and  will get the
contradiction if we choose $A$ small enough. So when $t=t_0$, we
assume
\begin{equation}\lab{610}
 \inf (e^{-2u}F)=\kappa.
\end{equation}
Applying (\ref{610}) to  (\ref{604}), we get
\begin{equation}\lab{611}
1-2\alpha \{1-\alpha(1+\sup f)(C_1A)^2\}e^{-u}\mid\bigtriangledown
u\mid^2+C_2(C_1A)^2\geq \kappa.
\end{equation}
Then (\ref{608}) and  (\ref{611}) imply
\begin{equation}\lab{612}
 \begin{aligned}
   e^{-u}\mid \bigtriangledown u\mid^2
  \leq &\frac
 {1-\kappa+C_2(C_ 1A)^2}{2\alpha\{1-\alpha(1+\sup
 f)(C_1A)^2\}}\\
 =&
\frac{1-\kappa}{2\alpha}+\frac{C_2(C_1A)^2+(1-\kappa)\alpha(1+\sup
f)(C_1A)^2}
 {2\alpha\{1-\alpha(1+\sup f)(C_1A)^2\}}\\
 \leq &\frac{1-\kappa}{2\alpha}+\left(\frac {C_2} {\alpha}+1+\sup
 f\right)(C_1A)^2.
 \end{aligned}
 \end{equation}
  We  apply the  maximum principle to the function
\begin{equation}\lab{613}
G=1-2\alpha e^{-u}\mid \bigtriangledown u\mid^2+2\alpha
e^{-\epsilon u}-2\alpha e^{-\varepsilon\inf u},
\end{equation}
where $\varepsilon$ is some constant satisfying $0<\varepsilon<1$
which will be determined later. Comparing $G$ (\ref{613}) to
$e^{-2u}F$ (\ref{602}), we get
\begin{equation}\lab{614}
e^{-2u}F-G=e^{-2u}O(1)-2\alpha e^{-\varepsilon u}+2\alpha
e^{-\varepsilon\inf u}
\end{equation}
and from $\inf (e^{-2u}F)=\kappa$, we  see
\begin{equation}\lab{615}
\kappa-\sup (e^{-2u}\mid O(1)\mid)-2\alpha e^{-\varepsilon \inf
u}\leq \inf G\leq \kappa+\sup (e^{-2u}\mid O(1)\mid)+2\alpha
e^{-\varepsilon \inf u}.
\end{equation}
We can use (\ref{605}) and (\ref{609}) to estimate
\begin{equation*}
\begin{aligned}
\sup\mid O(1)\mid\leq &2\alpha \sup f +\alpha^2(\sup
f)^2(C_1A)^2+2\alpha\sup f\left\{(1+C_2)\right\}\\
&+2\alpha\left\{(1+C_2)\right\}+2\alpha^2(C_1A)\sup\mid\bigtriangledown
f\mid^2\\
&+2\alpha^2(C_1A)\sup\mid\bigtriangleup f\mid+2\alpha
\sup\mid\mu\mid\\
\leq &2\alpha\sup f+\alpha^2(\sup f)^2+2\alpha(1+\sup f)\left(1+C_2\right)\\
&+2\alpha^2\sup\mid\bigtriangledown f\mid^2+2\alpha^2\sup
\mid\bigtriangleup f\mid+2\alpha \sup\mid\mu\mid\\
\leq &C_2+2\alpha(1+\sup f)(1+C_2).
\end{aligned}
\end{equation*}
So
\begin{equation}\lab{617}
\begin{aligned}
&\sup(e^{-2u}\mid O(1)\mid)+2\alpha e^{-\varepsilon \inf u}\\
\leq &(C_1A)^2\{C_2+2\alpha (1+\sup f)(1+C_2)\}+2\alpha
(C_1A)^\varepsilon\\
\leq &2\{C_2+2\alpha(1+\sup f)\}(C_1A)^\varepsilon\\
=&C_2'(C_1A)^\varepsilon,
\end{aligned}
\end{equation}
where
\begin{equation*}
C_2'=2\{C_2+2\alpha(1+\sup f)\}
\end{equation*}
 depends only on $\alpha$, $f$  and  $\mu$. Combining (\ref{615})
and (\ref{617}), we get
\begin{equation}\lab{619}
\kappa-C_2'(C_1A)^\varepsilon\leq \inf G\leq
\kappa+C_2'(C_1A)^\varepsilon.
\end{equation}
Let $G$ achieve the minimum at the point $q_2\in S$. At the point
$q_2$, we apply (\ref{617}) and (\ref{619}) to (\ref{614}) to
estimate
\begin{equation}\lab{620}
\begin{aligned}
e^{-2u(q_2)}F(q_2)=&G(q_2)+e^{-2u(q_2)}O(1)(q_2)-2\alpha
e^{-\varepsilon
u(q_2)}+2\alpha^{-\varepsilon\inf u}\\
\leq &\inf G+\sup(e^{-2u}\mid O(1)\mid)+2\alpha e^{-\varepsilon
\inf
u}\\
\leq &\kappa+2C_2'(C_1A)^\varepsilon.
\end{aligned}
\end{equation}
We  apply (\ref{619}) to (\ref{613}) to estimate
\begin{equation}\lab{621}
\begin{aligned}
e^{-u(q_2)}\mid\bigtriangledown
u\mid^2(q_2)=&(2\alpha)^{-1}\{1-G(q_2)+2\alpha e^{-\varepsilon
u(q_2)}-2\alpha
e^{-\varepsilon \inf u}\}\\
\geq &(2\alpha)^{-1} \{1-\inf G-2\alpha e^{-\varepsilon\inf u}\}\\
\geq &(2\alpha)^{-1}\{1-\kappa-C_2'(C_1A)^\varepsilon-2\alpha(C_1A)^\varepsilon\}\\
=&(1-\kappa)/(2\alpha)-(1+(2\alpha)^{-1}C_2')(C_1A)^\varepsilon.
\end{aligned}
\end{equation}
Take
\begin{equation*}
C_3=\max \{\alpha^{-1}C_2+1+\sup f, 2C_2', 1+(2\alpha)^{-1}C_2'\}.
\end{equation*}
Then (\ref{610}) and (\ref{620}) imply
\begin{equation}\lab{623}
\kappa\leq e^{-2u(q_2)}F(q_2)\leq \kappa+C_3(C_1A)^\varepsilon;
\end{equation}
(\ref{612}) and (\ref{621}) imply
\begin{equation}\lab{624}
(1-\kappa)/(2\alpha)-C_3(C_1A)^\varepsilon\leq
e^{-u(q_2)}\mid\bigtriangledown u\mid^2(q_2)\leq
(1-\kappa)/(2\alpha)+C_3(C_1A)^\varepsilon.
\end{equation}
We now compute $P(G)F$ at the point $q_2$. In the following we
replace $t\alpha f$ by $f$ and $t\mu$ by $\mu$. At the point
$q_2$, from $\bigtriangledown G(q_2)=0$, we have
\begin{equation}\lab{625}
\bigtriangledown(\mid\bigtriangledown
u\mid^2)=(\mid\bigtriangledown u\mid^2-\varepsilon
e^{(1-\varepsilon)u})\bigtriangledown u.
\end{equation}
Because $\omega_S$ is K\"ahler,  we can choose the normal
coordinate $(z_1,z_2)$ at the point $q_2$, i.e., $g_{i\bar
j}=\delta_{ij}$ and $dg_{i\bar j}=0$.  At the same time , we can
assume $\frac{\partial u}{\partial z_1}\neq 0$ and $\frac{\partial
u}{\partial z_2}=0$. Because $u$ is real, we can further  assume
that $\frac {\partial u}{\partial x_1}\geq 0$ and $\frac{\partial
u}{\partial y_1}=0$. So at the point $q_2$, $u_1=u_{\bar 1}$ and
\begin{equation}\lab{627}
2u_1u_1=2u_1u_{\bar 1}=2u_{\bar 1}u_{\bar 1}=\mid\bigtriangledown
u\mid^2.
\end{equation}
If we assume
\begin{equation*}
A<\left(\frac{1-\kappa}{2\alpha C_3}\right)^{\frac
1\varepsilon}C_1^{-1},
\end{equation*}
then
$$\frac {1-\kappa}{2\alpha}-C_3(C_1A)^{\varepsilon}>0.$$
Hence (\ref{624}) implies $\mid\bigtriangledown u\mid^2>0$ and
(\ref{625}) implies
\begin{equation}\lab{628}
\begin{aligned}
&u_{11}+u_{1\bar1}=u_{1\bar1}+u_{\bar 1\bar 1}=(\mid
\bigtriangledown
u\mid^2-\varepsilon^{(1-\varepsilon)u})/2\\
&u_{12}+u_{\bar 12}=u_{1\bar 2}+u_{\bar 1\bar 2}=0.
\end{aligned}
\end{equation}
From (\ref{503}) and (\ref{601}), we can see
\begin{equation}\lab{629}
\begin{aligned}
P(u)F=&(e^u+fe^{-u})\bigtriangleup u+16\alpha\frac{\det u_{i\bar
j}}{\det g_{i\bar j}}\\
=&\frac{1}{\alpha}\left\{(e^u+fe^{-u})^2+2\alpha(e^u+fe^{-u})\bigtriangleup
u+16\alpha^2\frac{\det u_{i\bar j}}{\det g_{i\bar
j}}\right\}\\
&-\alpha^{-1}(e^u+fe^{-u})^2-(e^u+fe^{-u})\bigtriangleup
u\\
=&\alpha^{-1}F-\alpha^{-1}(e^u+fe^{-u})(e^u+fe^{-u}+\alpha\bigtriangleup
u).
\end{aligned}
\end{equation}
We then compute
\begin{equation}\lab{630}
\begin{aligned}
P(2\alpha e^{-\varepsilon u})F=&-2\alpha \varepsilon
e^{-\varepsilon u}P(u)F+2\alpha\varepsilon^2e^{-\varepsilon
u}\cdot (2g'^{i\bar
j}u_iu_{\bar j})\cdot F\\
=&-2\alpha \varepsilon e^{-\varepsilon u}P(u)F+2\alpha
\varepsilon^2 e^{-\varepsilon u}\cdot g'^{1\bar
1}\mid \bigtriangledown u\mid^2\cdot F\\
=&-2\varepsilon e^{-\varepsilon u}F+2\varepsilon e^{-\varepsilon
u}(e^u+fe^{-u})(e^u+fe^{-u}+\alpha \bigtriangleup u)\\
&+2\alpha \varepsilon^2 e^{-\varepsilon u}\mid \bigtriangledown
u\mid^2(e^u+fe^{-u}+4\alpha u_{2\bar 2}).
\end{aligned}
\end{equation}
Using (\ref{625}), we derive
\begin{equation}\lab{631}
\begin{aligned}
&P(-2\alpha e^{-u}\mid\bigtriangledown u\mid^2)F\\
=&2\alpha e^{-u}\mid\bigtriangledown u\mid^2P(u)F-2\alpha
e^{-u}P(\mid\bigtriangledown u\mid^2)F\\
&-2\alpha e^{-u}\mid\bigtriangledown u\mid^2\cdot 2g'^{i\bar j}u_iu_{\bar j}F\\
&+2\alpha e^{-u}\cdot 2 g'^{i\bar j}\{\partial_i u\partial_{\bar
j}(\mid\bigtriangledown u\mid^2)+\partial_{\bar j}u\partial_i(\mid
\bigtriangledown
u\mid^2)\}\cdot F\\
=&2e^{-u}\mid\bigtriangledown
u\mid^2F-2e^{-u}(e^u+fe^{-u})\mid\bigtriangledown
u\mid^2(e^u+fe^{-u}+\alpha \bigtriangleup u)\\
&+\{2\alpha e^{-u}\mid\bigtriangledown u\mid^4-4\alpha\varepsilon
e^{-\varepsilon u}\mid\bigtriangledown u\mid^2\}(e^u+fe^{-u}+4\alpha u_{2\bar 2})\\
&-2\alpha e^{-u}P(\mid\bigtriangledown u\mid^2)F.
\end{aligned}
\end{equation}
Combining (\ref{630}) and (\ref{631}), we get
\begin{equation}\lab{632}
\begin{aligned}
P(G)F=&P(1-2\alpha e^{-u}\mid\bigtriangledown u\mid^2+2\alpha
e^{-\varepsilon
u}-2\alpha e^{-\varepsilon\inf u})F\\
=&\{2e^{-u}\mid\bigtriangledown u\mid^2-2\varepsilon
e^{-\varepsilon
u}\}F\\
&-\{2e^{-u}(e^u+fe^{-u})\mid\bigtriangledown
u\mid^2-2\varepsilon e^{-\varepsilon u}(e^u+fe^{-u})\}(e^u+fe^{-u}+\alpha \bigtriangleup u)\\
&+\{2\alpha e^{-u}\mid\bigtriangledown
u\mid^4+(2\alpha\varepsilon^2-4\alpha\varepsilon)e^{-\varepsilon
u}\mid\bigtriangledown u\mid^2\}(e^u+fe^{-u}+4\alpha
u_{2\bar 2})\\
&-2\alpha e^{-u}P(\mid\bigtriangledown u\mid^2)F.
\end{aligned}
\end{equation}
We now compute the term
\begin{equation}\lab{633}
\begin{aligned}
\alpha P(\mid\bigtriangledown u\mid^2)F =&4\alpha g'^{i\bar
j}(g^{k\bar l}u_ku_{\bar l})_{i\bar j}F\\
=& 4\alpha g'^{i\bar j}\{u_{i\bar
jk}u_{\bar k}+u_{i\bar j\bar k}u_{k}\}F \\
&+4\alpha g'^{i\bar j}\{u_{i\bar k}u_{k\bar j}+u_{ik}u_{\bar k\bar
j}+\partial_i\partial_{\bar j}(g^{1\bar 1})u_1u_{\bar 1}\}F\\
\geq &4\alpha g'^{i\bar j}\{u_{i\bar jk}u_{\bar k}+u_{i\bar j\bar
k}u_{k}\}F +4\alpha g'^{i\bar j}\{u_{i\bar k}u_{k\bar
j}\}F\\
&+4\alpha g^{i\bar j}\{u_{i1}u_{\bar 1\bar j}\} F+4\alpha
g'^{i\bar j}\{\partial_i\partial_{\bar j}(g^{1\bar 1})u_1u_{\bar
1}\}F.
\end{aligned}
\end{equation}
We deal with the first term in (\ref{633}) by applying the
definition of $g'_{i\bar j}$,
\begin{equation*}
\begin{aligned}
&4\alpha g'^{i\bar j}\{u_{i\bar jk}u_{\bar k}+u_{i\bar j\bar
k}u_k\}F\\
=&4\alpha \{g'_{1\bar 1}u_{2\bar 2k}+g'_{2\bar 2}u_{1\bar
1k}-g'_{1\bar 2}u_{2\bar 1k}-g'_{2\bar 1}u_{1\bar 2k}\}u_{\bar
k}(\det g_{i\bar j})^{-1}\\
&+4\alpha \{g'_{1\bar 1}u_{2\bar 2\bar k}+g'_{2\bar 2}u_{1\bar
1\bar k}-g'_{1\bar 2}u_{2\bar 1\bar k}-g'_{2\bar 1}u_{1\bar 2\bar
k}\}u_{k}(\det g_{i\bar j})^{-1}\\
=&4\alpha (e^u+fe^{-u})\{g^{i\bar j}u_{i\bar jk}u_{\bar
k}+g^{i\bar
j}u_{i\bar j\bar k}u_k\}\\
&+16\alpha^2\{u_{1\bar 1}u_{2\bar 2k}+u_{2\bar 2}u_{1\bar 1
k}-u_{1\bar 2}u_{2\bar 1 k}-u_{2\bar 1}u_{1\bar 2k}\}u_{\bar
k}(\det g_{i\bar j})^{-1}\\
&+16\alpha^2\{u_{1\bar 1}u_{2\bar 2\bar k}+u_{2\bar 2}u_{1\bar 1
\bar k}-u_{1\bar 2}u_{2\bar 1\bar k}-u_{2\bar 1}u_{1\bar 2\bar
k}\}u_{k}\det g_{i\bar j})^{-1}\\
=&2\alpha(e^u+fe^{-u})\bigtriangledown\bigtriangleup u\cdot
\bigtriangledown u+16\alpha^2\bigtriangledown\left(\frac{\det
u_{i\bar j}}{\det g_{i\bar j}}\right)\cdot \bigtriangledown u.
\end{aligned}
\end{equation*}
Using the equation to the last term of above equality, we find
\begin{equation*}
\begin{aligned}
&4\alpha g'^{i\bar j}\{u_{i\bar jk}u_{\bar k}+u_{i\bar j\bar k}u_k\}F \\
=&2\alpha(e^u+fe^{-u})\bigtriangledown\bigtriangleup
u\cdot\bigtriangledown
u-2\alpha\bigtriangledown\bigtriangleup(e^u-fe^{-u})\cdot
\bigtriangledown u-2\alpha \bigtriangledown\mu\cdot
\bigtriangledown u\\
=&-2\alpha (e^u-fe^{-u})\mid\bigtriangledown u\mid^2\bigtriangleup
u-2\alpha(e^u+fe^{-u})\mid\bigtriangledown
u\mid^4\\
&-2\alpha(e^u-fe^{-u})\bigtriangledown\mid\bigtriangledown
u\mid^2\cdot \bigtriangledown u-4\alpha
e^{-u}\bigtriangledown(\bigtriangledown u\cdot \bigtriangledown
f)\cdot \bigtriangledown u\\
&+6\alpha e^{-u}\mid\bigtriangledown u\mid^2\bigtriangledown
u\cdot\bigtriangledown f-2\alpha e^{-u}\mid\bigtriangledown
u\mid^2\bigtriangleup f\\
&+2\alpha e^{-u}\bigtriangledown\bigtriangleup f\cdot
\bigtriangledown u-2\alpha\bigtriangledown\mu\cdot
\bigtriangledown u-2\alpha e^{-u}(\bigtriangledown u\cdot
\bigtriangledown f)\bigtriangleup u.
\end{aligned}
\end{equation*}
From (\ref{609}), we see $e^{-u}\mid\bigtriangledown u\mid^2<C_4$,
where $C_4$ only depends on $\alpha$, $f$ and $\mu$ and does not
depend on  $A$.  In the following we use $C_4$ in the generic
sense. We have gotten $\mid \bigtriangledown u\mid^2\leq
C_4e^{u}$. Our assumptions of $A$ implies $e^u>1$, $\mid
\bigtriangledown u\mid\leq C_4 e^u$. In the following we will deal
with such small terms. So  we have
\begin{equation}\lab{636}
\begin{aligned}
&4\alpha g'^{i\bar j}\{u_{i\bar jk}u_{\bar k}+u_{i\bar j\bar k}u_k\}F \\
\geq&-2\alpha (e^u-fe^{-u})\mid\bigtriangledown
u\mid^2\bigtriangleup u-2\alpha(e^u+fe^{-u})\mid\bigtriangledown
u\mid^4\\
&-2\alpha(e^u-fe^{-u})\bigtriangledown\mid\bigtriangledown
u\mid^2\cdot \bigtriangledown u-4\alpha
e^{-u}\bigtriangledown(\bigtriangledown u\cdot \bigtriangledown
f)\cdot \bigtriangledown u\\
&-C_4e^u-C_4(e^u+fe^{-u}+\alpha\bigtriangleup u)
\end{aligned}
\end{equation}
Applying (\ref{628}), we find
\begin{equation}\lab{638}
\begin{aligned}
&-4\alpha e^{-u}\bigtriangledown (\bigtriangledown
u\cdot \bigtriangledown f)\cdot \bigtriangledown u\\
=&-4\alpha e^{-u}\{(u_if_{\bar i}+u_{\bar i} f_i)_ku_{\bar
k}+(u_if_{\bar i}+u_{\bar i}f_i)_{\bar k}u_k\}\\
=&-4\alpha e^{-u}\{u_{ik}f_{\bar i}u_{\bar k}+u_{\bar i
k}f_iu_{\bar k}+u_{i\bar k}f_{\bar i}u_k+u_{\bar i\bar
k}f_iu_k\}\\
&-4\alpha e^{-u}\{u_if_{\bar i k}u_{\bar k}+u_{\bar
i}f_{ik}u_{\bar k}+u_if_{\bar i\bar
k}u_k+u_{\bar i}f_{i\bar k}u_k\}\\
=&-4\alpha e^{-u}\{u_{i1}f_{\bar i}+u_{\bar i\bar 1}f_i+u_{\bar
i1}f_i+u_{i\bar 1}f_{\bar i}\}u_1\\
&-2\alpha e^{-u}\{f_{1\bar 1}+f_{11}+f_{\bar 1\bar 1}+f_{1\bar
1}\}\mid\bigtriangledown u\mid^2\\
\geq & -4\alpha e^{-u}\{(u_{i1}+u_{i\bar 1})f_{\bar i}+(u_{\bar
i1}+u_{\bar i\bar 1})f_i\}u_1-C_4\\
= &-4\alpha e^{-u}\left\{\frac 12(\mid\bigtriangledown
u\mid^2-e^{(1-\varepsilon)u})f_{\bar 1}+\frac
12(\mid\bigtriangledown
u\mid^2-e^{(1\varepsilon)u})f_1\right\}u_1-C_4\\
\geq &-C_4e^u.
\end{aligned}
\end{equation}
Inserting (\ref{638}) into (\ref{636}) and applying (\ref{625}),
we find
\begin{equation}\lab{637}
\begin{aligned}
&4\alpha g'^{i\bar j}\{u_{i\bar jk}u_{\bar k}+u_{i\bar j\bar k}u_k\}F \\
\geq&-2\alpha (e^u-fe^{-u})\mid\bigtriangledown
u\mid^2\bigtriangleup u-2\alpha(e^u+fe^{-u})\mid\bigtriangledown
u\mid^4\\
&-2\alpha(e^u-fe^{-u})\mid\bigtriangledown
u\mid^4+2\alpha\varepsilon(e^u-fe^{-u})e^{(1-\varepsilon)u}\mid\bigtriangledown
u\mid^2\\
&-C_4e^u-C_4(e^u+fe^{-u}+\alpha\bigtriangleup u)\\
\geq &-2(e^u-fe^{-u})\mid\bigtriangledown
u\mid^2(e^u+fe^{-u}+\alpha \bigtriangleup
u)+2(e^u-fe^{-u})(e^{u}+fe^{-u})\mid\bigtriangledown u
\mid^2\\
&-4\alpha e^u\mid\bigtriangledown
u\mid^4+2\alpha\varepsilon(e^u-fe^{-u})e^{(1-\varepsilon)u}\mid\bigtriangledown
u\mid^2\\
&-C_4e^u-C_4(e^u+fe^{-u}+\alpha\bigtriangleup u)\\
=&-2(e^u-fe^{-u})\mid\bigtriangledown u\mid^2(e^u+fe^{-u}+\alpha
\bigtriangleup u)+2\mid\bigtriangledown
u\mid^2(F-O(1))\\
&-2f^2e^{-2u}\mid\bigtriangledown
u\mid^2+2\alpha\varepsilon(e^u-fe^{-u})e^{(1-\varepsilon)u}\mid\bigtriangledown
u\mid^2\\
&-C_4e^u-C_4(e^u+fe^{-u}+\alpha\bigtriangleup u)\\
\geq &2\mid\bigtriangledown
u\mid^2F-2(e^u-fe^{-u})\mid\bigtriangledown
u\mid^2(e^u+fe^{-u}+\alpha \bigtriangleup u)\\
&+2\alpha\varepsilon e^{(2-\varepsilon)u}\mid\bigtriangledown
u\mid^2-C_4e^u-C_4(e^u+fe^{-u}+\alpha\bigtriangleup u).
\end{aligned}
\end{equation}
Next we deal with the second term in (\ref{633}). We compute
\begin{equation}\lab{640}
\begin{aligned}
&4\alpha g'^{i\bar j}(u_{i\bar k}u_{k\bar j})F\\
=&4\alpha (e^u+fe^{-u})(u_{i\bar k}u_{k\bar i})\\
&+16\alpha^2 \{u_{1\bar1}u_{2\bar k}u_{k\bar 2}+u_{2\bar
2}u_{1\bar k}u_{k\bar 1}-u_{1\bar 2}u_{2\bar k}u_{k\bar
1}-u_{2\bar 1}u_{1\bar k}u_{k\bar 2}\}\\
=&4\alpha (e^u+fe^{-u})(u_{1\bar 1}+u_{2\bar 2})^2-8\alpha
(e^u+fe^{-u})(u_{1\bar 1}u_{2\bar 2}-u_{1\bar 2}u_{2\bar 1})\\
&+16\alpha^2(u_{1\bar 1}+u_{2\bar 2})(u_{1\bar 1}u_{2\bar
2}-u_{1\bar
2}u_{2\bar 1})\\
 =&\alpha
(e^u+fe^{-u})(\bigtriangleup u)^2-8\alpha
(e^u+fe^{-u})\left(\frac{\det u_{i\bar j}}{\det g_{i\bar
j}}\right)+8\alpha^2\bigtriangleup u\left(\frac{\det u_{i\bar
j}}{\det g_{i\bar j}}\right).
 \end{aligned}
\end{equation}
Using the equation, we have
\begin{equation}\lab{641}
\begin{aligned}
8\alpha^2\bigtriangleup u\left(\frac{\det u_{i\bar j}}{\det
g_{i\bar j}}\right)=&-\alpha\bigtriangleup
u\bigtriangleup(e^u-fe^{-u})-\alpha \mu\bigtriangleup u\\
=&-\alpha(e^u+fe^{-u})(\bigtriangleup
u)^2-\alpha(e^u-fe^{-u})\mid\bigtriangledown u\mid^2\bigtriangleup
u\\
&-\alpha\{2e^{-u}\bigtriangledown u\cdot \bigtriangledown
f-e^{-u}\bigtriangleup f+\alpha\mu\}\bigtriangleup u\\
\geq &-\alpha(e^u+fe^{-u})(\bigtriangleup
u)^2-\alpha(e^u-fe^{-u})\mid\bigtriangledown u\mid^2\bigtriangleup
u\\
&-C_4e^u-C_4(e^u+fe^{-u}+\alpha \bigtriangleup u),
\end{aligned}
\end{equation}
and
\begin{equation}\lab{642}
\begin{aligned}
&-8\alpha(e^u+fe^{-u})\frac{\det u_{i\bar j}}{\det g_{i\bar j}}\\
=&(e^u+fe^{-u})\bigtriangleup (e^u-fe^{-u})+
\mu(e^u+fe^{-u})\\
=&(e^u+fe^{-u})^2\bigtriangleup
u+(e^u+fe^{-u})(e^u-fe^{-u})\mid\bigtriangledown u\mid^2\\
&+(e^u+fe^{-u})\{2e^{-u}\bigtriangledown u\cdot \bigtriangledown
f-e^{-u}\bigtriangleup f+\mu\} \\
\geq &(e^u+fe^{-u})^2\bigtriangleup
u+(e^u+fe^{-u})(e^u-fe^{-u})\mid\bigtriangledown u\mid^2-C_4e^u.
\end{aligned}
\end{equation}
Inserting (\ref{641}) and (\ref{642}) into (\ref{640}), we get the
following estimate of the second term:
\begin{equation}\lab{643}
\begin{aligned}
&4\alpha g'^{i\bar j}(u_{i\bar k}u_{k\bar j})F\\
\geq &-\alpha(e^u-fe^{-u})\mid\bigtriangledown
u\mid^2\bigtriangleup u+(e^u+fe^{-u})^2\bigtriangleup
u\\
&+(e^u+fe^{-u})(e^u-fe^{-u})\mid\bigtriangledown
u\mid^2-C_4e^u-C_4(e^u+fe^{-u}+\alpha \bigtriangleup u)\\
\geq&\{\alpha^{-1}(e^u+fe^{-u})^2-(e^u-fe^{-u})\mid\bigtriangledown
u\mid^2\}(e^u+fe^{-u}+\alpha \bigtriangleup u)\\
&-\alpha^{-1}
(e^u+fe^{-u})^3+2(e^u+fe^{-u})(e^u-fe^{-u})\mid\bigtriangledown
u\mid^2\\
&-C_4e^u-C_4(e^u+fe^{-u}+\alpha \bigtriangleup u)\\
\geq&\{\alpha^{-1}(e^u+fe^{-u})^2-(e^u-fe^{-u})\mid\bigtriangledown
u\mid^2\}(e^u+fe^{-u}+\alpha \bigtriangleup u)\\
&-\alpha^{-1}(e^u+fe^{-u})F-C_4
e^{u}-C_4(e^u+fe^{-u}+\alpha\bigtriangleup u).
\end{aligned}
\end{equation}
Then we compute  the third  term in (\ref{633}). By denoting
$a=\frac{1}{2}(\mid\bigtriangledown u\mid^2-\varepsilon
e^{(1-\varepsilon)u})$, we can use (\ref{628}) to prove
\begin{equation}\lab{644}
\begin{aligned}
&4\alpha g'^{i\bar j}(u_{i1}u_{\bar 1\bar j})F\\
=&4\alpha (e^u+fe^{-u}+4\alpha u_{2\bar 2})u_{11}u_{\bar 1\bar
1}-16\alpha^2 u_{1\bar 2}u_{21}u_{\bar 1\bar 1}\\
&-16\alpha^2 u_{2\bar 1}u_{11}u_{\bar 1\bar
2}+4\alpha(e^u+fe^{-u}+4\alpha u_{1\bar
1})u_{21}u_{\bar 1\bar 2}\\
 =&4\alpha (e^u+fe^{-u}+4\alpha u_{2\bar
2})(a-u_{1\bar
1})^2+16\alpha^2 u_{1\bar 2}u_{2\bar 1}(a-u_{1\bar 1})\times 2\\
&+4\alpha(e^u+fe^{-u}+4\alpha u_{1\bar
1})u_{1\bar 2}u_{\bar 21}\\
=&4\alpha(e^u+fe^{-u})a^2-8\alpha a(e^u+fe^u)u_{1\bar
1}+4\alpha(e^u+fe^{-u})u_{1\bar 1}^2+16\alpha^2 a^2u_{2\bar
2}\\
&+4\alpha(e^u+fe^{-u})u_{1\bar 2}u_{2\bar 1}+16\alpha^2u_{1\bar
1}\frac{\det u_{i\bar j}}{\det g_{i\bar j}}-32\alpha^2 a\frac{\det
u_{i\bar j}}{\det g_{i\bar j}}.
\end{aligned}
\end{equation}
Using the equation again, we have
\begin{equation}\lab{645}
\begin{aligned}
16\alpha^2u_{1\bar 1}\frac {\det u_{i\bar j}}{\det g_{i\bar j}}=&
-2\alpha u_{1\bar 1}\{\bigtriangleup(e^u-fe^{-u})+\mu\}\\
=&-2\alpha u_{1\bar 1}(e^u+fe^{-u})\bigtriangleup
u-2\alpha(e^u-fe^{-u})\mid\bigtriangledown u\mid^2u_{1\bar 1}\\
&-2\alpha\{2e^{-u}\bigtriangledown u\cdot \bigtriangledown
f-e^{-u}\bigtriangleup f+\mu\}u_{1\bar 1}\\
\geq &-4\alpha(e^u+fe^{-u})u_{1\bar
1}^2-4\alpha(e^u+fe^{-u})u_{1\bar 1}u_{2\bar
2}\\
&-\alpha(e^{u}-fe^{-u})\mid\bigtriangledown u\mid^2\bigtriangleup
u+2\alpha(e^u-fe^{-u})\mid\bigtriangledown u\mid^2u_{2\bar
2}\\
&-C_4e^u-C_4(e^u+fe^{-u}+4\alpha u_{1\bar 1})
\end{aligned}
\end{equation}
and
\begin{equation}\lab{646}
\begin{aligned}
-32\alpha^2 a\frac{\det u_{i\bar j}}{\det g_{i\bar j}}=&4\alpha a
\{\bigtriangleup(e^u-fe^{-u})+\mu\} \\
=&4\alpha a (e^u+fe^{-u})\bigtriangleup u+4\alpha a
(e^u-fe^{-u})\mid\bigtriangledown u\mid^2\\
&+4\alpha a\{2e^{-u}\bigtriangledown u\cdot \bigtriangledown
f-e^{-u}\bigtriangleup f+\mu\}\\
\geq&8\alpha a (e^u+fe^{-u})u_{1\bar 1}+8\alpha a
(e^u+fe^{-u})u_{2\bar 2}\\
&+4\alpha a (e^u-fe^{-u})\mid\bigtriangledown u\mid^2-C_4e^u
\end{aligned}
\end{equation}
Inserting (\ref{645}) and (\ref{646}) into (\ref{644}) and
simplifying, we get
\begin{equation*}
\begin{aligned}
 &4\alpha g'^{i\bar j}u_{i1}u_{\bar 1\bar j}F\\
\geq &-\alpha(e^u-fe^{-u})\mid\bigtriangledown
u\mid^2\bigtriangleup
u\\
&+\{16\alpha^2a^2+8\alpha a(e^u+fe^{-u})+2\alpha (e^u-fe^{-u})\mid
\bigtriangledown u\mid^2\}u_{2\bar
2}\\
&+4\alpha a^2(e^u+fe^{-u})+4\alpha
a(e^u-fe^{-u})\mid\bigtriangledown u\mid^2\\
&-4\alpha (e^u+fe^{-u})\frac {\det u_{i\bar j}}{\det g_{i\bar
j}}-C_4 e^u-C_4(e^u+fe^{-u}+\alpha \bigtriangleup u)\\
\geq&\frac 12 (e^u+fe^{-u})^2\bigtriangleup
u-\alpha(e^u-fe^{-u})\mid\bigtriangledown u\mid^2\bigtriangleup
u\\
&+\{16\alpha^2a^2+8\alpha a(e^u+fe^{-u})+2\alpha (e^u-fe^{-u})\mid
\bigtriangledown u\mid^2\}u_{2\bar
2}\\
&+4\alpha a^2(e^u+fe^{-u})+4\alpha
a(e^u-fe^{-u})\mid\bigtriangledown u\mid^2\\
&+\frac 1{2} (e^u+fe^{-u})(e^u-fe^{-u})\mid\bigtriangledown u\mid^2
-C_4 e^u-C_4(e^u+fe^{-u}+\alpha \bigtriangleup u)\\
\geq &\frac{1}{2\alpha}F(e^u+fe^{-u}+\alpha\bigtriangleup
u)-\frac{1}{2\alpha}(e^u+fe^{-u})F-2aF\\
&+\{4\alpha a^2+2a(e^u+fe^{-u})+\frac
12(e^u-fe^{-u})\mid\bigtriangledown u\mid^2\}(e^u+fe^{-u}+4\alpha
u_{2\bar 2})\\
&-C_4 e^u-C_4(e^u+fe^{-u}+\alpha\bigtriangleup u)\\
\end{aligned}
\end{equation*}
Now putting $a=\frac 12 (\mid\bigtriangledown u\mid^2-\varepsilon
e^{(1-\varepsilon)u})$ into above inequality and simplifying, we
conclude an  estimate of the third term:
\begin{equation}\lab{648}
\begin{aligned}
&4\alpha g'^{i\bar j}u_{i1}u_{\bar 1\bar j}F\\
\geq &\frac{1}{2\alpha}F(e^u+fe^{-u}+\alpha\bigtriangleup u)\\
&+\{\varepsilon
e^{(1-\varepsilon)u}-\frac{1}{2\alpha}(e^u+fe^{-u})-\mid\bigtriangledown
u\mid^2\}F\\
&+\{\frac 32 e^u\mid \bigtriangledown u\mid^2+\alpha \mid
\bigtriangledown u\mid^4-2\alpha \varepsilon
e^{(1-\varepsilon)u}\mid \bigtriangledown u\mid^2-\varepsilon
e^{(2-\varepsilon)u}\}(e^u+fe^{-u}+4\alpha u_{2\bar 2})\\
&-C_4e^u-C_4(e^u+fe^{-u}+\alpha \bigtriangleup u).
\end{aligned}
\end{equation}
The last term of (\ref{633}) is
\begin{equation}\lab{649}
4\alpha g'^{i\bar j}\partial_{i\bar j}(g^{1\bar 1})u_1u_{\bar
1}F\geq -C_4\mid \bigtriangledown
u\mid^2(e^u+fe^{-u}+\alpha\bigtriangleup u)
\end{equation}
where $C_4$ also depends on the curvature bound of the given
metric $\omega_S$. Inserting (\ref{637}, \ref{643}, \ref{648},
\ref{649}) into (\ref{633}) and simplifying, we get
\begin{equation}\lab{650}
\begin{aligned}
&\alpha P(\mid\bigtriangledown u\mid^2)F\\
\geq &\{\mid \bigtriangledown
u\mid^2-\frac{3}{2\alpha}e^u+\varepsilon
e^{(1-\varepsilon)u}\}F+2\alpha\varepsilon e^{(2-\varepsilon)u}\mid\bigtriangledown u\mid^2\\
&+\{\frac{1}{2\alpha}F+\frac
1\alpha(e^u+fe^{-u})^2-3(e^u-fe^{-u})\mid\bigtriangledown
u\mid^2\}(e^u+fe^{-u}+\alpha\bigtriangleup u)\\
&+\{\frac{3}{2}e^u\mid \bigtriangledown
u\mid^2+\alpha\mid\bigtriangledown u\mid^4-2\alpha\varepsilon
e^{(1-\varepsilon)u}\mid\bigtriangledown
u\mid^2-\varepsilon e^{(2-\varepsilon)u}\}(e^u+fe^{-u}+4\alpha u_{2\bar 2})\\
&-C_4e^u-C_4e^u(e^u+fe^{-u}+\alpha\bigtriangleup u).
\end{aligned}
\end{equation}
In order to get above inequality, we have used  $e^{-u}F\leq C_4
e^u$. Then inserting (\ref{650}) into (\ref{632}),  we find
finally
\begin{equation}\lab{651}
\begin{aligned}
P(G)F&\leq \{\frac 3\alpha-4\varepsilon e^{-\varepsilon
u}\}F-4\alpha
\varepsilon e^{(1-\varepsilon)u}\mid\bigtriangledown u\mid^2+C_4\\
&-\{\frac 3\alpha e^{-u}F-2\varepsilon
e^{(1-\varepsilon)u}-C_4\}(e^u+fe^{-u}+\alpha\bigtriangleup u)\\
&-\{3\mid\bigtriangledown u\mid^2-2\varepsilon
e^{(1-\varepsilon)u}-2\alpha\varepsilon^2e^{-\varepsilon
u}\mid\bigtriangledown u\mid^2\}(e^u+fe^{-u}+4\alpha u_{2\bar
2})\\
\leq &\frac 3\alpha F-2\varepsilon e^{-\varepsilon u}F-2\varepsilon e^{(2-\varepsilon)u}+C_4\\
&-\{\frac 3\alpha e^{-u}F-2\varepsilon
e^{(1-\varepsilon)u}-C_4\}\frac{e^u+fe^{-u}+4\alpha u_{1\bar 1}}{2}\\
&-\{\frac{3}{\alpha}e^{-u}F-6\varepsilon
e^{(1-\varepsilon)u}+6\mid\bigtriangledown
u\mid^2-4\alpha\varepsilon^2 e^{-\varepsilon
u}\mid\bigtriangledown u\mid^2-C_4\}\frac{e^u+fe^{-u}+4\alpha
u_{2\bar
2}}{2}.\\
\end{aligned}
\end{equation}
Let
\begin{equation}\lab{652}
\begin{aligned}
a_1&=\frac 3\alpha F-2\varepsilon e^{-\varepsilon u}F-2\varepsilon e^{(2-\varepsilon)u}+C_4\\
a_2&=\frac{3}{\alpha}e^{-u}F-2\varepsilon e^{(1-\varepsilon)u}-C_4\\
a_3&=\frac{3}{\alpha}e^{-u}F-6\varepsilon
e^{(1-\varepsilon)u}+6\mid\bigtriangledown
u\mid^2-4\alpha\varepsilon^2 e^{-\varepsilon
u}\mid\bigtriangledown u\mid^2-C_4.
\end{aligned}
\end{equation}
Because at the point $q_2$, $P(G)F\geq 0$. Then (\ref{651})
implies
\begin{equation}\lab{653}
a_1\geq a_2\frac {e^u+ fe^{-u}+4\alpha u_{1\bar
1}}{2}+a_3\frac{e^u+ fe^{-u}+4\alpha u_{2\bar 2}}{2}.
\end{equation}
We fix $\kappa$ such that  $0<\kappa<1$. We choose
$\varepsilon>0$ satisfying
\begin{equation}\lab{654}
\varepsilon <\min
\left\{1,\alpha^{-1/2},(2\alpha)^{-1}\kappa\right\}
\end{equation}
Then \begin{equation*} 3-2\alpha\varepsilon^2>0
\end{equation*}
and
\begin{equation*}
\frac 3 \alpha \kappa -6\varepsilon>0.
\end{equation*}
We assume
\begin{equation}\lab{655}
A<\frac{\frac 3 \alpha  \kappa -6\varepsilon}{C_4}C_1^{-1}.
\end{equation}
Then  $\kappa,\varepsilon$ and $A$ satisfy
\begin{equation*}
\frac{3}{\alpha}\kappa-6\varepsilon -C_4C_1A> 0.
\end{equation*}
We find
\begin{equation*}
\begin{aligned}
a_2\geq& e^u\{\frac{3}{\alpha}e^{-2u}F-2\varepsilon
e^{-\varepsilon
u}-C_4e^{-u}\}\\
\geq &e^u\{\frac 3\alpha \kappa-2\varepsilon
(C_1A)^\varepsilon-C_4C_1A\}\\
\geq &e^u\{\frac 3\alpha \kappa-6\varepsilon-C_4C_1A\} > 0
\end{aligned}
\end{equation*}
 and
\begin{equation*}
\begin{aligned}
a_3\geq &e^u\{\frac 3\alpha \kappa-6\varepsilon
(C_1A)^\varepsilon-C_4C_1A\}+2\mid\bigtriangledown u\mid^2(
3-2\alpha\varepsilon^2(C_1A)^\varepsilon)\\
\geq &e^u\{\frac 3\alpha
\kappa-6\varepsilon-C_4C_1A\}+2\mid\bigtriangledown u\mid^2(
3-2\alpha\varepsilon^2)> 0.
\end{aligned}
\end{equation*}
Applying arithmetic-geometric inequality to (\ref{653}), we find
\begin{eqnarray}\lab{659}
\begin{aligned}
a_1^2\geq&\left(a_2\frac {e^u+ fe^{-u}+4 u_{1\bar
1}}{2}+a_3\frac{e^u+fe^{-u}+4\alpha u_{2\bar
2}}{2}\right)^2\\
\geq &a_2a_3 (e^u+ fe^{-u}+4 u_{1\bar 1})(e^u+
fe^{-u}+4\alpha u_{2\bar 2})\\
\geq&a_2a_3F.
\end{aligned}
\end{eqnarray}
Using (\ref{602}), we can write $a_3$ as
\begin{equation}\lab{660}
\begin{aligned}
a_3=&\frac 3 \alpha  e^{-u}F+6\mid\bigtriangledown
u\mid^2-6\varepsilon e^{(1-\varepsilon)u}-4\alpha\varepsilon^2
e^{-\varepsilon
u}\mid\bigtriangledown u\mid^2-C_4\\
=&\frac 3 \alpha e^u-6\varepsilon
e^{(1-\varepsilon)u}-4\alpha\varepsilon^2 e^{-\varepsilon
u}\mid\bigtriangledown u\mid^2-e^{-u}O(1)-C_4.
\end{aligned}
\end{equation}
Inserting (\ref{652}) and (\ref{660}) into (\ref{659}) and
simplifying , we can get
\begin{equation}\lab{663}
\begin{aligned}
&4\varepsilon^2 e^{-2\varepsilon
u}F^2+4\varepsilon^2e^{2(2-\varepsilon)u}+12\varepsilon^2
e^{-(1+\varepsilon)u}\mid\bigtriangledown u\mid^2F^2+C_4'e^{2u}\\
\geq&\frac 6 \alpha \varepsilon e^{(2-\varepsilon)u}F-\frac 6
\alpha \varepsilon e^{-\varepsilon u}
F^2+4\varepsilon^2e^{2(1-\varepsilon)u}F\\
\geq &\frac 6\alpha \varepsilon e^{-\varepsilon
u}F(e^{2u}-F)\\
=&\frac 6\alpha \varepsilon e^{-\varepsilon u}F(2\alpha
e^u\mid\bigtriangledown u\mid^2-O(1))\\
\geq&12\varepsilon e^{(1-\varepsilon)u}\mid\bigtriangledown
u\mid^2F-C_4'e^{2u},
\end{aligned}
\end{equation}
where $C_4'$ may be bigger than $C_4$ and we shall denote  it by
$C_4$.  Dividing (\ref{663}) by $4\varepsilon e^{-\varepsilon
u}e^{2u}F$, we  get
\begin{equation}\lab{662}
\begin{aligned}
&\varepsilon e^{-\varepsilon u}(e^{-2u}F)+\varepsilon
\frac{e^{-\varepsilon u}}{e^{-2u}F}+3\varepsilon
(e^{-u}\mid\bigtriangledown u\mid^2)(e^{-2u}F)+C_4\frac{
e^{-(2-\varepsilon)u}}{\varepsilon e^{-2u}F}\geq 3
(e^{-u}\mid\bigtriangledown u\mid^2).
\end{aligned}
\end{equation}
Using the inequalities (\ref{623}) and (\ref{624}) to two sides of
above inequality, we obtain
\begin{equation}\lab{663}
\begin{aligned}
&\varepsilon e^{-\varepsilon u}(e^{-2u}F)+\varepsilon
\frac{e^{-\varepsilon u}}{e^{-2u}F}+3\varepsilon
(e^{-u}\mid\bigtriangledown u\mid^2)(e^{-2u}F)+C_4\frac{
e^{-(2-\varepsilon)u}}{\varepsilon e^{-2u}F}\\ \leq&
\varepsilon(C_1A)^\varepsilon
(\kappa+C_3(C_1A)^\varepsilon)+\varepsilon \frac
{(C_1A)^\varepsilon}{\kappa}\\
&+3\varepsilon(\kappa+C_3(C_1A)^\varepsilon)
\left(\frac{1-\kappa}{2\alpha}+C_3(C_1A)^\varepsilon\right)+
C_4\frac{(C_1A)^{2-\varepsilon}}{\varepsilon
\kappa}\\
\leq&\left\{\kappa\varepsilon+\varepsilon C_3+\frac \varepsilon
\kappa+3\varepsilon \kappa C_3+3\varepsilon \frac
{1-\kappa}{2\alpha}C_3+3\varepsilon C_3^2+\frac {C_4}{\varepsilon
\kappa}\right\}(C_1A)^\varepsilon+\frac{3\varepsilon
\kappa}{2\alpha}(1-\kappa)\\
\leq&\left\{1+\frac \varepsilon\kappa
+\varepsilon\left(1+3\kappa+\frac 3{2\alpha}+3C_3\right)C_3+\frac
{C_4}{\varepsilon\kappa}\right\}(C_1A)^\varepsilon+\frac{3\varepsilon
\kappa}{2\alpha}(1-\kappa)
\end{aligned}
\end{equation}
and
\begin{equation}\lab{664}
3(e^{-u}\mid\bigtriangledown u\mid^2)\geq \frac 3
{2\alpha}(1-\kappa)-3C_3(C_1A)^\varepsilon.
\end{equation}
Applying (\ref{663}) and (\ref{664}) to (\ref{662}), we see
\begin{equation*}
\begin{aligned}
\left\{1+\frac \varepsilon\kappa
+3C_3+\varepsilon\left(1+3\kappa+\frac
3{2\alpha}+3C_3\right)C_3+\frac
{C_4}{\varepsilon\kappa}\right\}(C_1A)^\varepsilon\geq \frac 3
{2\alpha}(1-\kappa)(1-\varepsilon\kappa).
\end{aligned}
\end{equation*}
So at last we get at the point $(t_0,q_2)$,
\begin{equation}\lab{666}
A\geq\left(\frac{\frac 3
{2\alpha}(1-\kappa)(1-\varepsilon\kappa)}{\left\{1+\frac
\varepsilon\kappa +3C_3+\varepsilon\left(1+3\kappa+\frac
3{2\alpha}+3C_3\right)C_3+\frac
{C_4}{\varepsilon\kappa}\right\}(C_1A)^\varepsilon}\right)^{\frac
1\varepsilon}C_1^{-1}.
\end{equation}
Now it is easy to prove  the following
\begin{prop}\lab{prop 3}
Let $t\in {\bf T}$ and $u$ is a solution of equation (\ref{2801})
under the elliptic condition $(e^u+t\alpha
fe^{-u})\omega_S+2\alpha\sqrt{-1}\partial\bar\partial u>0$ and the
normalization $\left(\int e^{-4u}\right)^{\frac 14}=A$ and $\int
1\frac {\omega_S^2}{2!}=1$. Given any constant $\kappa\in (0,1)$,
we fix some positive constant $\varepsilon$ satisfying
\begin{equation}\lab{667}
\varepsilon<\min\{1,\alpha^{-\frac 1 2},(2\alpha)^{-1}\kappa\}.
\end{equation}
Suppose that $A$ satisfies
\begin{equation}\lab{668}
A<\min\left\{1, C_1^{-1},\{2\alpha(1+\sup f)\}^{-\frac
12}C_1^{-1},\left(\frac {1-\kappa}{2\alpha C_3}\right)^{\frac
1\varepsilon}C_1^{-1},\frac {\frac
{3}{\alpha}-6\varepsilon}{C_4}C_1^{-1}\right\}
\end{equation}
and
\begin{equation}\lab{669}
A<\left(\frac{\frac 3
{2\alpha}(1-\kappa)(1-\varepsilon\kappa)}{\left\{1+\frac
\varepsilon\kappa +3C_3+\varepsilon\left(1+3\kappa+\frac
3{2\alpha}+3C_3\right)C_3+\frac
{C_4}{\varepsilon\kappa}\right\}(C_1A)^\varepsilon}\right)^{\frac
1\varepsilon}C_1^{-1},
\end{equation}
where $C_1$ is determined in above section and depends on
$\alpha$, $f$ and $\mu$, and also depends on the Sobolev constant;
$C_3$ and $C_4$ are determined in above discussion and depend on
$\alpha$, $f$, $\mu$, and $C_4$ also depends the curvature bound
of $\omega_S$. Then $F>\kappa e^{2u}\geq \kappa(C_1A)^{-2}$.
\end{prop}
\begin{proof}
When $t=0$, the equation has an unique solution $u=-\ln A$ and so
$e^{-2u}F(0,\cdot)\equiv 1$. According to our continuity
assumption, we claim that for any $t\in {\bf T}$, $e^{-2u}F(t,
\cdot)>\kappa$. Otherwise  if there is a $t_0\in {\bf T}$ such
that the equation has a solution $u$ and $\inf (e^{-2u}F)=\kappa$.
Fix this $t_0$ and apply the  maximum principle to the function
$G=1-2\alpha e^{-u}\mid\bigtriangledown u\mid^2+2\alpha
e^{-\varepsilon u}-2\alpha e^{-\varepsilon \inf u}$. Let $G$
achieve the minimum at the point $q_2$. Then at point $q_2$,
$P(G)F>0$. From above discussion, we have gotten the inequality
(\ref{666}) at point $q_2$ under assumptions (\ref{667}) and
(\ref{668}), which contradicts to the assumption (\ref{669}). So
$e^{-2u}F>\kappa$ and then $F>\kappa e^{2u}>\kappa(C_1A)^{-2}$.
\end{proof}

\section{Second order estimate}\lab{sec: 8}
We now consider the  second order a priori estimate of $u$. Since
we have proved $F>\kappa(C_1A)^{-2}>0$, $e^u+fe^{-u}+\alpha
\bigtriangleup u\geq F^{\frac{1}{2}}>\kappa^{\frac
12}(C_1A)^{-1}>0$. It is sufficient to have an upper  estimate of
$e^u+fe^{-u}+\alpha \bigtriangleup u$. We fix some point and
choose the normal coordinate $(z_1,z_2)$ at this point for the
given  metric $g_{i\bar j}$, i.e., at this point, $g_{i\bar
j}=\delta_{ij}$ and $dg_{i\bar j}=0$. We replace $t\alpha f$ by
$f$ and $t\mu$ by $\mu$. We can rewrite the equation as
\begin{equation}\lab{701}
\frac{\det g'_{i\bar j}}{\det g_{i\bar j}}=F,
\end{equation}
where \begin{equation*}
F=(e^u+fe^{-u})^2-2\alpha(e^u-fe^{-u})\mid\bigtriangledown
u\mid^2-4\alpha e^{-u}\bigtriangledown u\cdot\bigtriangledown
f+2\alpha e^{-u}\bigtriangleup f-2\alpha\mu.
\end{equation*}
Differentiating  (\ref{701}), we have
\begin{equation}\lab{703}
g'^{i\bar j}\frac{\partial g'_{i\bar j}}{\partial z_k}=g^{i\bar
j}\frac{\partial g_{i\bar j}}{\partial
z_k}+\frac{1}{F}\frac{\partial F}{\partial z_k}.
\end{equation}
We differentiate (\ref{703}) again to obtain
\begin{equation*}
\begin{aligned}
&-g'^{i\bar q}g'^{p\bar j}\frac{\partial g'_{p\bar q}}{\partial
\bar{z}_l}\frac{\partial g'_{i\bar j}}{\partial z_k}+g'^{i\bar
j}\frac{\partial^2 g'_{i\bar j}}{\partial z_k\partial\bar z_l}\\
=&-g^{i\bar q}g^{p\bar j}\frac{\partial g_{p\bar q}}{\partial \bar
z_l}\frac{\partial g_{i\bar j}}{\partial z_k}+g^{i\bar
j}\frac{\partial^2 g_{i\bar j}}{\partial z_k\partial \bar
z_l}+\frac 1F\frac{\partial^2F}{\partial z_k\partial \bar
z_l}-\frac {1}{F^2}\frac{\partial F}{\partial z_k}\frac{\partial
F}{\partial \bar z_l},
\end{aligned}
\end{equation*}
or
\begin{equation}\lab{705}
\begin{aligned}
4\alpha g'^{i\bar j}\frac{\partial^4u}{\partial z_i\partial \bar
z_j\partial z_k\partial \bar z_l}&=g'^{i\bar q}g'^{p\bar
j}\frac{\partial g'_{p\bar q}}{\partial \bar{z}_l}\frac{\partial
g'_{i\bar j}}{\partial z_k}-g'^{i\bar
j}\frac{\partial^2[(e^u+fe^{-u})g_{i\bar j}]}{\partial
z_k\partial\bar z_l}\\
 &-g^{i\bar
q}g^{p\bar j}\frac{\partial g_{p\bar q}}{\partial \bar
z_l}\frac{\partial g_{i\bar j}}{\partial z_k}+g^{i\bar
j}\frac{\partial^2 g_{i\bar j}}{\partial z_k\partial \bar
z_l}\\
&+\frac 1F\frac{\partial^2F}{\partial z_k\partial \bar z_l}-\frac
{1}{F^2}\frac{\partial F}{\partial z_k}\frac{\partial F}{\partial
\bar z_l}.
\end{aligned}
\end{equation}
Contracting (\ref{705}) with $g^{k\bar l}$ and using the fact that
the metric $\omega_S$ is Ricci-flat and the coordinate is normal,
we have
\begin{equation*}
\begin{aligned}
4\alpha g'^{i\bar j}\frac{\partial^2}{\partial z_i\partial\bar
z_j}\left(g^{k\bar l}\frac{\partial^2u}{\partial z_k\partial \bar
z_l}\right)=& g^{k\bar l}g'^{i\bar q}g'^{p\bar j}\frac{\partial
g'_{p\bar q}}{\partial \bar{z}_l}\frac{\partial g'_{i\bar
j}}{\partial z_k}-g^{k\bar l}g'^{i\bar
j}\frac{\partial^2(e^u+fe^{-u})}{\partial
z_k\partial\bar z_l}g_{i\bar j}\\
&+g^{k\bar l}\frac 1F\frac{\partial^2F}{\partial z_k\partial \bar
z_l}-g^{k\bar l}\frac {1}{F^2}\frac{\partial F}{\partial
z_k}\frac{\partial F}{\partial
\bar z_l}\\
&+4\alpha g'^{i\bar j}\frac{\partial^2g^{k\bar l}}{\partial
z_i\partial \bar z_j}\frac{\partial^2u}{\partial z_k\partial\bar
z_l}.
\end{aligned}
\end{equation*}
Timing $F$ to above equation, we see
\begin{equation}\lab{707}
\begin{aligned}
\alpha P(\bigtriangleup u)F=&-2^{-1}\bigtriangleup
(e^u+fe^{-u})\sum g'^{i\bar i}\cdot F+
4\alpha g'^{i\bar j}(g^{k\bar l})_{i\bar j}\cdot u_{k\bar l}\cdot F\\
&+2^{-1} \bigtriangleup F-(2F)^{-1}\mid\bigtriangledown
F\mid^2+g^{k\bar l}g'^{i\bar q}g'^{p\bar j}g'_{i \bar jk}g'_{p\bar
q\bar l}\cdot F\\
 =&-(e^u+fe^{-u}+\alpha\bigtriangleup u)\bigtriangleup
(e^u+fe^{-u})+4\alpha g'^{i\bar j}(g^{k\bar l})_{i\bar j}\cdot u_{k\bar l}\cdot F\\
&+2^{-1} \bigtriangleup F-(2F)^{-1}\mid\bigtriangledown
F\mid^2+g^{k\bar l}g'^{i\bar q}g'^{p\bar j}g'_{i \bar jk}g'_{p\bar
q\bar l}\cdot F.
\end{aligned}
\end{equation}
We shall apply the maximum principle to the  function
\begin{equation*}
e^{-\lambda_1u+\lambda_2\mid\bigtriangledown
u\mid^2}\cdot(e^u+fe^{-u}+\alpha\bigtriangleup u),
\end{equation*}
where $\lambda_1$ and $\lambda_2$ are some positive constants
which will be determined later.  By  computation,
\begin{equation}\lab{709}
\begin{aligned}
&P(e^{-\lambda_1u+\lambda_2\mid\bigtriangledown
u\mid^2}\cdot(e^u+fe^{-u}+\alpha\bigtriangleup u))
\cdot e^{-(-\lambda_1 u+\lambda_2\mid\bigtriangledown u\mid^2)}\\
=&(e^u+fe^{-u}+\alpha\bigtriangleup
u)\cdot(-\lambda_1P(u)+\lambda_2P(\mid\bigtriangledown u\mid^2))\\
&+P(e^u+fe^{-u}+\alpha\bigtriangleup u)\\
&+(e^u+fe^{-u}+\alpha\bigtriangleup u)\cdot
\mid\bigtriangledown'(-\lambda_1 u+\lambda_2\mid\bigtriangledown
u\mid^2)\mid^2_{g'}\\
&+2\bigtriangledown'(-\lambda_1 u+\lambda_2\mid\bigtriangledown
u\mid^2)\cdot_{g'}\bigtriangledown'(e^u+fe^{-u}+\alpha\bigtriangleup
u),
\end{aligned}
\end{equation}
where we denote  $2g'^{i\bar j}\psi_i\psi_{\bar j}$ by
$\mid\bigtriangledown' \psi\mid^2_{g'}$ and $g'^{i\bar
j}(\psi_i\varphi_{\bar j}+\psi_{\bar j}\varphi_i)$ by
$\bigtriangledown'\psi\cdot_{g'}\bigtriangledown'\varphi$.
Applying the Schwarz' inequality to the last term of (\ref{709}),
we have
\begin{equation}\lab{710}
\begin{aligned}
&2\bigtriangledown'(-\lambda_1 u+\lambda_2\mid\bigtriangledown
u\mid^2)\cdot_{g'}\bigtriangledown'(e^u+fe^{-u}+\alpha\bigtriangleup
u)\\
\geq&-2\mid\bigtriangledown'(-\lambda_1
u+\lambda_2\mid\bigtriangledown
u\mid^2)\mid_{g'}\cdot\mid\bigtriangledown'(e^u+fe^{-u}+\alpha\bigtriangleup
u)\mid_{g'}\\
\geq&-(e^u+fe^{-u}+\alpha\bigtriangleup
u)\mid\bigtriangledown'(-\lambda_1 u+\lambda_2\mid\bigtriangledown
u\mid^2)\mid_{g'}^2\\
&-(e^u+fe^{-u}+\alpha\bigtriangleup
u)^{-1}\mid\bigtriangledown'(e^u+fe^{-u}+\alpha\bigtriangleup
u)\mid_{g'}^2.
\end{aligned}
\end{equation}
Inserting (\ref{710}) into (\ref{709}), we have
\begin{equation}\lab{711}
\begin{aligned}
&P(e^{-\lambda_1u+\lambda_2\mid\bigtriangledown
u\mid^2}\cdot(e^u+fe^{-u}+\alpha\bigtriangleup u))
\cdot e^{-(-\lambda_1 u+\lambda_2\mid\bigtriangledown u\mid^2)}\\
\geq&(e^u+fe^{-u}+\alpha\bigtriangleup
u)\cdot(-\lambda_1P(u)+\lambda_2P(\mid\bigtriangledown u\mid^2))\\
&+P(e^u+fe^{-u}+\alpha\bigtriangleup u)\\
&-(e^u+fe^{-u}+\alpha\bigtriangleup
u)^{-1}\mid\bigtriangledown'(e^u+fe^{-u}+\alpha\bigtriangleup
u)\mid_{g'}^2.
\end{aligned}
\end{equation}
In computing the last term of (\ref{711}), we assume that
$g_{i\bar j}=\delta_{ij}$ and $u_{i\bar j}=u_{i\bar i}\delta_{ij}$
at a point. Then using the method of \cite{Yau}, we find
\begin{equation}\lab{712}
\begin{aligned}
&(e^u+fe^{-u}+\alpha\bigtriangleup
u)^{-1}\mid\bigtriangledown'(e^u+fe^{-u}+\alpha\bigtriangleup
u)\mid_{g'}^2\\
&=(e^u+fe^{-u}+\alpha\bigtriangleup u)^{-1}\cdot 2g'^{i\bar
j}(e^u+fe^{-u}+\alpha\bigtriangleup
u)_i(e^u+fe^{-u}+\alpha\bigtriangleup u)_{\bar j}\\
&=\frac 12(e^u+fe^{-u}+\alpha\bigtriangleup u)^{-1}\sum_ig'^{i\bar
i}\left(\sum_k g_{k\bar ki}\right)\left(\sum_l g_{l\bar
l\bar i}\right)\\
&=\frac 12 (e^u+fe^{-u}+\alpha\bigtriangleup u)^{-1}\sum_i
g'^{i\bar i}\sum_k\left(\frac{g'_{k\bar ki}}{g'^{\frac
 12}_{k\bar k}}\cdot
g'^{\frac 12}_{k\bar k}\right)\sum_l\left(\frac{g'_{l\bar l\bar
i}}{g'^{\frac 12}_{l\bar l}}\cdot g'^{\frac 12}_{l\bar l}\right)\\
&\leq\frac 12 (e^u+fe^{-u}+\alpha\bigtriangleup u)^{-1}\sum_i
g'^{i\bar i}\sum_k(g'_{k\bar ki}g'_{k\bar k\bar i} g'^{k\bar
k})\sum_lg'_{l\bar l}\\
&=\sum_{ik}g'^{i\bar i}g'^{k\bar k}g'_{k\bar k i}g'_{k\bar k\bar
i}.
\end{aligned}
\end{equation}
Note that when $i\neq k$,
\begin{equation}\lab{713}
g'_{k\bar ki}=g'_{i\bar kk}+(e^u+fe^{-u})_i-[(e^u+fe^{-u})g_{i\bar
k}]_k=g'_{i\bar kk}+(e^u+fe^{-u})_i
\end{equation}
and
\begin{equation}\lab{714}
g'_{k\bar k\bar i}=g'_{k\bar i\bar k}+(e^u+fe^{-u})_{\bar
i}-[(e^u+fe^{-u})g_{k\bar i}]_{\bar k}=g'_{k\bar i\bar
k}+(e^u+fe^{-u})_{\bar i}.
\end{equation}
Inserting (\ref{713}) and (\ref{714}) into (\ref{712}), we see
\begin{equation*}
\begin{aligned}
&(e^u+fe^{-u}+\alpha\bigtriangleup
u)^{-1}\mid\bigtriangledown'(e^u+fe^{-u}+\alpha\bigtriangleup
u)\mid_{g'}^2\\
\leq & g'^{i\bar i}g'^{k\bar k}g'_{i\bar kk}g'_{k\bar i\bar k}\\
&+g'^{1\bar 1}g'^{2\bar 2}(g'_{1\bar 22}(e^u+fe^{-u})_{\bar
1}+g'_{2\bar 1\bar 2}(e^u+fe^{-u})_1)\\
&+g'^{1\bar 1}g'^{2\bar 2}(g'_{2\bar 11}(e^u+fe^{-u})_{\bar
2}+g'_{1\bar 2\bar 1}(e^u+fe^{-u})_2)\\
&+g'^{1\bar 1}g'^{2\bar 2}\{(e^u+fe^{-u})_1(e^u+fe^{-u})_{\bar
1}+(e^u+fe^{-u})_2(e^u+fe^{-u})_{\bar 2})\\
\leq & g'^{i\bar i}g'^{k\bar k}g'_{i\bar kk}g'_{k\bar i\bar k}\\
&+g'^{1\bar 1}g'^{2\bar 2}(g'_{2\bar 21}(e^u+fe^{-u})_{\bar
1}+g'_{2\bar 2\bar 1}(e^u+fe^{-u})_1)\\
&+g'^{1\bar 1}g'^{2\bar 2}(g'_{1\bar 12}(e^u+fe^{-u})_{\bar
2}+g'_{1\bar 1\bar 2}(e^u+fe^{-u})_2)\\
&-g'^{1\bar 1}g'^{2\bar 2}\{(e^u+fe^{-u})_1(e^u+fe^{-u})_{\bar
1}+(e^u+fe^{-u})_2(e^u+fe^{-u})_{\bar 2})\\
\leq& g'^{i\bar i}g'^{k\bar k}g'_{i\bar kk}g'_{k\bar i\bar
k}\\
&+(g'^{2\bar 2}g'_{2\bar 21})(g'^{1\bar 1}(e^u+fe^{-u})_{\bar
1})+(g'^{2\bar 2}g'_{2\bar 2\bar 1})(g'^{1\bar
1}(e^u+fe^{-u})_{1})\\
 &+(g'^{1\bar 1}g'_{1\bar
12})(g'^{2\bar 2}(e^u+fe^{-u})_{\bar 2})+(g'^{1\bar 1}g'_{1\bar
1\bar 2})(g'^{2\bar 2}(e^u+fe^{-u})_{2}).
\end{aligned}
\end{equation*}
By the Schwarz inequality, we can estimate
\begin{equation}\lab{715}
\begin{aligned}
&(e^u+fe^{-u}+\alpha\bigtriangleup
u)^{-1}\mid\bigtriangledown'(e^u+fe^{-u}+\alpha\bigtriangleup
u)\mid_{g'}^2\\
 \leq &g'^{i\bar i}g'^{k\bar k}g'_{i\bar
kk}g'_{k\bar i \bar k}+g'^{2\bar 2}g'^{2\bar 2}g'_{2\bar
21}g'_{2\bar 2\bar
1}+g'^{1\bar 1}g'^{1\bar 1}g'_{1\bar 12}g'_{1\bar 1\bar 2}\\
&+g'^{1\bar 1}g'^{1\bar 1}(e^u+fe^{-u})_1(e^u+fe^{-u})_{\bar
1}+g'^{2\bar 2}g'^{2\bar 2}(e^u+fe^{-u})_2(e^u+fe^{-u})_{\bar
2}\\
\leq &g'^{i\bar i}g'^{k\bar k}g'_{i\bar kj}g'_{k\bar i
\bar j}+C_5(g'^{1\bar 1}g'^{1\bar 1}+g'^{2\bar 2}g'^{2\bar 2})\\
\leq &g'^{i\bar i}g'^{k\bar k}g'_{i\bar kj}g'_{k\bar i \bar
j}+C_5\frac{g'^2_{1\bar 1}+g'^2_{2\bar 2}}{g'^2_{1\bar
1}g'^2_{2\bar 2}}\\
\leq &g'^{i\bar i}g'^{k\bar k}g'_{i\bar kj}g'_{k\bar i
\bar j}+C_5F^{-2}(g'_{1\bar 1}+g'_{2\bar 2})^2\\
\leq &g'^{i\bar i}g'^{k\bar k}g'_{i\bar kj}g'_{k\bar i \bar
j}+C_5(e^u+fe^{-u}+\alpha\bigtriangleup u)^2,
\end{aligned}
\end{equation}
where $C_5$ is some constant. In this section we will use the
constant $C_5$ in the generic sense which depends on $f$,
$\alpha$, $\mu$, the curvature bound of the metric $\omega_S$, and
$u$ up to first order derivation. It  can also depend on the lower
bound of $F$ as we have proven that $F\geq \kappa e^{2u}\geq
\kappa (C_1A)^{-2}$. Note when we assume that $g_{i\bar
j}=\delta_{ij}$ and $u_{i\bar j}=u_{i\bar i}\delta_{ij}$, the last
term of (\ref{707}) is $g'^{i\bar i}g'^{k\bar k}g'_{i\bar
kj}g'_{k\bar i\bar j}$. Multiplying (\ref{711}) by $F$ and then
inserting (\ref{707}) and (\ref{715}) into it, we obtain
\begin{equation}\lab{716}
\begin{aligned}
&P(e^{-\lambda_1u+\lambda_2\mid\bigtriangledown
u\mid^2}\cdot(e^u+fe^{-u}+\alpha\bigtriangleup u))
\cdot e^{-(-\lambda_1 u+\lambda_2\mid\bigtriangledown u\mid^2)}\cdot F\\
\geq&-\lambda_1(e^u+fe^{-u}+\alpha \bigtriangleup u)P(u)\cdot F\\
&+\lambda_2(e^u+fe^{-u}+\alpha\bigtriangleup
u)P(\mid\bigtriangledown u\mid^2)\cdot F\\
&-(e^u+fe^{-u}+\alpha\bigtriangleup u)\bigtriangleup
(e^u+fe^{-u})+4\alpha g'^{i\bar j}(g^{k\bar l})_{i\bar j}u_{k\bar l}\cdot F\\
&+2^{-1} \bigtriangleup F-(2F)^{-1}\mid\bigtriangledown
F\mid^2+P(e^u+fe^{-u})\cdot F\\
 &-C_5(e^u+fe^{-u}+\alpha\bigtriangleup
u)^2.
\end{aligned}
\end{equation}
We assume that $e^{-\lambda_1u+\lambda_2\mid\bigtriangledown
u\mid^2}(e^u+fe^{-u}+\alpha\bigtriangleup u)$ achieve  the maximum
at the point $q_3$. Taking the normal coordinate $(z_1,z_2)$ at
the point $q_3$ with respect  to the given  metric $\omega_S$, we
 estimate every term in (\ref{716}).   At the point $q_3$,
$\bigtriangledown\{e^{(-\lambda_1u+\lambda_2\mid\bigtriangledown
u\mid^2)}\cdot(e^u+fe^{-u}+\alpha\bigtriangleup u)\}=0$. We can
get
\begin{equation}\lab{717}
\bigtriangledown\bigtriangleup u=\alpha^{-1} (e^u+fe^{-u}+\alpha
\bigtriangleup u)(\lambda_1\bigtriangledown
u-\lambda_2\bigtriangledown\mid\bigtriangledown
u\mid^2)-\alpha^{-1} \bigtriangledown(e^u+fe^{-u}).
\end{equation}
At first we derive some inequalities which will be used to
estimate terms in (\ref{716}). Using the equation we compute
\begin{equation}\lab{718}
\begin{aligned}
&4\alpha^2g^{i\bar j}g^{k\bar l}u_{i\bar l}u_{k\bar j}
=4\alpha^2\sum_{i,j}\mid u_{i\bar j}\mid^2\\
=&4\alpha^2(u^2_{1\bar 1}+u^2_{2\bar 2}+2u_{1\bar 2}u_{2\bar 1})\\
=&4\alpha^2(u_{1\bar 1}+u_{2\bar 2})^2-8\alpha^2\det u_{i\bar j}\\
=&\alpha^2(\bigtriangleup
u)^2+\alpha\bigtriangleup(e^u-fe^{-u})+\alpha \mu\\
\leq &(e^u+fe^{-u}+\alpha\bigtriangleup
u)^2-2\alpha(e^u+fe^{-u})\bigtriangleup
u\\
&-(e^u+fe^{-u})^2+\alpha(e^u+fe^{-u})\bigtriangleup u+C_5\\
=&(e^u+fe^{-u}+\alpha\bigtriangleup
u)^2-(e^u+fe^{-u})(e^u+fe^{-u}+\alpha\bigtriangleup u)+C_5\\
\leq& (e^u+fe^{-u}+\alpha\bigtriangleup u)^2+C_5.
\end{aligned}
\end{equation}
Let
\begin{equation*}
\Gamma=4g^{i\bar j}g^{k\bar l}u_{,ik}u_{,\bar j\bar l}
\end{equation*}
where indices preceded by a comma, e.g., $u_{,ik}$ indicate
covariant differentiation with respect to the given metric
$\omega_S$. At the point $q_3$,  we use the
 normal coordinate. Therefore at $q_3$,
 $u_{,ik}=u_{ik}$ and $u_{,\bar j\bar l}=u_{\bar j\bar
l}$ (see p.345 of \cite{Yau} paper or the next section). Hence
 $\Gamma=4u_{ik}u_{\bar i\bar k}=4\sum_{ik}\mid
u_{ik}\mid^2$. We  use the inequality (\ref{718}) to estimate
\begin{equation}\lab{720}
\begin{aligned}
\mid\bigtriangledown\mid\bigtriangledown u\mid^2\mid^2=&2g^{p\bar
q}(\mid\bigtriangledown u\mid^2)_p(\mid\bigtriangledown
u\mid^2)_{\bar q}\\
=&2g^{p\bar q}(2g^{i\bar j}u_iu_{\bar j})_p(2g^{k\bar l}u_ku_{\bar
l})_{\bar q}\\
=&8(u_{ip}u_{\bar i}+u_iu_{\bar ip})(u_{k\bar p}u_{\bar
k}+u_ku_{\bar k\bar p})\\
=&8(u_{ip}u_{k\bar p}u_{\bar i}u_{\bar k}+u_{ip}u_{\bar k\bar
p}u_{\bar i}u_k+u_{\bar ip}u_{k\bar p}u_iu_{\bar k}+u_{\bar
ip}u_{\bar k\bar p}u_iu_k).
\end{aligned}
\end{equation}
As was done in above  section,  we take the normal coordinate at
the point $q_3$ such that $u_1=u_{\bar 1}$ and $u_2=u_{\bar 2}=0$.
Then applying the Schwarz inequality  and (\ref{718}) to
(\ref{720}), we get
\begin{equation}\lab{721}
\begin{aligned}
\mid\bigtriangledown\mid\bigtriangledown u\mid^2\mid^2=&
4(u_{1p}u_{1\bar p}+u_{\bar 1\bar p}u_{\bar 1p}+u_{1p}u_{\bar
1\bar p}+u_{\bar 1p}u_{1\bar p})\mid\bigtriangledown u\mid^2\\
\leq&4(\mid u_{1p}\mid^2+\mid u_{1\bar p}\mid^2+\mid
u_{1p}\mid^2+\mid
u_{1\bar p}\mid^2)\mid\bigtriangledown u\mid^2\\
=&8(\mid u_{1p}\mid^2+\mid u_{1\bar p}\mid^2)\mid\bigtriangledown
u\mid^2\\
\leq&2\mid\bigtriangledown
u\mid^2\left\{\Gamma+\alpha^{-2}(e^u+fe^{-u}+\alpha\bigtriangleup
u)^2\right\}+C_5.
\end{aligned}
\end{equation}
So,
\begin{equation}\lab{722}
\mid\bigtriangledown\mid\bigtriangledown u\mid^2\mid\leq
\sqrt{2}\mid\bigtriangledown u\mid\left\{\Gamma^{\frac
12}+\alpha^{-1}(e^u+fe^{-u}+\alpha\bigtriangleup u)\right\}+C_5.
\end{equation}
We also need to estimate
\begin{equation*}
\begin{aligned}
\mid\bigtriangledown(\bigtriangledown u\cdot\bigtriangledown
f)\mid^2=&2g^{p\bar q}(\bigtriangledown u\cdot\bigtriangledown
f)_p(\bigtriangledown u\cdot\bigtriangledown f)_{\bar q}\\
=&2g^{p\bar q}(g^{i\bar j}(u_i f_{\bar j}+u_{\bar j} f_i))_p\cdot
(g^{k\bar
l}(u_kf_{\bar l}+u_{\bar l}f_k)_{\bar q}\\
=&2(u_{ip}f_{\bar i}+u_if_{\bar ip}+u_{\bar ip}f_i+u_{\bar
i}f_{ip})(u_{k\bar p}f_{\bar k}+u_kf_{\bar k\bar p}+u_{\bar k\bar
p}f_k+u_{\bar k}f_{k\bar p})\\
=&2(u_{ip}u_{k\bar p}f_{\bar i}f_{\bar k}+u_{\bar ip}u_{\bar k\bar
p}f_if_k+u_{ip}u_{\bar k\bar p}f_{\bar i}f_k+u_{\bar ip}u_{k\bar
p}f_if_{\bar k})\\
&+2(u_{ip}f_{\bar i}u_kf_{\bar k\bar p}+u_{\bar k\bar p}f_ku_{\bar
i}f_{ip}+u_{ip}f_{\bar i}u_{\bar k}f_{k\bar p}+u_{\bar k\bar
p}f_ku_if_{\bar ip})\\
&+2(u_{\bar ip}f_iu_kf_{\bar k\bar p}+u_{k\bar p}f_{\bar k}u_{\bar
i}f_{ip}+u_{\bar ip}f_iu_{\bar k}f_{k\bar p}+u_{k\bar p}f_{\bar
k}u_if_{\bar ip})\\
&+2(u_iu_kf_{\bar ip}f_{\bar k\bar p}+u_iu_{\bar k}f_{\bar
ip}f_{k\bar p}+u_{\bar i}u_kf_{ip}f_{\bar k\bar p}+u_{\bar
i}u_{\bar k}f_{ip}f_{k\bar p}).\\
\end{aligned}
\end{equation*}
Changing the indices $i$ and $k$ in some terms and then applying
the  Schwarz inequality, we can get
\begin{equation}\lab{723}
\begin{aligned}
\mid\bigtriangledown(\bigtriangledown u\cdot \bigtriangledown
f)\mid^2=&2(u_{ip}u_{k\bar p}f_{\bar i}f_{\bar k}+u_{\bar k
p}u_{\bar
i\bar p}f_if_k)\\
&+(u_{ip}u_{\bar k\bar p}f_{\bar i}f_k+u_{kp}u_{\bar i\bar
p}f_{\bar k}f_i)+(u_{\bar ip}u_{k\bar p}f_if_{\bar k}+u_{i\bar
p}u_{\bar
kp}f_kf_{\bar i})\\
&+2(u_{ip}u_kf_{\bar i}f_{\bar k\bar p}+u_{\bar i\bar p}u_{\bar
k}f_{kp}f_i+u_{ip}u_{\bar k}f_{k\bar p}f_{\bar i}+u_{\bar i\bar p}u_kf_{\bar kp}f_i)\\
&+2(u_{\bar ip}u_kf_{\bar k\bar p}f_i+u_{i\bar p}u_{\bar
k}f_{kp}f_{\bar i}+u_{\bar i\bar p}u_{\bar k}f_{k\bar
p}f_i+u_{i\bar p}u_kf_{\bar kp}f_{\bar i})+C_5\\
\leq& C_5(2\mid u_{ip}\mid\mid u_{k\bar p}\mid+\mid u_{ip}\mid\mid
u_{\bar k\bar p}\mid +\mid u_{\bar ip}\mid\mid
u_{k\bar p}\mid)\\
&+C_5(\mid u_{ip}\mid+\mid u_{i\bar p}\mid)+C_5\\
\leq &C_5(\mid u_{ip}\mid^2+\mid u_{k\bar p}\mid^2)+C_5\\
 \leq&C_5\Gamma+C_5(e^u+fe^{-u}+\alpha\bigtriangleup u)^2+C_5.
\end{aligned}
\end{equation}
Then,
\begin{equation}\lab{724}
\mid\bigtriangledown(\mid\bigtriangledown u\cdot \bigtriangledown
f)\mid\leq C_5\Gamma^{\frac
12}+C_5(e^u+fe^{-u}+\alpha\bigtriangleup u)+C_5.
\end{equation}
Applying (\ref{718}), (\ref{717}) and (\ref{722}), we can estimate
\begin{equation}\lab{725}
\begin{aligned}
\bigtriangleup\mid\bigtriangledown u\mid^2=&2g^{i\bar j}(2g^{k\bar
l}u_ku_{\bar l})_{i\bar j}\\
=&4g^{i\bar j}g^{k\bar l}(u_{i\bar jk}u_{\bar l}+u_ku_{i\bar j\bar
l}+u_{ik}u_{\bar j \bar l}+u_{i\bar l}u_{k\bar j})+4g^{i\bar
j}(g^{k\bar l})_{i\bar j}u_ku_{\bar l}\\
=&2g^{k\bar l}\{(2g^{i\bar j}u_{i\bar j})_ku_{\bar l}+(2g^{i\bar
j}u_{i\bar j})_{\bar l}u_k\}+4g^{i\bar j}g^{k\bar l}u_{i\bar
l}u_{k\bar j}+4g^{i\bar j}g^{k\bar l}u_{ik}u_{\bar j\bar l}\\
\leq&2\bigtriangledown\bigtriangleup u\cdot \bigtriangledown
u+\Gamma+\alpha^{-2}(e^u+fe^{-u}+\alpha\bigtriangleup u)^2+C_5\\
=&2\alpha^{-1}(e^u+fe^{-u}+\alpha\bigtriangleup
u)(\lambda_1\mid\bigtriangledown
u\mid^2-\lambda_2\bigtriangledown\mid\bigtriangledown
u\mid^2\cdot\bigtriangledown
u)\\
&-2\alpha^{-1}\bigtriangledown(e^u+fe^{-u})\cdot \bigtriangledown
u+\Gamma+\alpha^{-2}(e^u+fe^{-u}+\alpha \bigtriangleup
u)^2+C_5\\
\leq&2\alpha^{-1} \mid\bigtriangledown
u\mid^2\lambda_1(e^u+fe^{-u}+\alpha\bigtriangleup u)\\
&+ 2\alpha^{-1} \lambda_2(e^u+fe^{-u}+\alpha\bigtriangleup
u)\mid\bigtriangledown\mid\bigtriangledown u\mid^2\mid\cdot \mid
\bigtriangledown u\mid\\
&+\Gamma+\alpha^{-2}(e^u+fe^{-u}+\alpha\bigtriangleup u)^2+C_5\\
\leq&\left\{\alpha^{-2}+2\sqrt{2}\alpha^{-2}\mid\bigtriangledown
u\mid^2\lambda_2\right\}(e^u+fe^{-u}+\alpha\bigtriangleup u)^2\\
&+2\sqrt{2}\alpha^{-1}\mid\bigtriangledown
u\mid^2\lambda_2(e^u+fe^{-u}+\alpha\bigtriangleup u)\Gamma^{\frac12}+\Gamma\\
&+C_5\lambda_1(e^u+fe^{-u}+\alpha\bigtriangleup u)+C_5\\
\leq&(C_5\lambda^2_2+C_5\lambda_2+C_5)(e^u+fe^{-u}+\alpha\bigtriangleup
u)^2+2\Gamma \\
&+C_5\lambda_1(e^u+fe^{-u}+\alpha\bigtriangleup u)+C_5.
\end{aligned}
\end{equation}
For the same reason we can also estimate
\begin{equation}\lab{726}
\begin{aligned}
\bigtriangleup(\bigtriangledown u\cdot \bigtriangledown
f)=&2(u_if_{\bar i}+u_{\bar i}f_i)_{k\bar k}+2(g^{i\bar j})_{k\bar
k}(u_if_{\bar j}+u_{\bar j}f_i)\\
\leq &2(u_{k\bar ki}f_{\bar i}+u_{k\bar k\bar
i}f_i)+2(u_{ik}f_{\bar i\bar k}+u_{\bar i\bar k}f_{ik})\\
&+2(u_{i\bar k}f_{\bar ik}+u_{\bar ik}f_{i\bar k})+C_5\\
\leq&\bigtriangledown\bigtriangleup u\cdot \bigtriangledown
f+2\mid
u_{ik}\mid^2+2\mid u_{i\bar k}\mid^2+C_5\\
\leq&\alpha^{-1}(e^u+fe^{-u}+\alpha\bigtriangleup
u)(\lambda_1\bigtriangledown u\cdot\bigtriangledown
f-\lambda_2\bigtriangledown\mid\bigtriangledown u\mid^2\cdot
\bigtriangledown f)\\
&+2^{-1}\Gamma+(2\alpha^2)^{-1}(e^u+fe^{-u}+\alpha\bigtriangleup
u)^2+C_5\\
\leq&\Gamma+(C_5\lambda_2^2+C_5\lambda_2+C_5)(e^u+fe^{-u}+\alpha\bigtriangleup
u)^2\\
&+C_5\lambda_1(e^u+fe^{-u}+\alpha\bigtriangleup u)+C_5.
\end{aligned}
\end{equation}
We now deal with every term in (\ref{716}). For the first term, we
use (\ref{615}) to obtain
\begin{equation}\lab{727}
\begin{aligned}
&-\lambda_1(e^u+fe^{-u}+\alpha\bigtriangleup u)P(u)F\\
=&-\lambda_1(e^u+fe^{-u}+\alpha\bigtriangleup u)(\alpha^{-1}
F-\alpha^{-1}(e^u+fe^{-u})(e^u+fe^{-u}+\alpha\bigtriangleup
u))\\
\geq&\alpha^{-1}\lambda_1(e^u+fe^{-u})(e^u+fe^{-u}+\alpha\bigtriangleup
u)^2-C_5\lambda_1(e^u+fe^{-u}+\alpha\bigtriangleup u)\\
\geq &(\alpha C_1A)^{-1}\lambda_1(e^u+fe^{-u}+\alpha\bigtriangleup
u)^2-C_5\lambda_1(e^u+fe^{-u}+\alpha\bigtriangleup u).
\end{aligned}
\end{equation}
Next we deal with the second term
$\lambda_2(e^u+fe^{-u}+\alpha\bigtriangleup
u)P(\mid\bigtriangledown u\mid^2)F$:
\begin{equation}\lab{728}
\begin{aligned}
P(\mid\bigtriangledown u\mid^2)F=&4g'^{i\bar j}(u_{i\bar
jk}u_{\bar k}+u_{i\bar j\bar k}u_k+u_{i\bar k}u_{k\bar
j}+u_{ik}u_{\bar k\bar j})F+4g'^{i\bar j}(g^{k\bar
l})_{i \bar j}u_ku_{\bar l}F\\
\geq&4g'^{i\bar j}(u_{i\bar jk}u_{\bar k}+u_{i\bar j\bar
k}u_k)F+4g'^{i\bar j}u_{ik}u_{\bar k\bar
j}F-C_5(e^u+fe^{-u}+\alpha\bigtriangleup u).
\end{aligned}
\end{equation}
From (\ref{636}), we know
\begin{equation}\lab{729}
\begin{aligned}
&4g'^{i\bar j}(u_{i\bar jk}u_{\bar k}+u_{i\bar j\bar k}u_k)F\\
\geq&-2(e^u-fe^{-u})\bigtriangledown\mid\bigtriangledown
u\mid^2\cdot \bigtriangledown u-4e^{-u}\bigtriangledown
(\bigtriangledown u\cdot \bigtriangledown f)\cdot \bigtriangledown
u\\
&-C_5(e^u+fe^{-u}+\alpha\bigtriangleup u)-C_5\\
\geq &-C_5\mid\bigtriangledown\mid\bigtriangledown
u\mid^2\mid\cdot\mid\bigtriangledown
u\mid-C_5\mid\bigtriangledown(\bigtriangledown
u\cdot\bigtriangledown f)\mid\cdot\mid\bigtriangledown u\mid\\
&-C_5(e^u+fe^{-u}+\alpha\bigtriangleup u)-C_5.
\end{aligned}
\end{equation}
Applying (\ref{722}) and (\ref{724}), we get
\begin{equation}
4g'^{i\bar j}(u_{i\bar jk}u_{\bar k}+u_{i\bar j\bar k}u_k)F\geq
-C_5\Gamma^\frac12-C_5(e^u+fe^{-u}+\alpha\bigtriangleup u)-C_5.
\end{equation}
Inserting (\ref{729}) into (\ref{728}), we obtain
\begin{equation}\lab{730}
\begin{aligned}
&\lambda_2(e^u+fe^{-u}+\alpha\bigtriangleup u)P(\mid
\bigtriangledown u\mid^2)F\\
\geq&\lambda_2(e^u+fe^{-u}+\alpha\bigtriangleup u)(4g'^{i\bar
j}u_{ik}u_{\bar k\bar j})F\\
&-C_5\lambda_2(e^u+fe^{-u}+\alpha\bigtriangleup u)\Gamma^{\frac
12}-C_5\lambda_2(e^u+fe^{-u}+\alpha\bigtriangleup
u)^2\\
&-C_5\lambda_2(e^u+fe^{-u}+\alpha\bigtriangleup u)\\
\geq &\lambda_2(e^u+fe^{-u}+\alpha\bigtriangleup u)(4g'^{i\bar
j}u_{ik}u_{\bar k\bar j})F\\
&-\Gamma-C_5(\lambda_2^2+\lambda_2)(e^u+fe^{-u}+\alpha\bigtriangleup
u)^2\\
&-C_5\lambda_2(e^u+fe^{-u}+\alpha\bigtriangleup u).
\end{aligned}
\end{equation}
We assume that $g_{i\bar j}=\delta_{ij}$ and $u_{i\bar j}=u_{i\bar
i}\delta_{ij}$ at the point $q_3$. Then
\begin{equation}\lab{731}
\begin{aligned}
&\lambda_2(e^u+fe^{-u}+\alpha\bigtriangleup u)(4g'^{i\bar
j}u_{ik}u_{\bar k\bar j})F\\
=&4\lambda_2 F \cdot g'^{i\bar i}(e^u+fe^{-u}+\alpha\bigtriangleup u)(u_{ik}u_{\bar i\bar k})\\
=&4\lambda_2F\frac{1}{g'_{i\bar i}}\left(\frac{g'_{1\bar
1}+g'_{2\bar 2}}{2}\right)u_{ik}u_{\bar k\bar i}\\
=&2\lambda_2 F\left\{\left(1+\frac{g'_{2\bar 2}}{g'_{1\bar
1}}\right)u_{1k}u_{\bar k\bar 1}+\left(1+\frac{g'_{1\bar
1}}{g'_{2\bar 2}}\right)u_{2k}u_{\bar k\bar 2}\right\}\\
\geq&2\lambda_2Fu_{ik}u_{\bar k\bar i}\geq\frac 1 2 \lambda_2
F\Gamma\\
 \geq&\frac 1 2 (C_1A)^{-2}\kappa\lambda_2\Gamma.
\end{aligned}
\end{equation}
Inserting (\ref{731}) into (\ref{730}), we find an  estimate of
the second term in (\ref{716})
\begin{equation}\lab{732}
\begin{aligned}
&\lambda_2(e^u+fe^{-u}+\alpha\bigtriangleup
u)P(\mid\bigtriangledown u\mid^2)F\\
\geq &\left(2^{-1}(C_1A)^{-2} \kappa
\lambda_2-1\right)\Gamma-C_5(\lambda_2^2+\lambda_2)(e^u+fe^{-u}+\alpha\bigtriangleup
u)^2\\
&-C_5\lambda_2(e^u+fe^{-u}+\alpha\bigtriangleup u).
\end{aligned}
\end{equation}
The third term is
\begin{equation}\lab{733}
\begin{aligned}
&-(e^u+fe^{-u}+\alpha\bigtriangleup u)\bigtriangleup(e^u+fe^{-u})\\
\geq&-(e^u+fe^{-u}+\alpha\bigtriangleup
u)\{(e^u-fe^{-u})\bigtriangleup
u+C_5\}\\
\geq &-\alpha^{-1} (e^u-fe^{-u})(e^u+fe^{-u}+\alpha\bigtriangleup
u)^2-C_5(e^u+fe^{-u}+\alpha\bigtriangleup u)\\
\geq &-C_5(e^u+fe^{-u}+\alpha\bigtriangleup
u)^2-C_5(e^u+fe^{-u}+\alpha\bigtriangleup u)
\end{aligned}
\end{equation}
and the forth term is
\begin{equation}\lab{734}
\begin{aligned}
&4\alpha g'^{i\bar j}(g^{k\bar l})_{i\bar j}u_{k\bar l}F\\
=&4\alpha\left(g'_{1\bar 1}(g^{k\bar l})_{2 \bar 2}+g'_{2\bar
2}(g^{k\bar l})_{1\bar 1}-g'_{1\bar 2}(g^{k\bar l})_{2\bar
1}-g'_{2\bar 1}(g^{k\bar l})_{1\bar 2}\right)u_{k\bar l}\\
=&4\alpha(e^u+fe^{-u})g^{i\bar j}(g^{k\bar l})_{i\bar j}u_{k\bar
l}\\
&+16\alpha^2\left(u_{1\bar 1}(g^{k\bar l})_{2 \bar 2}+u_{2\bar
2}(g^{k\bar l})_{1\bar 1}-u_{1\bar 2}(g^{k\bar l})_{2\bar
1}-u_{2\bar 1}(g^{k\bar l})_{1\bar 2}\right)u_{k\bar l}\\
\geq&-64\alpha^2\max\mid R_{i\bar j k\bar l}\mid\sum\mid u_{i\bar
j}\mid^2\\
\geq&-C_5(e^u+fe^{-u}+\alpha\bigtriangleup u)^2-C_5,
 \end{aligned}
\end{equation}
where $C_5$ depends the curvature of $\omega_S$.  Next we deal
with the fifth term. From the definition of $F$, we have
\begin{equation}\lab{737}
\begin{aligned}
2^{-1}\bigtriangleup
F=&2^{-1}\bigtriangleup\{(e^u+fe^{-u})^2-2\alpha[(e^u-fe^{-u})\mid\bigtriangledown
u\mid^2-2\alpha e^{-u}\bigtriangledown u\cdot\bigtriangledown
f+\alpha e^{-u}\bigtriangleup f- \mu]\}\\
=&-\alpha(e^u-fe^{-u})\bigtriangleup\mid\bigtriangledown
u\mid^2+2\alpha^{-u}\bigtriangleup(\bigtriangledown u\cdot
\bigtriangledown f)\\
&-\alpha\bigtriangledown (e^u-fe^{-u})\cdot
\bigtriangledown\mid\bigtriangledown
u\mid^2-2\alpha^{-u}\bigtriangledown u\cdot
\bigtriangledown(\bigtriangledown u \cdot \bigtriangledown f)\\
&-C_5(e^u+fe^{-u}+\alpha\bigtriangleup u)-C_5\\
=&-C_5\mid\bigtriangleup\mid\bigtriangledown
u\mid^2\mid-C_5\mid\bigtriangleup(\bigtriangledown u\cdot
\bigtriangledown
f)\mid-C_5\mid\bigtriangledown\mid\bigtriangledown u\mid^2\mid-C_5\mid\bigtriangledown
(\bigtriangledown u\cdot \bigtriangledown f)\mid\\
&-C_5(e^u+fe^{-u}+\alpha\bigtriangleup u)-C_5.
\end{aligned}
\end{equation}
We note that the inequalities (\ref{725}) and (\ref{726}) are also
true for $\mid\bigtriangleup\mid\bigtriangledown u\mid^2\mid$ and
$\mid\bigtriangleup(\bigtriangledown u\cdot \bigtriangledown
f)\mid$. Applying (\ref{725}), (\ref{726}), (\ref{722}) and
(\ref{724}), we get
\begin{equation}\lab{736}
\begin{aligned}
2^{-1}\bigtriangleup F\geq&
-C_5\Gamma-(C_5\lambda_2^2+C_5\lambda_2+C_5)(e^u+fe^{-u}+\alpha\bigtriangleup
u)^2\\
&-(C_5\lambda_1+C_5)(e^u+fe^{-u}+\alpha\bigtriangleup u)-C_5.
\end{aligned}
\end{equation}
 We also observe that
\begin{equation*}
\bigtriangledown F=-C_5\bigtriangledown\mid\bigtriangledown
u\mid^2-C_5\bigtriangledown(\bigtriangledown
u\cdot\bigtriangledown f)-C_5\bigtriangledown u-C_5.
\end{equation*}
Then applying the Schwarz inequality,
\begin{equation*}
\begin{aligned}
-(2F)^{-1}\mid\bigtriangledown F\mid^2\geq
&-C_5\mid\bigtriangledown\mid\bigtriangledown
u\mid^2\mid^2-C_5\mid\bigtriangledown\mid\bigtriangledown
u\mid^2\mid\cdot \mid\bigtriangledown(\bigtriangledown
u\cdot\bigtriangledown
f)\mid\\
&-C_5\mid\bigtriangledown(\bigtriangledown u\cdot \bigtriangledown
f)\mid^2-C_5\mid\bigtriangledown\mid\bigtriangledown
u\mid^2\mid\\
&-C_5\mid\bigtriangledown(\bigtriangledown u\cdot \bigtriangledown
f)\mid-C_5.
\end{aligned}
\end{equation*}
Then applying (\ref{721})-(\ref{724}), we can get
\begin{equation}\lab{739}
-(2F)^{-1}\mid\bigtriangledown F\mid^2\geq
-C_5\Gamma-C_5(e^u+fe^{-u}+\alpha\bigtriangleup
u)^2-C_5(e^u+fe^{-u}+\alpha\bigtriangleup u)-C_5.
\end{equation}
The last term is
\begin{equation}\lab{740}
\begin{aligned}
P(e^u+fe^{-u})F =& (e^u-fe^{-u})P(u)F+(e^u+fe^{-u})\cdot
2g'^{i\bar j}u_iu_{\bar j}F\\
&-e^{-u}\cdot 2g'^{i\bar j}(u_if_{\bar j}+u_{\bar
j}f_i)F+e^{-u}P(f)F\\
\geq &-C_5(e^u+fe^{-u}+\alpha\bigtriangleup u)-C_5.
\end{aligned}
\end{equation}
Inserting (\ref{727}), (\ref{732}), (\ref{733}),
(\ref{734}),(\ref{736}), (\ref{739}) and (\ref{740}) into
(\ref{716}), at last we get
\begin{equation}\lab{741}
\begin{aligned}
&P(e^{-\lambda_1 u+\lambda_2\mid\bigtriangledown u\mid^2}\cdot
(e^u+fe^{-u}+\alpha\bigtriangleup u))F\cdot e^{-(-\lambda_1
u+\lambda_2\mid\bigtriangledown u\mid^2)}\\
\geq &\left\{(\alpha
C_1A)^{-1}\lambda_1-C_5(\lambda_2^2+\lambda_2+1)\right\}(e^u+fe^{-u}+\alpha\bigtriangleup
u)^2\\
&-\{C_5\lambda_1+C_5\lambda_2+C_5\}(e^u+fe^{-u}+\alpha\bigtriangleup
u)\\
&+(2^{-1}(C_1A)^{-2} \kappa\lambda_2-C_5)\Gamma-C_5\Gamma^{\frac
12}-C_5.
\end{aligned}
\end{equation}
Fix the constant $C_5$. Take $\lambda_2$ big enough such that
\begin{equation*}
2^{-1}(C_1A)^{-2}\kappa\lambda_2-C_5>0
\end{equation*}
and then take $\lambda_1$ big enough such that
\begin{equation*}
(\alpha C_1A)^{-1}\lambda_1-C_5(\lambda_2^2+\lambda_2+1)>0.
\end{equation*}
Fix $\lambda_1$ and $\lambda_2$. Then we can now estimate
$e^{-\lambda_1 u+\lambda_2\mid\bigtriangledown
u\mid^2}(e^u+fe^{-u}+\alpha\bigtriangleup u)$. In fact, it must
achieve its maximum at some point $q_3$ so the  right-hand  side
of (\ref{741}) is non-positive. At this point,
\begin{equation*}
\begin{aligned}
0\geq&\left\{(\alpha
C_1A^{-1}\lambda_1-C_5(\lambda_2^2+\lambda_2+1)\right\}(e^u+fe^{-u}+\alpha\bigtriangleup
u)^2\\
&-\{C_5\lambda_1+C_5\lambda_2+C_5\}(e^u+fe^{-u}+\alpha\bigtriangleup
u)\\
&+(2^{-1}(C_1A)^{-2}\kappa\lambda_2-C_5)\Gamma-C_5\Gamma^{\frac 12}-C_5\\
\geq &\left\{(\alpha
C_1A)^{-1}\lambda_1-C_5(\lambda_2^2+\lambda_2+1)\right\}(e^u+fe^{-u}+\alpha\bigtriangleup
u)^2\\
&-C_5(\lambda_1+\lambda_2+1)(e^u+fe^{-u}+\alpha\bigtriangleup u)\\
&-\frac {C_5} {4(2^{-1}(C_1A)^{-2}\kappa \lambda_2-C_5)}-C_5.
\end{aligned}
\end{equation*}
Hence $(e^u+fe^{-u}+\alpha\bigtriangleup u)(q_3)$ has an upper
bound $C_5'$ depending on $\alpha$, $f$, $\mu$, the curvature
bound of metric $\omega_S$, $A$. Since $e^{-\lambda_1
u+\lambda_2\mid\bigtriangledown
u\mid^2}(e^u+fe^{-u}+\alpha\bigtriangleup u)$ achieves its maximum
at the point $q_3$, we  get the estimate
\begin{equation*}
(e^u+fe^{-u}+\alpha\bigtriangleup u)\leq
C_5'\frac{\sup(e^{-\lambda_1 u+\lambda_2\mid\bigtriangledown
u\mid^2})}{\inf(e^{-\lambda_1 u+\lambda_2\mid\bigtriangledown
u\mid^2})}\leq C_5'\frac {e^{-\lambda_1\inf
u+\lambda_2\sup\mid\bigtriangledown u\mid^2}}{e^{-\lambda_1\sup
u}}.
\end{equation*}
As $\mid\bigtriangledown u\mid^2$ has the upper bound (7.8), we
get an upper bound of $e^u+fe^{-u}+\alpha\bigtriangleup u$. In
conclusion, we have proved the following
\begin{prop}\lab{prop 4}
Let $S$ be a $K3$ surface with Calabi-Yau metric $\omega_S$ such
that $\int_S 1\frac {\omega_S^2}{2!}=1$. Let $u\in C^{4}(S)$ be
the solution of the equation $\bigtriangleup(e^u-t\alpha
fe^{-u})+8\alpha \frac {\det g'_{i\bar j}}{\det g_{i\bar
j}}+t\mu=0$ which satisfies  the condition $(e^u+t\alpha
fe^{-u})\omega_S+2\alpha\sqrt{-1}\partial\bar\partial u>0$ and
$\left(\int_S e^{-4u}\right)^{\frac 14}=A<<1$ (see (\ref{668}) and
(\ref{669})). Then $e^u+t\alpha fe^{-u}+\alpha \bigtriangleup u$
has an upper bound depending only on $\alpha$, $f$, $\mu$,
$\omega_S$ and $A$. Moreover, combing with  the Proposition
\ref{prop 3},  $e^u+t\alpha fe^{-u}+4\alpha u_{i\bar i}$, for
$i=1,2$, have the positive lower and upper bounds depending only
on $\alpha$, $f$, $\mu$, $\omega_S$ (both Sobolev constant and
curvature bound) and $A$.
\end{prop}

\section{Third order estimate}\lab{sec: 9}
In this section we use indices to denote partial derivatives,
e.g.,  $u_i=\partial_i u=\frac {\partial u}{\partial z_i}$,
$u_{i\bar j}=\partial_{i\bar j}u=\frac {\partial^2u}{\partial
z_i\partial\bar z_j}$. Indices preceded by a comma, e.g.,
$u_{,ik}$ indicate covariant differentiation with respect to the
given metric $\omega_S$.  Let
\begin{eqnarray*} \Gamma&=&g^{i\bar j}g^{k\bar l}u_{,ik}u_{,\bar
j\bar l}\\
\Theta&=&g'^{i\bar r}g'^{s\bar j}g'^{k\bar t}u_{,i\bar jk}u_{,\bar
r s\bar
t}\\
\Xi&=&g'^{i\bar j}g'^{k\bar l}g'^{p\bar q}u_{,ikp}u_{,\bar j\bar l
\bar  q}\\
\Phi&=&g'^{i\bar j}g'^{k\bar l}g'^{p\bar q}g'^{r\bar s}u_{,i\bar l
pr}u_{,\bar jk\bar q
\bar  s}\\
\Psi&=&g'^{i\bar j}g'^{k\bar l}g'^{p\bar q}g'^{r\bar s}u_{,i\bar l
p\bar s}u_{,\bar jk\bar qr}.
\end{eqnarray*}
We shall apply the maximum  principle to the function
\begin{equation}\lab{801}
(\lambda_3+\alpha\bigtriangleup
u)\Theta+\lambda_4(m+\alpha\bigtriangleup
u)\Gamma+\lambda_5\mid\bigtriangledown
u\mid^2\Gamma+\lambda_6\Gamma,
\end{equation}
where all $\lambda_i$ for $i=3,4,5,6$ are positive constants and
will be determined later;  $m$ is a fixed constant such that
$m+\alpha\bigtriangleup u>0$. At first we assume that
$\lambda_3+\alpha\bigtriangleup u>1$.   We shall  use $C_6$ as a
constant in generic sense which depends only on $\alpha$, $f$,
$\mu$, $\omega_S$ and $u$ up to the second order derivations. Let
the function (\ref{801}) achieve the maximum at a point $q_4\in
S$. Before computing $P((\ref{801}))$ at $q_4$, we need to derive
some relations between partial derivatives and covariant
differentiations. Pick a normal coordinate at $q_4$ such that
$g_{i\bar j}=\delta_{ij}$, $\partial g_{i\bar j}/\partial
z_k=\partial g_{i\bar j}/\partial \bar z_k=0$.  Then  at $q_4$, we
have
\begin{eqnarray*}
\begin{array}{cccc}
u_{,i\bar j}=u_{i\bar j}, & u_{,ij}=u_{ij}, & u_{,\bar i\bar
j}=u_{\bar i\bar j}, &\\
u_{,i\bar jk}=u_{i\bar jk},& u_{,\bar ij\bar k}=u_{\bar ij\bar
k},& u_{,i\bar j\bar k}=u_{i\bar j\bar k}, & u_{,\bar i\bar
jk}=u_{\bar i\bar jk}\\
\partial_{\bar k l}(u_{,ij})=u_{,ij\bar kl},& \ldots\ldots\\
\end{array}
\end{eqnarray*}
We also have \begin{equation*}
\begin{aligned}
u_{,ik\bar \gamma}=u_{,i\bar \gamma k}+u_sR^s_{ik\bar \gamma},\ \
\  u_{,\bar j\bar l \delta}=u_{,\bar j\delta\bar l}+u_{\bar
s}R^{\bar s}_{\bar j\bar l\delta}.
\end{aligned}
\end{equation*}
Now we compute every term in $P((\ref{801}))$.
\begin{equation}\lab{802}
\begin{aligned}
P(\mid\bigtriangledown
u\mid^2)&=4g'^{\delta\bar\gamma}\partial_\delta\partial_{\bar\gamma}(g^{i\bar
j}u_iu_{\bar
j})\\
&\geq4g'^{\delta\bar\gamma}g^{i\bar
j}(u_{i\bar\gamma\delta}u_{\bar j}+u_{i}u_{\bar
j\delta\bar\gamma}+u_{i\bar\gamma}u_{\bar
j\delta}+u_{i\delta}u_{\bar
j\bar \gamma})-C_6\\
&\geq 4g'^{\delta\bar\gamma}g^{i\bar j}\{u_{,i\delta}u_{,\bar
j\bar\gamma}+u_{,i\bar\gamma\delta}u_{\bar j}+u_iu_{,\bar
j\delta\bar\gamma}\}-C_6\\
&\geq m_1\Gamma-C_6\sum\mid u_{,i\bar
\gamma\delta}\mid\mid u_{\bar j}\mid-C_6\\
 &\geq
m_1\Gamma-C_6\Theta^{\frac{1}{2}}-C_6.
\end{aligned}
\end{equation}
Since Proposition \ref{prop 4} shows that the metric $\omega'$ is
uniformly equivalent to $\omega_S$, we see that such an $m_1>0$
exists. Next we estimate $P(\alpha \bigtriangleup u)$. From
(\ref{707}) we know
\begin{equation}\lab{803}
\alpha P(\bigtriangleup u)\geq g^{i\bar j}g'^{\delta \bar
p}g'^{q\bar \gamma}g'_{\delta\bar\gamma i}g'_{\bar p q\bar
j}+(2F)^{-1}\bigtriangleup F-(2F^2)^{-1}\mid\bigtriangledown
F\mid^2-C_6.
\end{equation}
We compute
\begin{equation}\lab{804}
\begin{aligned}
&g^{i\bar j}g'^{\delta \bar p}g'^{q\bar
\gamma}g'_{\delta\bar\gamma
i}g'_{\bar pq\bar j}\\
\geq& g^{i\bar j}g'^{\delta \bar p}g'^{q\bar \gamma}(4\alpha
u_{\delta\bar\gamma i})(4\alpha u_{\bar
pq\bar j})\\
&+g'^{\delta\bar p}g'^{q\bar\gamma}g^{i\bar
j}\{((e^u+fe^{-u})g_{q\bar p})_{\bar j}(4\alpha
u_{\delta\bar\gamma i})+(4\alpha u_{q\bar p\bar
j}((e^u+fe^{-u})g_{\delta\bar\gamma})_{i})\}\\
\geq &16\alpha^2  g^{i\bar j}g'^{\delta \bar p}g'^{q\bar
\gamma}u_{,\delta\bar\gamma i}u_{,\bar p q\bar
j}-C_6\sum\mid(e^u+fe^{-u})_{\bar
j}\mid\mid u_{\delta\bar\gamma i}\mid\\
\geq &m_2\Theta-C_6\Theta^{\frac 12}.
\end{aligned}
\end{equation}
From (\ref{737}),
\begin{equation}\lab{805}
\bigtriangleup F\geq -C_6 \sum\mid u_{,k \bar l i}\mid\mid u_{\bar
j}\mid-C_6\Gamma-C_6\geq -C_6\Theta^{\frac 12}-C_6\Gamma-C_6.
\end{equation}
Inserting (\ref{804}), (\ref{805}) and (\ref{739}) into
(\ref{803}), we get
\begin{equation}\lab{806}
\begin{aligned}
P(\alpha\bigtriangleup u)\geq m_2\Theta-C_6\Theta^{\frac
12}-C_6\Gamma-C_6\geq m_2\Theta-C_6\Gamma-C_6,
\end{aligned}
\end{equation}
where we have used  $m_2$ in the generic sense. We  also
calculate:
\begin{equation}\lab{807}
\begin{aligned}
P(\Gamma)&=2g'^{\delta\bar\gamma}\partial_\delta\partial_{\bar\gamma}\Gamma \\
&\geq 2g'^{\delta\bar \gamma}g^{i\bar j}g^{k\bar
l}\{(u_{,ik})_{\bar\gamma \delta}u_{,\bar j\bar l}+u_{,ik}
(u_{,\bar j\bar l})_{\delta\bar\gamma}\}\\
&+2g'^{\delta\bar \gamma}g^{i\bar j}g^{k\bar
l}\{(u_{,ik})_{\delta} (u_{,\bar j\bar
l})_{\bar\gamma}+(u_{,ik})_{\bar\gamma}(u_{,\bar j\bar
l})_{\delta}\}-C_6\Gamma
\\
&=2g'^{\delta\bar \gamma}g^{i\bar j}g^{k\bar l}\{u_{,ik\bar\gamma
\delta}u_{,\bar j\bar l}+u_{,ik} u_{,\bar j\bar
l\delta\bar\gamma}+u_{,ik\delta} u_{,\bar j\bar
l\bar\gamma}+u_{,ik\bar\gamma}u_{,\bar j\bar l\delta}\}-C_6\Gamma\\
&=2g'^{\delta\bar \gamma}g^{i\bar j}g^{k\bar l}\{u_{,ik\delta}
u_{,\bar j\bar l\bar\gamma}+u_{,ik\bar\gamma\delta}u_{,\bar j\bar
l}+u_{,ik} u_{,\bar j\bar l\delta\bar\gamma}\}\\
&+2g'^{\delta\bar \gamma}g^{i\bar j}g^{k\bar l}(u_{,i\bar\gamma
k}+u_{s}R^s_{ik\bar\gamma})
(u_{,\bar j\delta\bar l}+u_{\bar s}R^{\bar s}_{\bar j\bar l\delta})-C_6\Gamma\\
&\geq 2g'^{\delta\bar \gamma}g^{i\bar j}g^{k\bar
l}(u_{,ik\delta}u_{,\bar j\bar l\bar \gamma}+u_{i\bar\gamma
k}u_{\bar j\delta\bar l})\\
&-C_6\sum(\mid u_{,ik\bar\gamma\delta}\mid\mid u_{,\bar j\bar
l}\mid+\mid
u_{,i\bar\gamma k}\mid\mid u_s R^s_{ik\bar\gamma}\mid)-C_6\Gamma\\
&\geq m_3\Xi+m_3\Theta-C_6\Phi^{\frac
12}\Gamma^{\frac12}-C_6\Gamma\\
 &\geq
m_3\Xi+m_3\Theta-\epsilon_1\lambda_6^{-1}\Phi-C_6\lambda_6\epsilon_1^{-1}\Gamma.
\end{aligned}\end{equation}
Combining (\ref{802}) and (\ref{807}), we find
\begin{equation}\lab{808}
\begin{aligned}
&P(\mid\bigtriangledown u\mid^2\Gamma)\\
&=P(\mid\bigtriangledown u\mid^2)\Gamma+\mid\bigtriangledown
u\mid^2P(\Gamma)\\
 &+2g'^{\delta\bar\gamma}(\partial_\delta(\mid\bigtriangledown u\mid^2)
 \partial_{\bar\gamma}\Gamma
 +\partial_{\bar\gamma}(\mid\bigtriangledown u\mid^2)\partial_\delta \Gamma)
\\
&\geq
m_1\Gamma^2-C_6\Theta^{\frac{1}{2}}\Gamma-C_6\Gamma+\mid\bigtriangledown
u\mid^2(m_3\Xi+m_3\Theta-C_6\Phi^{\frac 12}\Gamma^{\frac 12}-C_6\Gamma)\\
&-C_6(\Gamma^{\frac 12}+1)(\Theta^{\frac{1}{2}}\Gamma^{\frac
12}+\Xi^{\frac{1}{2}}\Gamma^{\frac 12}+\Gamma^{\frac{1}{2}})
\\
&\geq
m_1\Gamma^2-\epsilon_1\lambda_5^{-1}\Phi-C_6\lambda_5\epsilon_1^{-1}
\Gamma-C_6\Xi-C_6\Theta-C_6.
 \end{aligned}
 \end{equation}
Combining (\ref{806}) and (\ref{807}), we get
\begin{equation}\lab{809}
\begin{aligned}
&\ \ \ \ P((m+\alpha\bigtriangleup u)\Gamma)\\
&=P(\alpha\bigtriangleup u)\Gamma+(m+\alpha\bigtriangleup
u)P(\Gamma)+2\alpha
g'^{\delta\bar\gamma}\{\partial_\delta(\bigtriangleup
u)\partial_{\bar\gamma}\Gamma+\partial_{\bar\gamma}(\bigtriangleup u)\partial_{\delta}\Gamma\}\\
&\geq (m_2\Theta-C_6\Gamma-C_6)\Gamma+(m+\alpha\bigtriangleup
u)(m_3\Xi+m_3\Theta-C_6\Phi^{\frac 12}\Gamma^{\frac 12}-C_6\Gamma)\\
&\ \ \ -C_6\Theta^{\frac{1}{2}}(\Theta^{\frac{1}{2}}+
\Xi^{\frac{1}{2}}+1)\Gamma^{\frac{1}{2}}\\
&\geq m_2\Theta\Gamma-C_6\Gamma^2-\epsilon_1\lambda_4^{-1}\Phi
-C_6\lambda_4\epsilon_1^{-1}\Gamma-C_6\Xi-C_6\Theta.
 \end{aligned}
\end{equation}
Now we deal with
 \begin{equation}\lab{810}
 \begin{aligned}
P((\lambda_3+\alpha\bigtriangleup
u)\Theta)&=P(\lambda_3+\alpha\bigtriangleup
u)\Theta+(\lambda_3+\alpha\bigtriangleup
u)P(\Theta)\\
&+2\alpha g'^{\delta\bar\gamma}\{\partial_{\delta}(\bigtriangleup
u)\partial_{\bar\gamma}\Theta+\partial_{\bar\gamma}(\bigtriangleup
u)\partial_{\delta}\Theta\}.
\end{aligned}
\end{equation}
Applying (\ref{806}), we get
\begin{equation}\lab{811}
P(\lambda_3+\alpha\bigtriangleup u)\Theta\geq
m_2\Theta^2-C_6\Gamma\Theta-C_6\Theta.
\end{equation}
Let $(\lambda_3+\alpha\bigtriangleup
u)\Theta+\lambda_4(m+\alpha\bigtriangleup
u)\Gamma+\lambda_5\mid\bigtriangledown
u\mid^2\Gamma+\lambda_6\Gamma$ achieve the maximum at the point
$q_4$. Then at the point $q_4$, we have,
\begin{eqnarray*}
\partial_{\bar\gamma}\Theta=-\frac{1}{\lambda_3+\alpha\bigtriangleup
u}\{\Theta\partial_{\bar\gamma}(\alpha\bigtriangleup
u)+\lambda_4\partial_{\bar\gamma}((m+\alpha\bigtriangleup
u)\Gamma)+\lambda_5\partial_{\bar\gamma}(\mid\bigtriangledown
u\mid^2\Gamma)+\lambda_6\partial_{\bar\gamma}\Gamma\}
\end{eqnarray*}
and
\begin{equation}\lab{812}
\begin{aligned}
&\ \ \ \ 2\alpha
g'^{\delta\bar\gamma}\{\partial_\delta(\lambda_3+\alpha\bigtriangleup
u)\partial_{\bar\gamma}\Theta+\partial_{\bar\gamma}(\lambda_3+\alpha\bigtriangleup
u)\partial_{\delta}\Theta\}\\
&=-\frac{4\alpha^2}{\lambda_3+\alpha\bigtriangleup u}\text{Re}\ \
\ g'^{\delta\bar\gamma}(\bigtriangleup
u)_{\delta}\{\alpha(\bigtriangleup
u)_{\bar\gamma}\Theta+\alpha\lambda_4(\bigtriangleup
u)_{\bar\gamma}\Gamma+\lambda_5(\mid\bigtriangledown
u\mid^2)_{\bar\gamma}\Gamma\\
&\ \ \ \ \ \ \ \ \ \ \ \ \ \ \ \ \ \ \ \ \ \ \ \ \ \ \ \
+[\lambda_4(m+\alpha\bigtriangleup
u)+\lambda_5\mid\bigtriangledown u\mid^2+\lambda_6]
\Gamma_{\bar\gamma}\}\\
&\geq \frac{-C_6}{\lambda_3+\alpha\bigtriangleup
u}\Theta^{\frac{1}{2}}\times\{\Theta^{\frac{3}{2}}+\lambda_4\Theta^{\frac{1}{2}}\Gamma+\lambda_5\Gamma^{\frac{3}{2}}\\
&\ \ \ \ \ \ \ \ \ \ \ \ \ \ \ \ \ \ \ \ \ \ \ \
 \ \
+(\lambda_4+\lambda_5+\lambda_6)(\Theta^{\frac{1}{2}}+\Xi^{\frac{1}{2}}+\Gamma^{\frac
12})
\Gamma^{\frac{1}{2}}\}\\
&\geq \frac{-C_6}{\lambda_3+\alpha\bigtriangleup
u}\{\Theta^2+(\lambda_4+\lambda_5+\lambda_6)
(\Theta\Gamma+\Theta+\Gamma+\Xi)+\lambda_5\Gamma^2\}-C_6.
\end{aligned}
\end{equation}
Inserting (\ref{811}) and (\ref{812}) into (\ref{810}), and then
combing (\ref{807})-(\ref{810}), we obtain
\begin{equation}\lab{813}
\begin{aligned}
&P((\lambda_3+\alpha\bigtriangleup
u)\Theta+\lambda_4(m+\alpha\bigtriangleup
u)\Gamma+\lambda_5\mid\bigtriangledown
u\mid^2\Gamma+\lambda_6\Gamma)\\
\geq &(\lambda_3+\alpha\bigtriangleup u)P(\Theta)+\left\{m_2-\frac
{C_6}{\lambda_3+\alpha\bigtriangleup
u}\right\}\Theta^2\\
&+\left\{\lambda_4m_2-C_6-\frac
{C_6}{\lambda_3+\alpha\bigtriangleup
u}(\lambda_4+\lambda_5+\lambda_6)\right\}\Theta\Gamma\\
&+\left\{\lambda_5 m_1-\frac
{C_6\lambda_5}{\lambda_3+\alpha\bigtriangleup
u}-C_6\lambda_4\right\}\Gamma^2\\
&+\left\{\lambda_6m_3-C_6(\lambda_4+\lambda_5)-
\frac{C_6(\lambda_4+\lambda_5+\lambda_6)}{\lambda_3+\alpha\bigtriangleup
u}\right\}\Xi\\
&-3\epsilon_1\Phi-C_7\Theta-C_7\Gamma-C_7,
\end{aligned}
\end{equation}
where $C_7$ depends also on $\lambda_i$ and $\epsilon_1$ at point
$q_4$. At last we can estimate  $P(\Theta)$. We follow paper
\cite{Yau} to obtain:
\begin{equation}\lab{814}
\begin{aligned}
P(\Theta)=&2g'^{\delta\bar\gamma}\partial_\delta\partial_{\bar
\gamma}(g'^{i\bar r}g'^{s\bar j}g'^{k\bar t}u_{,i\bar j
k}u_{,\bar r s\bar t})\\
=& 2g'^{\delta\bar\gamma}[\ \ \ 2g'^{i\bar a}g'^{b\bar p}g'^{q\bar
r}g'^{s\bar j}g'^{k\bar t}+2g'^{i\bar p}g'^{q\bar a}g'^{b\bar
r}g'^{s\bar j}g'^{k\bar t}\\
&\ \ \ \ \ \ \ +2g'^{i\bar p}g'^{q\bar r}g'^{s\bar a}g'^{b\bar
j}g'^{k\bar t}+2g'^{i\bar p}g'^{q\bar r}g'^{s\bar
j}g'^{k\bar a}g'^{b\bar t}\\
&\ \ \ \ \ \ \ +\ g'^{i\bar a}g'^{b\bar r}g'^{s\bar p}g'^{q\bar
j}g'^{k\bar t}+\ g'^{i\bar r}g'^{s\bar
a}g'^{b\bar p}g'^{q\bar j}g'^{k\bar t}\\
&\ \ \ \ \ \ \  +\ g'^{i\bar r}g'^{s\bar p}g'^{q\bar a}g'^{b\bar
j}g'^{k\bar t}+\ g'^{i\bar r}g'^{s\bar
p}g'^{q\bar j}g'^{k\bar a}g'^{b\bar t}]\\
 &\ \ \ \ \ \ \ \ \ \ \ \ \ \ \ \ \ \times \partial_\delta g'_{b\bar a}\partial_{\bar\gamma}
 g'_{\bar
pq}u_{,i\bar jk}u_{,\bar rs\bar t}\ \ (\text{first class})
\\
&-2g'^{\delta\bar\gamma}[2g'^{i\bar p}g'^{q\bar r}g'^{s\bar
j}g'^{k\bar t}+g'^{i\bar
r}g'^{s\bar p}g'^{q\bar j}g'^{k\bar t}]\\
&\ \ \ \ \times [\partial_{\bar\gamma}g'_{\bar pq}u_{,i\bar
jk\delta}u_{,\bar rs\bar t}+ \partial_{\delta}g'_{q\bar
p}u_{,\bar r s\bar t\bar\gamma}u_{,i\bar j k}] \ \ (\text{second class})\\
&-2g'^{\delta\bar\gamma}[2g'^{i\bar p}g'^{q\bar r}g'^{s\bar
j}g'^{k\bar t}+g'^{i\bar
r}g'^{s\bar p}g'^{q\bar j}g'^{k\bar t}]\\
&\ \ \ \times [\partial_{\bar\gamma}g'_{\bar pq}u_{,i\bar
jk}u_{,\bar rs\bar t\delta}+\partial_\delta g'_{q\bar
p}u_{,i \bar j k\bar\gamma}u_{,\bar r s\bar t}]\ \  (\text{third class})\\
&-2g'^{\delta\bar\gamma}[2g'^{i\bar p}g'^{q\bar r}g'^{s\bar
j}g'^{k\bar t}+g'^{i\bar r}g'^{s\bar p}g'^{q\bar j}g'^{k\bar
t}]\times
\partial_\delta\partial_{\bar\gamma}g'_{\bar pq}u_{,i\bar jk}u_{,\bar rs\bar
t}\ \ (\text{forth
class})\\
&+2g'^{\delta\bar\gamma}g'^{i\bar r}g'^{s\bar j}g'^{k\bar t}\times
[u_{,i\bar jk\bar \gamma\delta}u_{,\bar rs\bar t}+u_{,i\bar
jk}u_{,\bar rs\bar t\delta\bar \gamma}]\ \ (\text{fifth class})\\
&+2g'^{\delta\bar\gamma}g'^{i\bar r}g'^{s\bar j}g'^{k\bar t}\times
[u_{,i\bar jk\bar \gamma}u_{,\bar rs\bar t\delta}+u_{,i\bar
jk\delta}u_{,\bar rs\bar t\bar
\gamma}]\ \ (\text{sixth class})\\
&-C_6\Theta,
\end{aligned}
\end{equation}
where when we use normal coordinate so that at this point we have
$\partial_{\bar\beta}u_{,i\bar j k}=u_{,i\bar jk\bar\beta}$ and
$\partial_\alpha\partial_{\bar\beta}u_{,i\bar jk}=u_{,i\bar j
k\bar\beta\alpha}+u_{,i\bar sk}R^{\bar s}_{\bar j
\bar\beta\alpha}$. Comparing with (A.8) in \cite{Yau}, we should
deal with first five classes in (\ref{814}). The first class is:
\begin{equation}\lab{815}
\begin{aligned}
&\ \ \ \ 2g'^{\delta\bar\gamma}g'^{i\bar a}g'^{b\bar p}g'^{q\bar
r}g'^{s\bar j}g'^{k\bar t} \partial_\delta g'_{b\bar
a}\partial_{\bar\gamma}g'_{\bar
pq}u_{,i\bar jk}u_{,\bar rs\bar t}\\
&=2g'^{\delta\bar\gamma}g'^{i\bar a}g'^{b\bar p}g'^{q\bar
r}g'^{s\bar j}g'^{k\bar t} (4\alpha u_{b\bar a\delta})(4\alpha
u_{\bar pq\bar\gamma})u_{,i\bar jk}u_{,\bar rs\bar
t}\\
&+4\text{Re} \{g'^{\delta\bar\gamma}g'^{i\bar a}g'^{a\bar
p}g'^{q\bar r}g'^{s\bar j}g'^{k\bar t}((e^u+fe^{-u})_\delta\cdot
(4\alpha u_{\bar pq\bar \gamma})u_{,i\bar jk}u_{,\bar rs\bar
t})\}\\
&+2g'^{\delta\bar\gamma}g'^{i\bar a}g'^{a\bar p}g'^{p\bar
r}g'^{s\bar j}g'^{k\bar
t}((e^u+fe^{-u})_\delta(e^u+fe^{-u})_{\bar\gamma} u_{,i\bar
jk}u_{,\bar rs\bar
t})\\
&\geq 2g'^{\delta\bar\gamma}g'^{i\bar a}g'^{b\bar p}g'^{q\bar
r}g'^{s\bar j}g'^{k\bar t} (4\alpha u_{,b\bar a\delta})(4\alpha
u_{,\bar pq\bar\gamma})u_{,i\bar jk}u_{,\bar rs\bar
t}\\
&+4\text{Re} \{g'^{\delta\bar\gamma}g'^{i\bar a}g'^{a\bar
p}g'^{q\bar r}g'^{s\bar j}g'^{k\bar t}((e^u+fe^{-u})_\delta\cdot
(4\alpha u_{,\bar pq\bar \gamma})u_{,i\bar jk}u_{,\bar rs\bar
t})\}\\
&\geq 2g'^{\delta\bar\gamma}g'^{i\bar a}g'^{b\bar p}g'^{q\bar
r}g'^{s\bar j}g'^{k\bar t} (4\alpha u_{,b\bar a\delta})(4\alpha
u_{,\bar pq\bar\gamma})u_{,i\bar jk}u_{,\bar rs\bar
t}\\
 &\ \ \ -\epsilon_2/12(\lambda_3+\alpha\bigtriangleup
u) ^{-1}\Theta^2-C_6\epsilon_2^{-1}(\lambda_3+\alpha\bigtriangleup
u)\Theta.
\end{aligned}
\end{equation}
 The second class is:
\begin{equation}\lab{816}
\begin{aligned}
&\ \ \ \ -2g'^{\delta\bar\gamma}g'^{i\bar p}g'^{q\bar r}g'^{s\bar
j}g'^{k\bar t} \{\partial_{\bar\gamma}g'_{\bar pq}u_{,i\bar
jk\delta}u_{,\bar rs\bar t}+\partial_\delta g'_{q\bar p}
u_{,\bar rs\bar t\bar \gamma}u_{,i\bar j k}\}\\
&=-4\text{Re} \{g'^{\delta\bar\gamma}g'^{i\bar p}g'^{q\bar
r}g'^{s\bar j}g'^{k\bar t}
\partial_{\bar\gamma}((e^u+fe^{-u})g_{\bar pq}+4\alpha u_{\bar
pq})u_{,i\bar
jk\delta}u_{\bar rs\bar t}\}\\
&\geq-2g'^{\delta\bar\gamma}g'^{i\bar p}g'^{q\bar r}g'^{s\bar
j}g'^{k\bar t} \{(4\alpha u_{,\bar pq\bar\gamma})u_{,i\bar
jk\delta}u_{,\bar rs\bar t}+(4\alpha u_{,q\bar p\delta})u_{\bar
rs\bar t\bar \gamma}u_{,i\bar jk}\}
\\
&\ \ \ -\epsilon_1/3(\lambda_3+\alpha\bigtriangleup
u)^{-1}\Phi-C_6(\lambda_3+\alpha\bigtriangleup
u)\epsilon_1^{-1}\Theta.
 \end{aligned}
\end{equation}
The third class is:
\begin{equation}\lab{817}
\begin{aligned}
&\ \ \ \ -2g'^{\delta\bar\gamma}g'^{i\bar p}g'^{q\bar r}g'^{s\bar
j}g'^{k\bar t}\{\partial_\delta g'_{q\bar p}u_{,i \bar j
k\bar\gamma}u_{,\bar r s\bar
t}+\partial_{\bar \gamma} g'_{q\bar p}u_{,i\bar jk}u_{,\bar rs\bar t\delta}\}\\
&\geq-2g'^{\delta\bar\gamma}g'^{i\bar p}g'^{q\bar r}g'^{s\bar
j}g'^{k\bar t}\{(4\alpha u_{,q\bar p\delta})u_{,i \bar j
k\bar\gamma}u_{,\bar r s\bar
t}+(4\alpha u_{,\bar pq\bar\gamma})u_{,i\bar jk}u_{,\bar rs\bar t\delta}\}\\
&\ \ \ -\epsilon_1/3(\lambda_3+\alpha\bigtriangleup
u)^{-1}\Psi-C_6(\lambda_3+\alpha\bigtriangleup
u)\epsilon_1^{-1}\Theta.
\end{aligned}
\end{equation}
Next we deal with the forth class. By (\ref{705}),
\begin{equation}\lab{818}
\begin{aligned}
&-2g'^{\delta\bar\gamma}g'^{i\bar p}g'^{q\bar r}g'^{s\bar
j}g'^{k\bar t} \partial_\delta\partial_{\bar\gamma}g'_{\bar
pq}u_{,i\bar jk}u_{,\bar rs\bar t}\\
\geq&-2g'^{\delta\bar\gamma}g'^{i\bar p}g'^{q\bar r}g'^{s\bar
j}g'^{k\bar t} (4\alpha u_{\bar\gamma\delta\bar p q}
)u_{,i\bar jk}u_{,\bar rs\bar t}-C_6\Theta\\
\geq& -2g'^{\delta\bar a}g'^{b\bar\gamma} g'^{i\bar p}g'^{q\bar
r}g'^{s\bar j}g'^{k\bar t}\partial_{\bar p}g'_{\bar
ab}\partial_qg'_{\delta\bar\gamma}u_{,i\bar jk}u_{,\bar rs\bar t}-C_6\Theta\\
&-2g'^{i\bar p}g'^{q\bar r}g'^{s\bar j}g'^{k\bar t}\left\{
F^{-1}F_{q\bar p}-F^{-2}F_qF_{\bar p}\right\}u_{,i\bar jk}u_{,\bar
rs\bar t}.
\end{aligned}
\end{equation}
Then from (\ref{804}), (\ref{805}) and (\ref{741}), we can see
\begin{equation}\lab{819}
\begin{aligned}
& -2g'^{\delta \bar\gamma}g'^{i\bar p}g'^{q\bar r}g'^{s\bar
j}g'^{k\bar t}
\partial_\delta\partial_{\bar\gamma}g'_{\bar
pq}u_{,i\bar jk}u_{,\bar rs\bar t}\\
 \geq&-2g'^{i\bar p}g'^{q\bar r}g'^{s\bar
j}g'^{k\bar t}g'^{\delta\bar a}g'^{b\bar\gamma} (4\alpha u_{\bar
ab\bar p }) (4\alpha u_{,\delta\bar\gamma q})u_{,i\bar jk}u_{,\bar
rs\bar t}\\
&-C_6\Theta^{\frac 32}-C_6\Theta\Gamma^{\frac 12}-C_6\Gamma\Theta-C_6\Theta\\
\geq&-2g'^{i\bar p}g'^{q\bar r}g'^{s\bar j}g'^{k\bar
t}g'^{\delta\bar a}g'^{b\bar\gamma} (4\alpha u_{,\bar ab\bar p })
(4\alpha u_{,\delta\bar\gamma q})u_{,i\bar jk}u_{,\bar
rs\bar t}\\
& -C_6\Theta\Gamma-m_2/24(\lambda_3+\alpha\bigtriangleup
u)^{-1}\Theta^2-C_6(\lambda_3+\alpha\bigtriangleup
u)\Theta-C_6\Gamma.
\end{aligned}
\end{equation}
Now we deal with the fifth term. By direct calculation, we have
\begin{eqnarray*}
u_{,i\bar jk\bar\gamma\delta}&=&u_{i\bar j k\bar
\gamma\delta}+u_{,p\bar j\delta}R^p_{ik\bar\gamma}+u_{,i\bar
pk}R^{\bar p}_{\bar j\bar \gamma \delta}-u_{p\bar
j}\partial_{\delta}\partial_{\bar\gamma}(g^{p\bar
s}\partial_kg_{i\bar s})-u_{p\bar j\bar
\gamma}\partial_\delta(g^{p\bar s}\partial_kg_{i\bar s}).
\end{eqnarray*}
So the fifth class can be expressed
\begin{equation}\lab{820}
\begin{aligned}
&g'^{\delta\bar\gamma}g'^{i\bar r}g'^{s\bar j}g'^{k\bar
t}\{u_{,i\bar jk\bar \gamma\delta} u_{,\bar r s\bar t}+u_{,i\bar
jk}u_{,\bar rs\bar t\delta\bar\gamma}\}\\
\geq& g'^{\gamma\bar\delta}g'^{i\bar r}g'^{s\bar j}g'^{k\bar
t}\{u_{i\bar jk\bar \gamma\delta} u_{,\bar r s\bar t}+u_{,i\bar
jk}u_{\bar r s\bar t\delta\bar\gamma}\}-C_6\Theta.
\end{aligned}
\end{equation}
Differentiating (\ref {705}), we can get
\begin{equation}\lab{821}
\begin{aligned}
4\alpha g'^{\delta\bar\gamma}u_{\delta\bar\gamma i\bar
jk}=&4\alpha g'^{\delta \bar p}g'^{q\bar \gamma}g'_{q\bar
pk}u_{\delta\bar \gamma i\bar j}+(g'^{\delta \bar
p}g'^{q\bar\gamma}g'_{p\bar q\bar
j}g'_{\delta\bar\gamma i})_k\\
&+g'^{\delta\bar p}g^{q\bar \gamma}g'_{q\bar
pk}((e^u+fe^{-u})g_{\delta\bar\gamma})_{i\bar j}
-g'^{\delta\bar\gamma}((e^u+fe^{-u})g_{\delta\bar\gamma})_{i\bar jk}\\
&+F^{-1}F_{i\bar jk}-F^{-2}(F_kF_{i\bar j}+F_iF_{\bar jk}+F_{\bar
j}F_{ik})+2F^{-3}F_iF_{\bar j}F_k.
\end{aligned}
\end{equation}
Inserting (\ref{821}) into (\ref{820}), we get
\begin{equation}\lab{822}
\begin{aligned}
&g'^{\delta\bar\gamma}g'^{i\bar r}g'^{s\bar j}g'^{k\bar
t}\{u_{,i\bar jk\bar\gamma\delta}u_{,\bar rs\bar t}+u_{,i\bar
jk}u_{,\bar rs\bar t\delta\bar\gamma}\}\\
\geq&g'^{i\bar r}g'^{s\bar j}g'^{k\bar t}g'^{\delta \bar
p}g'^{q\bar\gamma}\{g'_{q\bar pk}u_{\delta\bar\gamma i\bar
j}u_{,\bar rs\bar t}+g'_{\bar qp\bar t}u_{\bar\gamma\delta\bar
rs}u_{,i\bar jk}\}-C_6\Theta\\
&+(4\alpha)^{-1}g'^{i\bar r}g'^{s\bar j}g'^{k\bar
t}\{(g'^{\delta\bar p}g'^{q\bar \gamma}g'_{\bar pq\bar
j}g'_{\delta\bar\gamma i})_ku_{\bar rs\bar t}+(g'^{\delta\bar
p}g'^{q\bar\gamma}g'_{\bar pq\bar s}g'_{\delta\bar\gamma r})_{\bar
t}u_{i\bar jk}\}\\
&+2(4\alpha)^{-1}\text{Re}\{g'^{i\bar r}g'^{s\bar j}g'^{k\bar
t}(F^{-1}F_{i\bar jk}-F^{-2}(F_{i\bar j}F_k+F_iF_{\bar jk}+F_{\bar
j}F_{ik})+2F^{-3}F_iF_{\bar j}F_k)u_{,\bar rs\bar t}\}.
\end{aligned}
\end{equation}
   We observe
\begin{equation}\lab{823}
\begin{aligned}
&g'^{i\bar r}g'^{s\bar j}g'^{k\bar t}g'^{\delta \bar
p}g'^{q\bar\gamma}\{g'_{q\bar pk}u_{\delta\bar\gamma i\bar
j}u_{,\bar rs\bar t}+g'_{\bar qp\bar t}u_{\bar\gamma\delta\bar
rs}u_{,i\bar jk}\}\\
\geq &g'^{i\bar r}g'^{s\bar j}g'^{k\bar t}g'^{\delta\bar
p}g'^{q\bar \gamma}\{(4\alpha u_{,q\bar p k})u_{,\delta\bar\gamma
i\bar j}u_{,\bar rs\bar t}+(4\alpha u_{,\bar p q\bar
t})u_{,\bar\gamma\delta\bar rs}u_{,i\bar jk}\}\\
&-C_6\sum\mid u_{,\delta\bar\gamma i\bar j}\mid\mid u_{,\bar
rs\bar t}\mid-C_6\sum \mid u_{,q\bar pk}\mid\mid u_{,\bar rs\bar
t}\mid-C_6\sum
\mid u_{,\bar rs\bar t}\mid\\
\geq &g'^{i\bar r}g'^{s\bar j}g'^{k\bar t}g'^{\delta\bar
p}g'^{q\bar \gamma}\{(4\delta u_{,q\bar p k})u_{,\delta\bar\gamma
i\bar j}u_{,\bar rs\bar t}+(4\alpha u_{,\bar p q\bar
t})u_{,\bar\gamma\delta\bar rs}u_{,i\bar jk}\}\\
&-C_6\Psi^{\frac 12}\Theta^{\frac 12}-C_6\Theta,
\end{aligned}
\end{equation}
and
\begin{equation}\lab{824}
\begin{aligned}
&(4\alpha)^{-1}g'^{i\bar r}g'^{s\bar j}g'^{k\bar
t}\{(g'^{\delta\bar p}g'^{q\bar \gamma}g'_{\bar pq\bar
j}g'_{\delta\bar\gamma i})_ku_{\bar rs\bar t}+(g'^{\delta\bar
p}g'^{q\bar\gamma}g'_{\bar pq\bar s}g'_{\delta\bar\gamma r})_{\bar
t}u_{i\bar jk}\}\\
\geq &(4\alpha)^{-1}g'^{i\bar r}g'^{s\bar j}g'^{k\bar
t}g'^{\delta\bar p}g'^{q\bar \gamma}\{(g'_{\bar pq\bar j
k}g'_{\delta\bar\gamma i}+g'_{\bar pq\bar j}g'_{\delta \bar\gamma
ik})u_{,\bar rs\bar t}\\
&\ \ \ \ \ \ \ \ \ \ \ \  \ \ \ \ \ \ \ \ \ \ \ \ \ \ \ \ \ \ \ \
+(g'_{\bar pq\bar s\bar t}g'_{\delta\bar\gamma r}+g'_{\bar pq\bar
s}g'_{\delta\bar\gamma
r\bar t})u_{,i\bar jk}\}\\
&-(4\alpha)^{-1}g'^{i\bar r}g'^{s\bar j}g'^{k\bar t}(g'^{\delta
\bar a}g'^{b\bar p}g'^{q\bar \gamma}+g'^{\delta\bar
p}g'^{q\bar a}g'^{b\bar \gamma})\\
&\ \ \ \ \ \ \ \ \ \ \ \ \ \ \ \ \ \ \ \ \ \ \ \ \ \ \ \ \cdot
(g'_{b\bar ak}g'_{\bar p q\bar j}g'_{\delta\bar\gamma i}u_{,\bar
rs\bar t}+g'_{\bar ab\bar t}g'_{\bar pq\bar s}g'_{\delta\bar\gamma
r}u_{,i\bar jk})\\
\geq & g'^{i\bar r}g'^{s\bar j}g'^{k\bar t}g'^{\delta\bar
p}g'^{q\bar \gamma}\{[u_{,\bar pq\bar j k}(4\alpha
u_{,\delta\bar\gamma i})+(4\alpha u_{,\bar pq\bar j})u_{,\delta
\bar\gamma
ik}]u_{,\bar rs\bar t}\\
&\ \ \ \ \ \ \ \ \ \ \ \  \ \ \ \ \ \ \ \ \ \ \ \ \ \ \ \ \
+[(u_{,\bar pq\bar s\bar t}(4\alpha u_{,\delta\bar\gamma
r})+(4\alpha u_{,\bar pq\bar s})u_{,\delta\bar\gamma
r\bar t}]u_{,i\bar jk}\}\\
&-g'^{i\bar r}g'^{s\bar j}g'^{k\bar t}\{g'^{\delta \bar
a}g'^{b\bar p}g'^{q\bar \gamma}+g'^{\delta\bar
p}g'^{q\bar a}g'^{b\bar \gamma}\}\\
&\ \ \ \ \ \ \ \ \ \ \ \ \ \ \ \ \ \cdot \{(4\alpha u_{,b\bar
ak})(4\alpha u_{,\bar p q\bar j})u_{,\delta\bar\gamma i}u_{,\bar
rs\bar t}+(4\alpha u_{,\bar ab\bar t})(4\alpha
u_{,\delta\bar\gamma r})u_{,\bar
pq\bar s}u_{,i\bar jk}\}\\
&-C_6\Theta^{\frac 32}-C_6\Psi^{\frac 12}\Theta^{\frac
12}-C_6\Phi^{\frac 12}\Theta^{\frac 12}-C_6\Theta-C_6.
\end{aligned}
\end{equation}
Then we estimate
\begin{equation}\lab{825}
\begin{aligned}
&(4\alpha)^{-1}\text{Re}\{g'^{i\bar r}g'^{s\bar j}g'^{k\bar
t}(F^{-1}F_{i\bar jk}-F^{-2}(F_{i\bar j}F_k+F_iF_{\bar jk}+F_{\bar
j}F_{ik})+2F^{-3}F_iF_{\bar j}F_k)u_{,\bar rs\bar t}\}\\
\geq&-C_6\Phi^{\frac 12}\Theta^{\frac 12}-C_6\Psi^{\frac
12}\Theta^{\frac 12}-C_6\Gamma^{\frac 12}\Theta-C_6\Gamma^{\frac
32}\Theta^{\frac 12}-C_6\Theta-C_6.
\end{aligned}
\end{equation}
Inserting (\ref{823})-(\ref{825}) into (\ref{822}), we get
\begin{equation}\lab{826}
\begin{aligned}
&g'^{\delta\bar\gamma}g'^{i\bar r}g'^{s\bar j}g'^{k\bar
t}\{u_{,i\bar jk\bar\gamma\delta}u_{,\bar rs\bar t}+u_{,i\bar
jk}u_{,\bar rs\bar t\delta\bar\gamma}\}\\
\geq &g'^{i\bar r}g'^{s\bar j}g'^{k\bar t}g'^{\delta\bar
p}g'^{q\bar \gamma}\{(4\alpha u_{,q\bar p k})u_{,\delta\bar\gamma
i\bar j}u_{,\bar rs\bar t}+(4\alpha u_{,\bar p q\bar
t})u_{,\bar\gamma\delta\bar rs}u_{,i\bar jk}\}\\
 & +g'^{i\bar r}g'^{s\bar j}g'^{k\bar t}g'^{\delta\bar p}g'^{q\bar
\gamma}\{[u_{,\bar pq\bar j k}(4\alpha u_{,\delta\bar\gamma
i})+(4\alpha u_{,\bar pq\bar j})u_{,\delta \bar\gamma
ik}]u_{,\bar rs\bar t}\\
&\ \ \ \ \ \ \ \ \ \ \ \  \ \ \ \ \ \ \ \ \ \ \ \ \ \ \ \ \
+[u_{,\bar pq\bar s\bar t}(4\alpha u_{,\delta\bar\gamma
r})+(4\alpha u_{,\bar pq\bar s})u_{,\delta\bar\gamma
r\bar t}]u_{,i\bar jk}\}\\
&-g'^{i\bar r}g'^{s\bar j}g'^{k\bar t}\{g'^{\delta \bar
a}g'^{b\bar p}g'^{q\bar \gamma}+g'^{\delta\bar
p}g'^{q\bar a}g'^{b\bar \gamma}\}\\
&\ \ \ \ \ \ \ \ \ \ \ \ \ \ \ \ \ \cdot \{(4\alpha u_{,b\bar
ak})(4\alpha u_{,\bar p q\bar j})u_{,\delta\bar\gamma i}u_{,\bar
rs\bar t}+(4\alpha u_{,\bar ab\bar t})(4\alpha
u_{,\delta\bar\gamma r})u_{,\bar
pq\bar s}u_{,i\bar jk}\}\\
& -\epsilon_1/2(\lambda_3+\alpha\bigtriangleup
u)^{-1}(\Phi+\Psi)-C_6\epsilon^{-1}_{1}(\lambda_3+\alpha\bigtriangleup
u)\Theta \\
&-m_2/16 (\lambda_3+\alpha\bigtriangleup
u)^{-1}\Theta^2-C_6(\lambda_3+\alpha\bigtriangleup
u)\Theta-C_6\Theta\Gamma-C_6\Gamma^2.
\end{aligned}
\end{equation}
Inserting (\ref{815})-(\ref{817}), (\ref{819}) and (\ref{826})
into (\ref{814}), diagonalizing and simplifying, then comparing to
(A.8) and (A.9) in \cite{Yau}, we obtain
\begin{equation}\lab{827}
\begin{aligned}
P(\Theta)&\geq\sum g'^{i\bar i}g'^{j\bar j}g'^{k\bar
k}g'^{\delta\bar\delta}\times\mid u_{i\bar j
k\bar\delta}-4\alpha\sum_p
u_{i\bar pk}u_{\bar jp\bar\delta}g'^{p\bar p}\mid^2\\
&\ \ \ \ +\sum g'^{i\bar i}g'^{j\bar j}g'^{k\bar
k}g'^{\delta\bar\delta}\times\mid u_{i\bar
jk\delta}-4\alpha\sum_p(u_{i\bar p\delta}u_{p\bar jk}+u_{i\bar
pk}u_{p\bar j\delta})g'^{p\bar p}\mid^2\\
&\ \ \ \ -\frac{1}{\lambda_3+\alpha\bigtriangleup
u}\left\{2\epsilon_1\Phi+2\epsilon_1\Psi+(\epsilon_2+\frac{m_{2}}{4})
\Theta^2\right\} -C_6\Theta\Gamma-C_7\Theta-C_7\Gamma\\
 &=\sum g'^{i\bar
i}g'^{j\bar j}g'^{k\bar
k}g'^{\delta\bar\delta}\times\mid\sqrt{1-2\epsilon_1(\lambda_3+\alpha\bigtriangleup
u)^{-1}} u_{i\bar j
k\bar\delta}\\
&\ \ \ \ \ \ \ \ \ \ \ \ \ \ \ \ \ \
 -4\alpha\left(\sqrt{1-2\epsilon_1(\lambda_3+\alpha\bigtriangleup
 u)^{-1}}\right)^{-1}
\sum_p
u_{i\bar pk}u_{\bar jp\bar\delta}g'^{p\bar p}\mid^2\\
&+\sum g'^{i\bar i}g'^{j\bar j}g'^{k\bar
k}g'^{\delta\bar\delta}\times\mid
\sqrt{1-5\epsilon_1(\lambda_3+\alpha\bigtriangleup
u)^{-1}}u_{i\bar
jk\delta}\\
 &\ \ \ \ \ \ \ -4\alpha\left(\sqrt{1-5\epsilon_1(\lambda_3+\alpha\bigtriangleup
u)^{-1}}\right)^{-1}\sum_p(u_{i\bar p\delta}u_{p\bar jk}+u_{i\bar
pk}u_{p\bar j\delta})g'^{p\bar p}\mid^2\\
&+ \frac{3\epsilon_1}{\lambda_3+\alpha\bigtriangleup
u}\Phi-C_6\Theta\Gamma-C_6\Gamma^2-\left(\frac{m_2/4+\epsilon_2}{\lambda_3+\alpha\bigtriangleup
u}+\frac {C_6
\epsilon_1}{\lambda_3+\alpha\bigtriangleup u-5\epsilon_1}\right)\Theta^2-C_7(\theta+\Gamma)\\
&\geq \frac{3\epsilon_1}{\lambda_3+\alpha\bigtriangleup
u}\Phi-C_6\Theta\Gamma-C_6\Gamma^2-\left(\frac{m_2/4+\epsilon_2}{\lambda_3+\alpha\bigtriangleup
u}+\frac {C_6 \epsilon_1}{\lambda_3+\alpha\bigtriangleup
u-5\epsilon_1}\right)\Theta^2-C_7(\Theta+\Gamma).
\end{aligned}
\end{equation}
 Inserting (\ref{827}) into
(\ref{813}),  at last we obtain
\begin{equation}\lab{829}
\begin{aligned}
&P((\lambda_3+\alpha\bigtriangleup
u)\Theta+\lambda_4(m+\alpha\bigtriangleup
u)\Gamma+\lambda_5\mid\bigtriangledown
u\mid^2\Gamma+\lambda_6\Gamma)\\
\geq &\left\{m_2-\frac {m_2} 4-\epsilon_2-\frac
{C_6}{\lambda_3+\alpha\bigtriangleup u}-C_6\epsilon_1\frac
{\lambda_3+\alpha\bigtriangleup u}
{\lambda_3+\alpha\bigtriangleup u-5\epsilon_1}\right\}\Theta^2\\
&+\left\{\lambda_4m_2-C_6-\frac
{C_6}{\lambda_3+\alpha\bigtriangleup
u}(\lambda_4+\lambda_5+\lambda_6)-C_6(\lambda_3+\alpha\bigtriangleup u)\right\}\Theta\Gamma\\
&+\left\{\lambda_5 m_1-\frac
{C_6\lambda_5}{\lambda_3+\alpha\bigtriangleup
u}-C_6\lambda_4-C_6(\lambda_3+\alpha\bigtriangleup u)\right\}\Gamma^2\\
&+\left\{\lambda_6m_3-C_6(\lambda_4+\lambda_5)-
\frac{C_6(\lambda_4+\lambda_5+\lambda_6)}{\lambda_3+\alpha\bigtriangleup
u}\right\}\Xi-C_7\Theta-C_7\Gamma-C_7.
\end{aligned}
\end{equation}
Note the generic constant $C_6$ does not depend on $\epsilon_i$
and $\lambda_i$. So we can fix it, because we can take the biggest
one. Fix $\epsilon_1$ and $\epsilon_2$ such that
$\epsilon_2+2C_6\epsilon_1<\frac{m_2}{4}$. Take $\lambda_3$ big
enough such that $\frac{C_6}{\lambda_3+\alpha\bigtriangleup
u}<\frac{m_2}{4}$ and $\frac {\lambda_3+\alpha\bigtriangleup u}
{\lambda_3+\alpha\bigtriangleup u-5\epsilon_1}<2$, then
\begin{equation}\lab{829}
\left\{m_2-\frac {m_2} 4-\epsilon_2-\frac
{C_6}{\lambda_3+\alpha\bigtriangleup u}-C_6\epsilon_1\frac
{\lambda_3+\alpha\bigtriangleup u} {\lambda_3+\alpha\bigtriangleup
u-5\epsilon_1}\right\}\Theta^2>\frac{m_2}{4}\Theta^2.
\end{equation}
Let
\begin{equation*}
\tilde{\lambda}_i=\frac{\lambda_i}{\lambda_3+\alpha\bigtriangleup
u} \ \ \ \ \text{for}\ \ \ i=4,5,6.
\end{equation*}
We  choose $\tilde{\lambda}_4$, $\tilde{\lambda}_5$ and
$\tilde{\lambda}_6$ such that
$$\tilde{\lambda}_4>\frac{C_6}{m_2}+1$$
$${\tilde{\lambda}_5}>\frac{C_6}{m_1}\tilde{\lambda}_4+\frac {C_6}{m_1}+1$$
and
$$\tilde{\lambda}_6>C_6\frac{\tilde{\lambda}_4+\tilde{\lambda}_5}{m_3}+1.$$
Then if we take $\lambda_3$ big enough such that
\begin{equation*}
 m_i(\lambda_3+\alpha\bigtriangleup u)-C_6(\tilde{\lambda}_4+\tilde{\lambda}_5+\tilde{\lambda}_6)-C_6>m_i,\ \
 \ \text{for}\ \ i=1,2,3,
\end{equation*}
we can estimate
\begin{equation}\lab{831}
\begin{aligned}
&\left\{\lambda_4m_2-C_6-\frac
{C_6}{\lambda_3+\alpha\bigtriangleup
u}(\lambda_4+\lambda_5+\lambda_6)-C_6(\lambda_3+\alpha\bigtriangleup u)\right\}\Theta\Gamma\\
\geq& \{m_2(\lambda_3+\alpha\bigtriangleup
u)-C_6(\tilde{\lambda}_4+\tilde{\lambda}_5+\tilde{\lambda}_6)-C_6\}\Theta\Gamma>
m_2\Theta\Gamma;
\end{aligned}
\end{equation}
\begin{equation}\lab{832}
\begin{aligned}
&\left\{\lambda_5 m_1-\frac
{C_6\lambda_5}{\lambda_3+\alpha\bigtriangleup
u}-C_6\lambda_4-C_6(\lambda_3+\alpha\bigtriangleup
u)\right\}\Gamma^2\\
&>\{m_1(\lambda_3+\alpha\bigtriangleup
u)-C_6\tilde{\lambda}_5\}\Gamma^2
>m_1\Gamma^2
\end{aligned}
\end{equation}
and
\begin{equation}\lab{833}
\begin{aligned}
&\left\{m_3\lambda_6-\frac{C_6}{\lambda_3+\alpha\bigtriangleup
u}(\lambda_4+\lambda_5+\lambda_6)-C_6(\lambda_4+\lambda_5)\right\}\Xi\\
&>\{m_3(\lambda_3+\alpha\bigtriangleup
u)-C_6(\tilde{\lambda}_4+\tilde{\lambda}_5+\tilde{\lambda}_6)\}\Xi>m_3\Xi.
\end{aligned}
\end{equation}
Inserting (\ref{829}), (\ref{831})-(\ref{833}) into (9.28), we see
that
\begin{equation*}\begin{aligned}
0&\geq P((\lambda_3+\alpha\bigtriangleup
u)\Theta+\lambda_4(m+\alpha\bigtriangleup
u)\Gamma+\lambda_5\mid\bigtriangledown u\mid^2\Gamma+\lambda_6\Gamma)\\
&\geq\frac{m_2}{4}\Theta^2+m_2\Theta\Gamma+
m_1\Gamma^2+m_3\Xi-C_7\Theta-C_7\Gamma-C_7.
\end{aligned}
\end{equation*}
Above inequality gives an estimate of the the quantity $\sup_S
\Theta$ and $\sup_S \Gamma$. This in turn gives the estimates of
$u_{i\bar jk}$ and $u_{ij}$ for all $i,j,k$.
\begin{prop}\lab{prop 5}
Let $\omega_S$ be a given Calabi-Yau metric on a $K3$ surface with
$\int_S 1\frac{\omega_S^2}{2!}=1$.  Let $t\in {\bf T}$ and $u\in
C^5(S)$ is a solution of the equation $
\bigtriangleup(e^{u}-t\alpha fe^{-u})+8\alpha \frac{\det u_{i\bar
j}}{\det g_{i\bar j}}+t\mu=0$ under the elliptic condition
$\omega'=(e^u+t\alpha
fe^{-u})\omega_S+2\alpha\sqrt{-1}\partial\bar\partial u>0$ and the
normalization $\left(\int_S e^{-4u}\right)^{\frac 14}=A<<1$ (see
(\ref{668}) and (\ref{669})). Then there is an estimate of the
derivatives $u_{i\bar j k}$ in terms of $\alpha$, $f$, $\mu$,
$\omega_S$ and $A$.
\end{prop}

\section{Estimates for the general case}
In general case, the equation is
\begin{equation*}
\sqrt{-1}\partial\bar\partial e^u\wedge
\omega_S-t\alpha\partial\bar\partial(e^{-u}\text{tr}(\bar\partial
B\wedge
\partial B^*\cdot g^{-1}))-\alpha\partial\bar\partial
u\wedge\partial\bar\partial u +t\mu\frac {\omega_S^2}{2!}=0.
\end{equation*}
Let
\begin{equation*}
\rho=-\sqrt{-1}\text{tr}(\bar\partial B\wedge
\partial B^*\cdot g^{-1}),
\end{equation*}
then $\rho$ is a  well-defined real $(1,1)$-form on $S$. We
replace $t\alpha\rho$ by $\rho$ and $t\mu$ by $\mu$. Then we can
rewrite the equation as
\begin{equation*}
\sqrt{-1}\partial\bar\partial e^u\wedge
\omega_S-\sqrt{-1}\partial\bar\partial(e^{-u}\rho)-\alpha\partial\bar\partial
u\wedge\partial\bar\partial u +\mu\frac {\omega_S^2}{2!}=0.
\end{equation*}
The elliptic condition is
\begin{equation*}
\omega'=e^u\omega_S+e^{-u}\rho+2\alpha\sqrt{-1}\partial\bar\partial
u>0.
\end{equation*}
If we let $\rho=\frac{\sqrt{-1}}{2}\rho_{i\bar j}dz_i\wedge d\bar
z_{ j}$, then $g'_{i\bar j}=e^ug_{i\bar j}+e^{-u}\rho_{i\bar
j}+4\alpha u_{i\bar j}$. Using the definition of $P$ and the
equation, we compute
\begin{equation*}
\begin{aligned}
&\int_SP(e^{-ku})\frac {\omega'^2}{2!}\geq
-k\int_Se^{-ku}P(u)\frac{\omega'^2}{2!}\\
=& -\sqrt{-1}k\int_S e^{-ku}\partial\bar\partial
u\wedge(e^u\omega_S+e^{-u}\rho+2\alpha\sqrt{-1}\partial\bar\partial
u)\\
=&-k\int_Se^{-(k-1)u}\bigtriangleup
u-\sqrt{-1}k\int_Se^{-(k+1)}\partial\bar\partial u\wedge
\rho+2k\int_Se^{-ku}\bigtriangleup
e^u\\
&-2k\sqrt{-1}\int_Se^{-ku}\partial\bar\partial(e^{-u}\rho)
+2k\int_Se^{-ku}\mu\\
=&k\int_Se^{-(k-1)u}\bigtriangleup
u+2k\int_Se^{-(k-1)u}\mid\bigtriangledown u\mid^2+\sqrt{-1}k\int_S
e^{-(k+1)u}\partial\bar\partial u\wedge \rho\\
&-2\sqrt{-1}k\int_Se^{-(k+1)u}\partial u\wedge \bar\partial
u\wedge \rho+2\sqrt{-1}k\int_Se^{-(k+1)u}\partial
u\wedge\bar\partial
\rho\\
&-2\sqrt{-1} k\int_Se^{-(k+1)u}\bar\partial u\wedge\partial
\rho-2\sqrt{-1}k\int_S e^{-(k+1)u}\partial\bar\partial
\rho+2k\int_S e^{-ku}\mu.
\end{aligned}
\end{equation*}
On the other hand, we can also compute
\begin{equation*}
\begin{aligned}
&\int_SP(e^{-ku})\frac{\omega'^2}{2!}=\sqrt{-1}\int_S\partial\bar\partial
e^{-ku}\wedge \omega'\\
=&\sqrt{-1}\int_S\partial\bar\partial e^{-ku}\wedge
(e^u\omega_S+e^{-u}\rho+2\sqrt{-1}\alpha\partial\bar\partial u)\\
=&-k\int_Se^{-(k-1)u}\bigtriangleup
u+k^2\int_Se^{-(k-1)u}\mid\bigtriangledown u\mid^2\\
&-\sqrt{-1} k\int_Se^{-(k+1)u}\partial\bar\partial u\wedge
\rho+\sqrt{-1}k^2\int_Se^{-(k+1)u}\partial u\wedge\bar\partial
u\wedge \rho.
\end{aligned}
\end{equation*}
Combing above two inequalities, we get
\begin{equation*}
\begin{aligned}
&k\int_Se^{-(k-1)u}\mid\bigtriangledown u\mid^2
+\sqrt{-1}k\int_Se^{-(k+1)u}\partial u\wedge\bar\partial u\wedge
\rho\\
\geq &2\int_Se^{-(k-1)u}\bigtriangleup
u+2\int_Se^{-(k-1)u}\mid\bigtriangledown u\mid^2+2\sqrt{-1}\int_S
e^{-(k+1)u}\partial\bar\partial u\wedge\rho\\
&-2\sqrt{-1}\int_Se^{-(k+1)u}\partial u\wedge\bar\partial u\wedge
\rho+2\sqrt{-1}\int_Se^{-(k+1)u}\partial u\wedge\bar\partial
\rho\\
&-2\sqrt{-1} \int_Se^{-(k+1)u}\bar\partial u\wedge\partial
\rho-2\sqrt{-1}\int_S e^{-(k+1)u}\partial\bar\partial
\rho+2\int_Se^{-ku}\mu.
\end{aligned}
\end{equation*}
Integrating by part and then simplifying it, when $k\geq 2$, we
get
\begin{equation}\lab{1008}
\begin{aligned}
&k\int_Se^{-(k-1)u}\mid\bigtriangledown
u\mid^2+\sqrt{-1}k\int_Se^{-(k+1)u}\partial u\wedge\bar\partial
u\wedge\rho\\
\leq &2\sqrt{-1}(1-\frac 1{1+k})\int_Se^{-(k+1)u}\partial \bar
\partial\rho+2k\int_Se^{-ku}\mu.
\end{aligned}
\end{equation}
Using the notation in section 3, we have
\begin{equation*}
\begin{aligned}
\rho=&-\sqrt{-1}\text{tr}(\bar\partial B\wedge \partial B^*\cdot
g^{-1})\\
=&-\sqrt{-1}\text{tr}\left(\begin{array}{c} \bar\partial f_1\\
\bar\partial f_2\end{array}\right)\wedge
\begin{array}{cc}(\partial \bar f_1 &
\partial \bar f_2)\end{array}\cdot\left(\begin{array}{cc}g^{1\bar 1}&
g^{2\bar 1}\\ g^{1\bar 2}& g^{2\bar 2}\end{array}\right)\\
=&\sqrt{-1}g^{i\bar j}\frac{\partial f_i}{\partial\bar z_l}\cdot
\overline{\frac{\partial f_j}{\partial \bar z_k}}dz_k\wedge d\bar
z_l
\end{aligned}
\end{equation*}
and
\begin{equation*}
\begin{aligned}
&\sqrt{-1}\partial u\wedge\bar\partial
u\wedge\rho\\
=&\frac{4}{\det g_{i\bar j}}g^{i\bar j}\left\{u_1u_{\bar
1}\frac{\partial f_i}{\partial \bar z_1}\overline{\frac{\partial
f_j}{\partial\bar z_1}}-u_1u_{\bar 2}\frac{\partial f_i}{\partial
\bar z_1}\overline{\frac{\partial f_j}{\partial\bar
z_2}}-u_2u_{\bar 1}\frac{\partial f_i}{\partial \bar
z_2}\overline{\frac{\partial f_j}{\partial\bar z_1}}+u_2u_{\bar
2}\frac{\partial f_i}{\partial \bar z_2}\overline{\frac{\partial
f_j}{\partial\bar z_2}}\right\}\frac{\omega_S^2}{2!}\\
=&4(\begin{array}{cc}u_1& u_2\end{array})\cdot
\left(\begin{array}{cc}\frac{\partial f_1}{\partial\bar
z_1}&\frac{\partial f_2}{\partial\bar z_1}\\ \frac{\partial
f_1}{\partial\bar z_2}&\frac{\partial f_2}{\partial\bar
z_2}\end{array}\right)\cdot
g\cdot\left(\begin{array}{cc}\frac{\partial
f_1}{\partial\bar z_1}&\frac{\partial f_2}{\partial\bar z_1}\\
\frac{\partial f_1}{\partial\bar z_2}&\frac{\partial
f_2}{\partial\bar z_2}\end{array}\right)^*\cdot
\left(\begin{array}{c} u_{\bar 1}\\ u_{\bar
2}\end{array}\right)\frac {\omega_S^2}{2!}.
\end{aligned}
\end{equation*}
So
\begin{equation*}
\sqrt{-1}k\int_Se^{-(k+1)u}\partial u\wedge\bar\partial
u\wedge\rho\geq 0.
\end{equation*}
Then  (\ref{1008}) implies the inequality (\ref{5009}) in section
6:
\begin{equation*}
\begin{aligned}
k\int_Se^{-(k-1)u}\mid\bigtriangledown u\mid^2 \leq&
2\sqrt{-1}(1-\frac 1{1+k})\int_Se^{-(k+1)u}\partial \bar
\partial\rho+2\int_Se^{-ku}\mu\\
\leq&C_0\int_Se^{-(k+1)u}+C_0\int_Se^{-ku}.
\end{aligned}
\end{equation*}
We follow the discussion in section 6 to get the estimate $\inf
u\geq -\ln(C_1A)$. If $A$ is small enough, we can get $\inf u$ big
enough. Then we can check  all other estimates can be derived
using the same method because the term $e^u$ can always control
terms such as $e^{-u}\mid \text{tr}(\bar \partial B\wedge
\partial B^*\cdot g)\mid$. Thus we get
\begin{prop}
Proposition 20, 21, 22 are also true for the equation of general
case:
\begin{equation}\lab{1002}
\sqrt{-1}\partial\bar\partial e^u\wedge
\omega_S-t\alpha\partial\bar\partial(e^{-u}\textup{tr}(\bar\partial
B\wedge
\partial B^*\cdot g^{-1}))-\alpha\partial\bar\partial
u\wedge\partial\bar\partial u +t\mu\frac {\omega_S^2}{2!}=0
\end{equation}
if we replace $f$ by $-\sqrt{-1}\textup{tr}(\bar\partial B\wedge
\partial B^*\cdot g^{-1})$.
\end{prop}
\begin{prop}
Proposition 23 is also true for the  equation (\ref{1002}).
\end{prop}

\section{Further remark--generalization}
Let $X$ be a (n+1)-dimensional complex manifold with Hermitian
metric $\omega$ and a nowhere vanishing holomorphic $(n+1,0)$-form
$\Omega$. As we state in the introduction, the string theorists
consider the following Strominger's  system:
\begin{equation}\label{1101}
F_H\wedge\omega^{n}=0; \ \ \  F^{2,0}_H=F^{0,2}_H=0;
\end{equation}
\begin{equation}\label{1102}
\sqrt{-1}\partial\dbar \omega= \frac{\alpha'}{4}(\text{tr} R\wedge
R-\text{tr}F_H\wedge F_H);
\end{equation}
\begin{equation}\label{1103}
d\sta \omega=\sqrt{-1}(\dbar-\partial)\ln\|\Omega\|_\omega.
\end{equation}
The third equation is equivalent to
\begin{equation}\label{1104}
d(\parallel\Omega\parallel_{\omega}\omega^{n})=0.
\end{equation}
Let $n\geq 2$. Motivated by the constructions in section 2 nd 4,
we propose to study the following system
\begin{equation}\label{1105}
F_H\wedge\omega^{n}=0; \ \ \  F^{2,0}_H=F^{0,2}_H=0;
\end{equation}
\begin{equation}\label{1106}
\left\{\sqrt{-1}\partial\dbar \omega-\frac{\alpha'}{4}(\text{tr}
R\wedge R-\text{tr}F_H\wedge F_H)\right\}\wedge\omega^{n-2}=0;
\end{equation}
\begin{equation}\label{1107}
d(\parallel\Omega\parallel^{2\frac{n-1}{n}}_{\omega}\omega^{n})=0.
\end{equation}

Then we can generalize our construction to complex manifolds with
$\dim\geq 3$.
 Let
$K$ be a Calabi-Yau $n$-fold with a Ricci-flat metric $\omega_K$
and a nowhere vanishing holomorphic $(n,0)$-form $\Omega_K$. Let
$\omega_1,\omega_2$ be a primitive harmonic $(1,1)$-forms such
that $\frac{\omega_1}{2\pi},\frac{\omega_2}{2\pi}\in
H^{1,1}(K,\mathbb{Z})$. Using these two forms, we can construct
an $(n+1)-$dimensional  complex manifold $X$: \\
1. $\pi:X\rightarrow K$ is a $T^2$-fibration over $K$. If we write
locally $\omega_1=d\alpha_1$ and $\omega_2=d\alpha_2$ for real
$1$-forms $\alpha_1$ and $\alpha_2$, then there is a coordinate
that $x$ and $y$ of fiber $T^2$ such that $dx+\sqrt{-1}dy$ is a
holomorphic $1$-form on $T^2$-fibers  and $dx+\alpha_1$ and
$dy+\alpha_2$ are
globally defined 1-forms on $X$. \\
2.
 Let
 \begin{equation*}
 \theta=(dx+\alpha_1)+\sqrt{-1}(dy+\alpha_2)
  \end{equation*}
  and
  let $$\Omega=\Omega_K\wedge\theta.$$
  Then $\Omega$ defines a nowhere vanishing
holomorphic $(n+1,0)$-form on $X$.\\
3. Let $u\in C^2(K)$ function on $K$ and
\begin{equation}
\omega_u=e^u\omega_K+\frac{\sqrt{-1}}{2}\theta\wedge\bar\theta.
\end{equation}
Then $(\Omega,\omega_u)$ satisfies equation (\ref{1107}).

As in section 4, we have
\begin{equation*}
\parallel\Omega\parallel^2_{\omega_u}
=\frac{\parallel\Omega\parallel^2_{\omega_u}}{\parallel\Omega\parallel^2_{\omega_0}}
=\frac{\omega_0^n}{\omega^n_u}=e^{-nu},
\end{equation*}
and
\begin{equation*}
\omega_u^{n}=e^{nu}\omega_K^{n}+\sqrt{-1}ne^{(n-1)u}
\omega_K^{n-1}\wedge\theta\wedge \bar\theta.
\end{equation*}
Then \begin{eqnarray*}
d(\parallel\Omega\parallel^{2\frac{n-1}{n}}\omega^{n})
&=&d(e^u\omega_K^{n})+\sqrt{-1}nd(\omega_K^{n-1}\wedge\theta\wedge\bar\theta)\\
&=&\sqrt{-1}n\omega_K^{(n-1)}(\wedge(\omega_1+\sqrt{-1}\omega_2)\wedge\bar\theta+\theta\wedge
(\omega_1-\sqrt{-1}\omega_2))=0,
\end{eqnarray*}
as $\omega_1,\omega_2$ are  primitive $(1,1)$-forms on $K$. So
$(\Omega,\omega_u)$ satisfies equation (\ref{1107}).

As $\omega_1,\omega_2$ are harmonic,  we can find $(1,0)$-forms
$\xi_1=\sum_{i=1}^{n}\xi_{1i}dz_i$  and
$\xi_2=\sum_{i=1}^{n}\xi_{2i}dz_i$, locally  where $\xi_{1i}$ and
$\xi_{2i}$ are smooth complex function on some open set of $K$,
such that $\omega_1=\bar\partial\xi_1$ and
$\omega_2=\bar\partial\xi_2$. Let
$$\phi_i=\xi_{1i}+\xi_{2i},\ \ \ \ \ \ \text{for}\ \ \ j=1,2,\cdots
n,$$ and let
$$B=(\phi_1,\phi_2,\cdots,\phi_{n}).$$
Let $R_u$ be the curvature of Hermitian connection of metric
$\omega_u$ of the holomorphic tangent bundle $T'X$ and $R_K$ be
the curvature of metric $\omega_K$. Then in section 3, we have
\begin{equation*}
\text{tr}R_u\wedge R_u=\text{tr}R_K\wedge
R_K+2\partial\bar\partial(e^{-u}\bar\partial B\wedge\partial
B^*\cdot g^{-1})+n\partial\bar\partial u\wedge\partial\bar\partial
u,
\end{equation*}
where $g$ is the Calabi-Yau metric associated to K\"ahler form
$\omega_K$. Let $E$ be the stable vector bundle over
$(K,\omega_K)$ with degree zero. According to the Uhlenbeck-Yau
theorem, there is a unique Hermitian-Yang-Mills metric $H$ up to
constants. Hence
$$(\pi^*E,\pi^*H,X,\omega_u)$$ satisfies the equation (\ref{1105}) and (\ref{1107}).
So we  only need to consider equation (\ref{1106}), which can be
decomposed to the following two equations
\begin{equation}
\frac{(n-2)!}{2}\int_K(\parallel\omega_1\parallel^2_{\omega_K}
+\parallel\omega_2\parallel^2_{\omega_K})\frac{\omega_K^{n}}{n!}
+\frac{\alpha'}{4}\int_K\text{tr}(F_H\wedge F_H-R_K\wedge
R_K)\wedge\omega_K^{n-2}=0
\end{equation}
and
\begin{equation}
\sqrt{-1}\partial\bar\partial
u\wedge\omega_K^{n-1}-2\partial\bar\partial(e^{-u}\text{tr}\bar\partial
B\wedge \partial B^*)\wedge K^{n-2}-n\partial\bar\partial u\wedge
\partial\bar\partial u\wedge K^{n-2}+\mu\frac{\omega_K^{n}}{n!}=0,
\end{equation}
where $\mu$ is a smooth function on $K$ and $\int_K
\mu\frac{\omega_K^{n}}{n!}=0$. In the next paper, we will continue
to consider this problem.

\end{document}